\documentclass[prd,nofootinbib,floats,aps,twocolumn,tightenlines,superscriptaddress,floatfix]{revtex4-1}

\usepackage{graphicx}
\usepackage[caption=false]{subfig}

\usepackage{dcolumn}
\usepackage{bm}
\usepackage{hyperref}

\usepackage{epsf,epsfig,amsmath,amssymb,dsfont}
\usepackage{amsthm,mathrsfs} 
\usepackage{multirow}
\usepackage{float}
\usepackage{colortbl}
\usepackage{float}
\usepackage{bm}
\usepackage{array}
\usepackage{mathtools}
\usepackage[normalem]{ulem}

\usepackage{comment}

\allowdisplaybreaks

\newcommand{\integral}[3]{\int_{#2}^{#3} \!\! \mathrm{d} #1 \,}
\newcommand{\difffrac}[2]{\frac{\mathrm{d} #1}{\mathrm{d} #2}}
\newcommand{\partialfrac}[2]{\frac{\partial #1}{\partial #2}}
\newcommand{\diff}{\text{d}}

\newcommand{\pv}{\text{PV}}
\newcommand{\ft}{\mathcal{F}}

\newcommand{\lP}{\left(}
\newcommand{\rP}{\right)}

\newcommand{\ci}{\text{Ci}}
\newcommand{\si}{\text{Si}}

\newcommand{\ii}{\mathrm{i}}
\newcommand{\ee}[1]{\mathrm{e}^{#1}} %

\newcommand{\sinc}{\mathrm{sinc}}

\newcommand{\GlCID}{g_\ell}

\newcommand{\tr}{\operatorname{Tr}}
\newcommand{\id}{\mathbb{I}}

\newcommand{\intH}{H_{\text{int}}}

\newcommand{\exptval}[1]{\left< {#1} \right>}
\newcommand{\comm}[2]{\left[{#1},{#2}\right]}

\newcommand{\ket}[1]{\left| {#1} \right\rangle}
\newcommand{\bra}[1]{\left\langle {#1} \right|}

\newcommand{\ketbra}[2]{\left| {#1}\vphantom{#2} \right\rangle\!\left\langle {#2}\vphantom{#1} \right|}
\newcommand{\nn}{\nonumber\\}

\newcommand{\metricN}{N}

\newcommand{\da}{\textsc{a}}
\newcommand{\db}{\textsc{b}}
\newcommand{\dd}{\textsc{d}}

\newcommand{\dket}[2]{\ket{#1_\textsc{#2}}}

\newcommand{\dketbra}[3]{\ket{\vphantom{#2}#1_\textsc{#3}}\!\bra{\vphantom{#1}#2_\textsc{#3}}}

\newcommand{\coord}[1]{\mathsf{#1}}
\newcommand{\spacoord}[1]{\bm{#1}}

\begin{document}

\title{Communication through quantum fields near a black hole} 

\author{Robert H. Jonsson}
\affiliation{Max Planck Institute of Quantum Optics, Hans-Kopfermann-Str. 1, 85748 Garching, Germany}
\affiliation{QMATH, Department of Mathematical  Sciences,  University  of  Copenhagen, Universitetsparken  5,  2100  Copenhagen,  Denmark}
\affiliation{Microtechnology and Nanoscience, MC2, Chalmers University of Technology, SE-412 96 G\"oteborg, Sweden}
\affiliation{Department of Applied Mathematics, University of Waterloo, Waterloo, Ontario, N2L 3G1, Canada}
\author{David Q. Aruquipa}
\affiliation{Centro Brasileiro de Pesquisas F\'isicas (CBPF), Rio de Janeiro, 
CEP 22290-180, 
Brazil.}
\author{Marc Casals}
\affiliation{Centro Brasileiro de Pesquisas F\'isicas (CBPF), Rio de Janeiro, 
CEP 22290-180, 
Brazil.}
\affiliation{School of Mathematics and Statistics, University College Dublin, Belfield, Dublin 4, Ireland.}
\author{Achim Kempf}
\affiliation{Institute for Quantum Computing, University of Waterloo, Waterloo, Ontario, N2L 3G1, Canada}
\affiliation{Department of Physics \& Astronomy, University of Waterloo, Waterloo, Ontario, Canada, N2L 3G1}
\affiliation{Perimeter Institute for Theoretical Physics, 31 Caroline St N, Waterloo, Ontario, N2L 2Y5, Canada}
\affiliation{Department of Applied Mathematics, University of Waterloo, Waterloo, Ontario, N2L 3G1, Canada}
\author{Eduardo Mart\'{i}n-Mart\'{i}nez}
\affiliation{Institute for Quantum Computing, University of Waterloo, Waterloo, Ontario, N2L 3G1, Canada}
\affiliation{Department of Physics \& Astronomy, University of Waterloo, Waterloo, Ontario, Canada, N2L 3G1}
\affiliation{Perimeter Institute for Theoretical Physics, 31 Caroline St N, Waterloo, Ontario, N2L 2Y5, Canada}
\affiliation{Department of Applied Mathematics, University of Waterloo, Waterloo, Ontario, N2L 3G1, Canada}

\begin{abstract}

We study the quantum channel between two localized first-quantized systems that communicate in 3+1 dimensional Schwarzschild spacetime via a quantum field. We analyze the information carrying capacity of direct and black hole-orbiting null geodesics as well as of the timelike contributions that arise because the strong Huygens principle does not hold on the Schwarzschild background. We find, in particular, that the  non-direct-null and timelike contributions, which do not possess an analog on Minkowski spacetime, can dominate over the direct null contributions. We cover the cases of both  geodesic and accelerated emitters. Technically, we apply tools previously designed for the study of wave propagation in curved spacetimes to a relativistic quantum information communication setup, first for generic spacetimes, and then for the case of Schwarzschild spacetime in particular.

\end{abstract}

\maketitle

\section{Introduction}
Spatially localized first-quantized systems that temporarily  couple to a quantum field have been used 
extensively in a plethora of contexts in quantum field theory in flat and curved spacetimes. They are useful, in particular, to describe the space-time localized absorption of field quanta and are, therefore, also known as `particle detectors'. 
As such, they provide, e.g., a useful operational formulation of the Unruh effect, see, e.g.,  \cite{Takagi,UnruhWald,achimaidaunruh,withouttherm} and help clarify its relationship with other similar phenomena, such as the Gibbons-Hawking effect \cite{Gibbons1977}. In particular, the ubiquitous Unruh-DeWitt model \cite{dewitt_quantum_1979} simplifies the detector to a classically-localized 2-level system and yet still captures most of the fundamental features of the light-matter interaction between atoms and molecules with the quantum electromagnetic field \cite{martin-martinez_relativistic_2018,Pozas2016}.

In the context of curved spacetimes, particle detectors allow one to better understand the notion of measurement in quantum field theory \cite{Earman2011,fewster2018} and have proven a powerful tool to define the elusive notion of particle in quantum field theory \cite{UnruhWald}. For example, particle detectors have been used in a number of curved spacetimes scenarios to characterize the particle content of different vacuum states. The applications of particle detectors range from cosmology \cite{Gibbons1977} to black hole scenarios such as Schwarzschild and Schwarzschild-AdS spacetimes (e.g., see  \cite{Jorma,KeithResponse}). 
 They have  also been used to study the entanglement structure of quantum field theory vacua both in flat spacetimes \cite{Valentini1991,Reznik1,Pozas2015}, cosmological backgrounds \cite{Nick,Kukita_2017} as well as in other simple curved scenarios such as the Anti-deSitter spacetimes \cite{Laura2,Keith2018} or in the presence of BTZ black holes \cite{Laura1}. 

Of course, first-quantized spacetime-localized systems can not only detect but also emit particles. Correspondingly, 
there have been a number of recent studies analyzing communication using particle detectors coupling to quantum fields, starting with \cite{cliche_relativistic_2010}, both in flat spacetime \cite{jonsson_quantum_2014,jonsson_information_2015,jonsson_information_2016,jonsson_quantum_2017,jonsson_transmitting_2018,landulfo_nonperturbative_2016,Petar2019,shockwavepaper}  and curved \cite{blasco_violation_2015,simidzija_information_2017} spacetimes. Among the results, it was shown, for example, that if there are multiple emitters, the choice of their entanglement can help shape their radiation field \cite{shockwavepaper}. 
Using the perhaps unfamiliar property that massless fields  propagate not only on the light cone but also via timelike paths (i.e., also at less than the speed of light) namely when the strong Huygens principle  \cite{mclenaghan_explicit_1969,mclenaghan_validity_1974,czapor_hadamards_2007} is violated in a curved spacetime,  the above studies showed that, in such spacetimes, particle detectors can communicate via timelike --as well as null-- signals carried by a massless quantum field.

In the present paper, we introduce spacetime curvature to the communication channel between particle detectors and we then focus on the special case of communication near a (Schwarzschild) black hole. The main challenge here is in the complexity of the calculations involved with the evaluation of two point functions in curved spacetime.

To address the complexity of the problem  we will employ a combination of traditional as well as new techniques which we next describe. As has been shown in~\cite{jonsson_quantum_2014,jonsson_information_2015,jonsson_quantum_2017}, to leading order in the coupling constant, the two-point function that the signal strength depends on is the retarded Green function. This is a classical Green function in the sense that it does not
depend on the quantum state of the field. In  recent years, several methods have been developed for, and applied to, the full calculation of the retarded Green function  in Schwarzschild space-time.

When the two spacetime  points are ``close", we calculate the retarded Green function using Hadamard form~\cite{Hadamard} techniques (e.g.,~\cite{Ottewill:2009uj,Ottewill:Wardell:2008,CDOWb}), which have already been used in the literature.

When the points are ``far" from each other, on the other hand, the Hadamard form is not valid and another method must be used for the calculation of the retarded Green function.
For example, it can be calculated
semi-analytically via a Fourier integral over real frequencies in~\cite{BUSS2018168} (where the quantum Feynman Green function was also calculated);
semi-analytically by deforming the Fourier integral into a contour on the complex-frequency plane (thus involving, in particular, a sum over quasinormal mode frequencies) in~\cite{Dolan:2011fh,CDOW13}; 
numerically via an approximation of a Dirac-delta distribution (appearing either in the source of the field equation or as initial data) 
by a narrow Gaussian distribution in~\cite{Zenginoglu:2012xe,PhysRevD.89.084021}.
In this paper, however, we follow a method introduced in~\cite{mark2017recipe} which we here apply,  for the first time to the best of our knowledge, to the full calculation of the retarded Green function when the points are not close. The method essentially consists of  the following. We first carry out a multipolar $\ell$-mode decomposition of the retarded Green function. The resulting $\ell$-modes satisfy a $(1+1)$-dimensional partial differential equation with  known characteristic initial data. We solve this  characteristic initial data problem using the finite difference scheme introduced in~\cite{Lousto:1997wf}, which we here develop to a higher order.
This calculation of the retarded Green function when the points are not close (using the characteristic initial data scheme) should agree with the calculation when the points are close (using Hadamard form techniques) in some region of overlap.
For this purpose, we further enhance our calculation
for  far-away points with the technique that has been recently introduced in~\cite{ModesDirectPart}.
This technique consists of subtracting from the  $\ell$-modes of the retarded Green function the $\ell$-modes of the divergence at coincident points that explicitly appears in the Hadamard form.
As a consequence, the calculation for far-away points 
becomes valid at closer distances and the region in Schwarzschild space-time where the desired region of overlap between methods exists is greatly increased (with respect to not using the technique introduced in~\cite{ModesDirectPart}).

For the case of the Schwarzschild black hole, we will show that communication is mediated by three different contributions: primary null light rays propagating directly from the sender to the receiver, secondary (and higher order) null light rays that orbit around the black hole before reaching the receiver, and timelike contributions to the communication that are due to violations of the strong Huygens principle on curved spacetime. 
We will study separately the strength of all three of these contributions to the signalling, as functions 
of the separation of the sender and receiver, their state of motion (static versus infalling), and of their distance from the event horizon. 

In particular, we will find that the strength of the non-direct  signalling contribution that is due to the violation of the strong Huygens principle in curved spacetime and that, therefore, possesses no analog in flat spacetime, can exceed the direct contributions that correspond to the usual null geodesics between sender and receiver. 

We will also find, for example, that when a static receiver is chosen to be increasingly close to the event horizon of the black hole, the receiver becomes less and less able to recover information from a sender further out, even when compensating for the blueshift of the sender's signal. This phenomenon is related to the fact that the proper time that the receiver has to resonate  (and thereby build up amplitude) with the blue-shifted signal diminishes with increasing blueshift of the sender's signal.

The paper is organized as follows: Sec.~\ref{sec:signaling_with_detectors} reviews communication between Unruh-DeWitt particle detectors, and gives a discussion of the quantum nature of signals between detectors and time dilation effects on them.
Sec.~\ref{sec:gf} reviews analytical features of the Green function and presents the methods used to efficiently evaluate it in the scenarios we consider.
Sec.~\ref{sec:staticspacetimes} discusses general features of signaling between particle detectors in static spacetimes.
This leads up to Sec.~\ref{sec:schwarzschild_static} where we calculate and analyze signaling between static detectors in the vicinity of a Schwarzschild black hole.
Furthermore, in Sec.~\ref{sec:infall} we analyze communication between a sender falling towards the black hole signaling to a static receiver outside of the black hole. We close with a summarizing discussion in Sec.~\ref{sec:discussion}.

We will use natural units ($c=G=\hbar=1$) and we denote the line-element of Schwarzschild spacetime in Schwarzschild coordinates
$\{t\in\mathbb{R},\, r\in (0,\infty),\, \theta\in [0,\pi],\, \varphi\in [0,2\pi)\}$ by
\begin{equation}
\diff s^2=-f \diff t^2+f^{-1}\diff r^2+r^2\left(\diff\theta^2+\sin^2\theta \diff\varphi^2\right),
\end{equation}
where $f:=1-2M/r$.

\section{Signaling with Unruh-DeWitt detectors}\label{sec:signaling_with_detectors}

To model the communication devices of sender and receiver, we use the Unruh-DeWitt particle detector model \cite{unruh_notes_1976,dewitt_quantum_1979}. The model was originally introduced to model the interaction of local observers with relativistic quantum fields in curved spacetime, and has since been widely used to model the light-matter interaction in quantum optics and in the general context of relativistic quantum information. 
It captures the relevant features of light-matter interaction when angular momentum exchange plays a negligible role in the detector's dynamics \cite{martin-martinez_relativistic_2018}.

The following three subsections review the detector model, the quantum channel between detectors and its leading order signal strength and introduce notation. 
The last two subsections concern aspects which, to the best of our knowledge, were not discussed in the literature before:
Sec.~\ref{sec:class_vs_quantum_signal} addresses the question of to what extent the leading order signal strength should be viewed as a quantum or classical effect; Sec.~\ref{sec:time_dilation_effects} addresses the impact of time dilation on the signal strength between detectors.

\subsection{Detector model and perturbative coupling}
The particle detector can  be viewed as modeling an atom  moving along   a predetermined  trajectory. Along its worldline, it couples to a background quantum field.
More specifically, for the purposes of this paper, we choose the detector as a two-level quantum system with Hamiltonian
\begin{align}
    H^{\tau_\textsc{d}}_\dd&= \frac{\Omega_\dd}2\left(\ketbra{e_\dd}{e_\dd}-\ketbra{g_\dd}{g_\dd}\right),
\end{align}
 where the superindex $\tau_\textsc{d}$ denotes that the Hamiltonian generates translations with respect to  the detector's proper time $\tau_\textsc{d}$. The states $\ket{g_\dd}$ and $\ket{e_\dd}$ (modelling, repectively, ground and excited  states of an atom) are orthogonal, and $\Omega_\dd\geq0$  is the energy gap of the detector.
We choose the quantum field to be a massless scalar Klein-Gordon field.
The interaction between the detector and the field is given by coupling the monopole operator of the detector to the field amplitude operator $\phi(\coord{x})$ along the worldline   of the detector, where $\coord{x}$ is a spacetime point.
Therefore, in the interaction picture, the interaction Hamiltonian (generating translations with respect to the detector's proper time  $\tau_\dd$) reads 
\begin{align}\label{eq:udwhamiltonian}
    H_{\text{int,\dd}}^{\tau_\textsc{d}}(\tau_\dd) &= \lambda_\dd \eta_\dd(\tau_\dd) \phi(\coord{x}_\dd(\tau_\dd)) \nn 
    &\qquad \otimes\left(\ketbra{e_\dd}{g_\dd} \ee{\ii \Omega_\dd \tau_\dd} +\ketbra{g_\dd}{e_\dd}\ee{-\ii\Omega_\dd\tau_\dd}\right).
\end{align}
Here, $\lambda_\dd$ is a coupling constant setting the overall strength of the coupling  between the detector and the field. In  (3+1)-dimensional spacetimes, $\lambda_\dd$ is dimensionless. 
The switching function $\eta_\dd(\tau_\dd)$ determines  when the detector couples to the field. It takes real values in $0\leq \eta_\dd \leq1$, is  generally assumed to be smooth and to be compactly supported. (However, as discussed later, we will not need to assume smoothness for the purpose of this article.) 
Finally, $\coord{x}_\dd(\tau_\dd)$ denotes the detector's worldline, parametrized by its own proper time.

The coupling between detector and field is assumed to be weak  enough so that  time-dependent perturbation theory applies.

\subsection{Quantum channel between detectors}

In order to study signaling, we equip both  Alice, the sender, and Bob, the  receiver, with particle detectors as communication devices. Alice and Bob can only prepare and measure their detectors, and control the coupling to the field through their switching functions. Other than that, they have no direct access to the field's observables.

In this setup, \cite{cliche_relativistic_2010}, to encode a message, Alice prepares her detector in an initial state  of her choice. For example, she could use two different states encoding bit values `0' and `1'. After this preparation, she couples her detector to the field. From the interaction with the detector, a signal emanates which propagates through the field towards Bob.

In order to receive the message, Bob initializes his detector at some fixed state, say the ground state $\ket{g_\db}$ (the same results can be achieved for any other initial state \cite{jonsson_quantum_2017}).
He then couples his detector to the field so that it interacts with Alice's signal that has propagated through the field. 
After the interaction is switched off, the final state of Bob's detector depends on Alice's signal, which in turn depends on Alice's initial state. Therefore, Bob may be able to infer Alice's message from the final state of his detector.

In the scenario just described, both Alice and Bob are equipped with detectors. Accordingly, the total Hilbert space of the system $\mathcal{H}=\mathcal{H}_A\otimes \mathcal{H}_F\otimes\mathcal{H}_B$ is the tensor product of Alice's and Bob's detector Hilbert spaces, $\mathcal{H}_A$ and $\mathcal{H}_B$ respectively,  and the Hilbert space of the field, $\mathcal{H}_F$. 
Correspondingly,  given a  time coordinate $t$, the total interaction Hamiltonian (generating translations with respect to $t$) is given by a sum of two interaction Hamiltonians of the detectors with the field,
\begin{align}\label{eq:sum Hint}
\intH^t=H^t_{\text{int},\da}\otimes\id_\db+\id_\da\otimes H^t_{\text{int},\db} 
\end{align}
Notice that the right-hand-side of \eqref{eq:sum Hint} contains two interaction terms that are not of the same form as those written in \eqref{eq:udwhamiltonian}. The reason is that the Hamiltonian in \eqref{eq:udwhamiltonian} generates translations with respect to the proper time of detector D, whereas in \eqref{eq:sum Hint} we are adding up two Hamiltonians corresponding to detectors with different proper times, thus the appropriate transformed Hamiltonians need to be considered.

As discussed in detail in \cite{martin-martinez_relativistic_2018}, the relationship between a detector-field Hamiltonian generating translations with respect to proper time $\tau_\dd$ and  one generating translations with respect to a different time parameter is, in general, complicated. However, as shown in \cite{martin-martinez_relativistic_2018}, for pointlike detectors the relationship simplifies: In the pointlike case, given the Hamiltonian $H^{\tau_\dd}_{\text{int},\dd}(\tau_\dd)$ generating  translations with respect to a time parameter $\tau_\dd$ (e.g., detector's proper time), the Hamiltonian $H^t_{\text{int},\dd}(t)$ generating translations with respect to a different time parameter $t$ is given by
\begin{equation}
    H^t_{\text{int},\dd}(t)=\frac{\text{d} \tau_\dd}{\text{d} t}H^{\tau_\dd}_{\text{int},\dd}[\tau_\dd(t)],
\end{equation}
which we use below for $\dd=\da,\db$.

The initial state of the total system is the product state of the field state  $\rho_\phi$, Alice's state $\rho_A$ and Bob's state $\dketbra{g}{g}b$, i.e., given by the density matrix
\begin{align}\label{eq:initialstate}
\rho_0=\rho_\da\otimes \rho_{\phi}\otimes \dketbra{g}{g}b.
\end{align}
In the interaction picture, this state evolves when Alice's and Bob's detectors couple to the field. 
The final state of Bob's detector is obtained by taking the partial trace over the field and Alice's detector of the total final state of detectors 
\begin{align}
\rho_\db= \tr_{A,F} \left(\hat U\rho_0 \hat U^\dagger \right).  
\end{align}
Here, $\hat U$ denotes the unitary operator, mapping the joint state of detectors and field before the coupling to  their state after the interaction.

As mentioned above we treat the time evolution of the detectors perturbatively. For this approach to work we need field states for which the Wightman function is regular enough for a perturbative approach to time evolution to work. In particular, we assume that the field starts out in a state which  we assume to be Hadamard, at least in the region of spacetime where Alice and Bob's worldlines are within the support of their switching functions. This assumption ensures that the detector transition probabilities we calculate below are well defined.  
Note that the assumption on the field state is  still  very general. It includes states such as the Hartle-Hawking, Unruh or Boulware vacua, unless Alice or Bob cross the regions where those states' Wightman function is not well-defined.

We use the Dyson series expansion to obtain a perturbative expansion of $\hat U$. 
Given arbitrary coordinate times $t_1,t_2,\dots$, it reads
\begin{align}\label{eq:dysonU}
    \hat U&=\id-\ii\integral{t_1}{-\infty}\infty \intH^{t_1}(t_1)\nn* 
    &\qquad\quad -\integral{t_1}{-\infty}\infty \integral{t_2}{-\infty}{t_1} \intH^{t_1}(t_1)\intH^{t_2}(t_2) + ...\, .
\end{align}
 Throughout the paper we will sometimes take integrals with respect to Schwarzschild coordinate time, and sometimes with respect to the detectors' proper time, in each case taking into account the appropriate $\difffrac{\tau_\dd}{t}$ factors in the perturbative time integrals where necessary.

The dependence of Bob's final state  $\rho_B$ on Alice's initial state $\rho_A$ is captured by the quantum channel map
\begin{align}
\xi:\rho_\da\mapsto\rho_\db,
\end{align}
i.e., the completely positive and trace-preserving map which maps the density operator of Alice's initial state to the density operator of Bob's final state. 
The quantum channel between detectors was first studied in \cite{cliche_relativistic_2010}. It has since been studied both in the perturbative regime \cite{jonsson_quantum_2014,jonsson_quantum_2017} as well as non-perturbatively \cite{landulfo_nonperturbative_2016,jonsson_transmitting_2018}.

\subsection{Leading order signal strength}\label{sec:leadingordersignal}

Treating the interaction between field and detectors perturbatively and assuming, as above, that Bob's detector starts out in the state $\dket{g}b$, and that the field's initial state is Hadamard and has vanishing one-point function (in the region where the detectors couple to the field), the perturbative expansion of Bob's final state is
\begin{align}\label{eq:rhoB}
\rho_B &= \left(1- \lambda_\db^2 P_2 \right)\dketbra{g}{g}b + \lambda_\db^2 P_2 \dketbra{e}e{b} \nn* 
&\quad + \lambda_\da\lambda_\db \left(\zeta C_2+ \zeta^* D_2^*\right) \dketbra{e}g{b} \nn* 
&\qquad + \lambda_\da\lambda_\db \left(\zeta^* C_2^*+ \zeta D_2\right) \dketbra{g}e{b} +\mathcal{O}( \lambda_\dd^4),
\end{align}
where the number 
 $\zeta\in\mathbb{C}$, together with $\theta,\beta\in\mathbb{R}$, denote  the coefficients of Alice's initial state
\begin{align}\label{eq:rhoA}
\rho_A= \theta \dketbra{e}e{a} + \beta \dketbra{g}g{a}+ \zeta \dketbra{e}g{a}+ \zeta^* \dketbra{g}e{a}.
\end{align}
The capital-letter coefficients arise from the Dyson series and partial trace, and  read \cite{jonsson_quantum_2014}:
\begin{widetext}
\begin{align}
P_2 &=\integral{t_1}{-\infty}{\infty}\difffrac{\tau_\db(t_1)}{t_1} \integral{t_2}{-\infty}{\infty} \difffrac{\tau_\db(t_2)}{t_2} \eta_\db(t_1)  \eta_\db(t_2)  \ee{\ii \Omega_\db \left( \tau_\db(t_1)- \tau_\db(t_2)\right)} \exptval{\phi\left(\coord{x_B}(t_2)\right) \phi\left( \coord{x_B}(t_1)\right) },\label{eq:P2}\\
C_2&=\integral{t_1}{-\infty}{\infty}\difffrac{\tau_\db(t_1)}{t_1}\integral{t_2}{-\infty}{t_1}  \difffrac{\tau_\da(t_2)}{t_2} \eta_\db\left(t_1\right) \eta_\da\left(t_2\right) \ee{\ii\left( \Omega_\db \tau_\db(t_1)- \Omega_\da \tau_\da(t_2) \right)} \comm{\phi\left(\coord{x_A}(t_2)\right)}{\phi\left(\coord{x_B}(t_1)\right)},\label{eq:C2} \\
D_2&=-\integral{t_1}{-\infty}{\infty}\difffrac{\tau_\db(t_1)}{t_1}\integral{t_2}{-\infty}{t_1}  \difffrac{\tau_\da(t_2)}{t_2} \eta_\db\left(t_1\right) \eta_\da\left(t_2\right) \ee{-\ii\left( \Omega_\db \tau_\db(t_1)+ \Omega_\da \tau_\da(t_2) \right)} \comm{\phi\left(\coord{x_A}(t_2)\right)}{\phi\left(\coord{x_B}(t_1)\right)}. \label{eq:D2}
\end{align}
\end{widetext}
Here we assume, that for the coordinate time used as integration variable,   $t_1>t_2$ implies that $(t_1,\spacoord x)$ cannot lie in the past light-cone of $(t_2,\bm y)$ for arbitrary spatial coordinates $\bm x,\bm y$.
The switching functions as functions of coordinate time are given by $\eta_\dd(t)=\eta_\dd(\tau_\dd(t))$.

The coefficient $P_2$  yields the leading order probability for Bob's detector to become excited, i.e., to be measured in the state $\ket{e_\db}$ due to the local interaction with the field and it is independent of Alice. It is the two off-diagonal coefficients $C_2$ and $D_2$ which mediate the leading order impact of Alice's initial state on Bob's final state.

For example, the role of $C_2$ and $D_2$ can be seen in a scenario where Alice has to transmit a bit to Bob. Here Alice is given a random bit `0' or `1' which she has to send to Bob in a single run of the experiment, i.e., in a single use of the communication channel. Bob then may perform a measurement on his detector, and has to tell from the outcome which bit Alice was trying to send him. When Alice uses the optimal choice of initial state, and Bob uses the optimal choice of measurement,
then the probability of the bit being transmitted correctly is \cite{jonsson_information_2015,jonsson_quantum_2017}
\begin{align}\label{eq:p}
    p=\frac12+\lambda_\da\lambda_\db \left( |C_2|+|D_2|\right)+\mathcal{O}(\lambda_\dd^4).
\end{align}
Here we can view the sum of the absolute values
\begin{align}
    \left|C_2\right|+\left|D_2\right|
\end{align}
as a measure for the leading order signal strength. In fact, it has been shown that this result generalises to arbitrary initial states of Bob's detector, as well as to other measures of the classical channel capacity, including the Holevo capacity \cite{jonsson_quantum_2017,jonsson_information_2015}. 

For the practical evaluation of the  signal strength it is helpful to note that one obtains $D_2$ from $C_2$ by changing the overall sign of the term, and the  sign of $\Omega_\db$, i.e.,
\begin{align}\label{eq:D2 from C2}
    D_2(\Omega_\da,\Omega_\db)=-C_2(\Omega_\da,-\Omega_\db).
\end{align}
Using smooth switching functions $\eta_\da$ and $\eta_\db$ ensures that the integrals in Eqs.~\eqref{eq:P2}, \eqref{eq:C2} and \eqref{eq:D2} for $P_2,C_2$ and $D_2$ converge. In particular, this is critical when studying the contribution to the single detector excitation probability $P_2$ which tends to exhibit ultraviolet divergences for non-smooth functions (see, e.g., \cite{satz_then_2007,louko_transition_2008,hodgkinson_how_2012}).
The signal terms $C_2$ and $D_2$ are less sensitive  to this issue. E.g., in Minkowski spacetime, the integrals converge even for sharp switching functions of the form
\begin{align}\label{eq:sharpswitchingfn} 
\eta_\textsc{d}(\tau_\textsc{d})=  \eta_{[A_1,A_2]}(\tau_\textsc{d})&\coloneqq\begin{cases} 1,\quad A_1\leq \tau_\textsc{d}\leq A_2 \\ 0,\quad \text{otherwise}\end{cases},
\end{align}
for some $A_1,A_2\in\mathbb{R}$ \cite{jonsson_quantum_2017}.
As we will see, the integrals in the expressions for $C_2$ and $D_2$ are also regular in the Schwarzschild case. Namely, we find in Section \ref{sec:schwarzschild_static} that the signal terms $C_2$ and $D_2$ can be evaluated for generic scenarios using such sharp switching functions.

The only field-theoretic object entering the terms $C_2$ and $D_2$  is the field's commutator. The commutator is given by the identity operator $\hat\id$ multiplied by the commutator function. The latter is obtained by subtracting the classical retarded Green function (as defined in detail in \eqref{eq:GF}) from the  advanced Green function of the field.
Thus, we have
\begin{align}\label{eq:comm_as_GF}
    \comm{\phi(\coord{x_1})}{\phi(\coord{x_2})}&=\frac{-\ii}{4\pi} G(\coord{x_1},\coord{x_2}) \hat\id\nn 
    &=\frac{-\ii}{4\pi} \left( G_{adv}(\coord{x_1},\coord{x_2}) - G_{ret}(\coord{x_1},\coord{x_2})\right)\hat\id,
\end{align}
where $\coord{x_{1,2}}$ are two spacetime points.
Inserting this into \eqref{eq:C2}, and using the detector proper times as integration variables, yields
\begin{align}\label{eq:C2Gret}
    C_2
    &=\frac{-\ii}{4\pi} \integral{\tau_\db}{-\infty}{\infty} \eta_\db\left(\tau_\db\right) \integral{\tau_\da}{-\infty}{\tau_\da(t(\tau_\db))}  \eta_\da\left(\tau_\da\right)\nn
    &\qquad\qquad \qquad \times
     \ee{\ii\, \Omega_\db \tau_\db}
        \ee{-\ii\, \Omega_\da \tau_\da} G_{ret}\left(\coord{x_B}(\tau_\db),\coord{x_A}(\tau_\da)\right)
\end{align}
and an analogous expression  for $D_2$ follows from \eqref{eq:D2 from C2}.

This shows that the leading order impact of Alice's detector on Bob's detector state is  determined by the retarded Green function, i.e., by the classical properties of the field. In particular, the leading order signal strength  is independent of the quantum state of the field. 
Effects due to the quantum properties of the field can only appear at subleading order in perturbation theory.

\subsection{In what sense are the leading order signals classical or quantum?}\label{sec:class_vs_quantum_signal}

Note that while the signalling between Alice and Bob is mediated through the field's retarded  (i.e., classical) Green function, this communication scenario  would not be strictly possible  in a classical scenario. The reason is that, in order to have a contribution to signalling that is of quadratic order in the coupling constants, i.e.,  $\mathcal{O}(\lambda_\da\lambda_\db)$, in this protocol, we need the emitter antenna to have non-zero non-diagonal elements in the energy eigenbasis (see Eqs.~\eqref{eq:rhoB} and \eqref{eq:rhoA}): the ability to prepare quantum superposition of observable states of the antenna (i.e., $\zeta\neq 0$) is crucial for this communication scenario to occur. If we did not consider quantum superpositions of antenna states, the leading order contribution to communication would happen at order $\mathcal{O}(\lambda_\dd^4)$ and would consist of the emission of a real quantum 
by Alice's detector and the absorption of this real photon by Bob's, and this would be subleading to the protocol studied here. 

However, this is a quirk derived from the fact that we are limiting ourselves to two-level antennas. We can prove that the protocol has indeed very little of quantum nature if we consider higher-dimensional detectors, for example, harmonic oscillators.
We shall see that, in this case, even when there is no quantum superposition at the start, communication happens at order $\mathcal{O}\left(\lambda_\dd^2\right)$.

In order to see this, let us consider two Unruh-DeWitt detectors  modelled as harmonic oscillators rather than two-level quantum systems (see, among many others,  \cite{Osc2,Osc3,Osc4,Osc5,Osc6,Osc7,Osc8,Osc9}). The interaction Hamiltonian describing the detectors-field coupling in $n+1$ dimensions in flat spacetime for two detectors comoving with the field quantization frame $(t,\bm x)$ is \begin{equation}
    H_\text{int}^t=\sum_{\dd\in\{\da,\db\}}\lambda_\dd \eta_\dd(t) Q_\dd(t)\int_{\mathbb{R}^n}\!\!\! \text{d}^n \bm x\, F(\bm x -\bm x_\dd)\phi(t,\bm x),
\end{equation}
which is analogous to the Hamiltonian \eqref{eq:udwhamiltonian} substituting the monopole moment by the $Q_\dd$ position quadrature of the harmonic oscillator. We recall that the index $\dd\in\{\da,\db\}$ labels the detectors. We have generalized to spatially extended detectors with a smearing function $F(\bm x)$ (the special case of a pointlike detector is recovered when $F(\bm x)=\delta(\bm x)$).

The interaction picture position operators of the oscillators, $Q_\dd (t)$ 
, are given in terms of ladder operators by
\begin{equation}
    Q_\dd(t)=a^\dagger_{\dd}e^{\ii {\Omega_\dd} t}+ a_\dd e^{-\ii {\Omega_\dd} t}
\end{equation}
where $\Omega_\dd$ is the energy gap between the energy levels of the $\dd$-th oscillator.

Let us assume that the two harmonic oscillator detectors and the field start at an arbitrary uncorrelated state: $\rho_0=\rho_\da\otimes\rho_\db\otimes \rho_\phi$. After time evolution, the detectors-field system evolves to a state $\rho_T=U\rho_0U^\dagger$, where $U=\mathcal{T}\exp[-\ii\int \text{d}t\, H_I]$ where $\mathcal{T}$ is the time ordering operator. The state of detector B after time evolution is obtained after tracing out detector A and the field from $\rho_T$. Expanding in Dyson series, we can write the post-interaction state of B as
\begin{equation}\label{Dysonette}
    \rho_{\db,T}=\rho_\db+\lambda_\db\rho^{(1)}_{\db,\text{noise}}+ \lambda_\db^2\rho^{(2)}_{\db,\text{noise}}+\lambda_\da\lambda_\db\rho_{\db,\text{sig}}^{(2)}+\mathcal{O}(\lambda_\dd^3).
\end{equation}
The two first corrections in \eqref{Dysonette} are local terms, independent of the initial state of the detector A, and even of whether the detector A couples to the field at all (and therefore constitute noise from the point of view of communication). The correction proportional to $\lambda_\da\lambda_\db$ constitutes the leading order signalling term, and tells us about the impact that the initial state of detector A has on the final state of detector B.

Let us consider that the detectors' switching functions are compactly supported and that their supports do not overlap in time. Without loss of generality, let us also assume that the detector A is switched on before B. This means that
\begin{align}\label{condos}
\nonumber \text{supp}[\eta_\da(t)]=[T_\da^{\text{on}},T_\da^{\text{off}}],\quad\text{supp}&[\eta_\db(t)]=[T_\db^{\text{on}},T_\db^{\text{off}}],\\
T_\db^{\text{on}}>T_\da^{\text{off}}.
\end{align}
Under this assumption~\cite{martin-martinez_2015}, the leading order contribution to the time evolved state of detector B from the presence of detector A is equal to
\begin{align}\label{traceda}
\rho_{\textsc{b},\text{sig}}^{(2)}\!\!=\!\! \iint_{\mathbb{R}^2}\!\!\!\!\text{d}t\,\text{d}t'   \eta_\da(t)\eta_\db(t')\!\tr\big( Q_\da(t)\rho_{\textsc{a}}\big) \mathcal{C}(t,t')  \big[Q_\db(t'),\rho_{\textsc{b}}\big]\!,
\end{align}
which is the analogue of \eqref{eq:C2} above, and where
\begin{align}\label{causalfunctional}
\mathcal{C}(t,t')\coloneqq&\int_{\mathbb{R}^n}\!\!\!\! \text{d}^n\bm x\!\!\int_{\mathbb{R}^n}\!\!\!\! \text{d}^n\bm x' F(\bm x\!-\!\bm x_\da) F(\bm x'\!-\!\bm x_\db)\\*
&\nonumber\qquad\qquad\qquad\qquad\times\left\langle\left[\phi\bigl(t,{\bm x}\bigr), \phi\bigl({t',\bm x'}\bigr)\right]\right\rangle_{\rho_{_\phi}}
\end{align}
is a purely imaginary function that corresponds to the pull-back of the commutator expectation value (The Pauli-Jordan functional) to the smeared  trajectories of the detectors. Notice that this is independent of the field state since the commutator is a c-number. A full derivation of \eqref{traceda} can be found following step by step the analogous two-level system calculation that yields equation (15) in \cite{martin-martinez_2015}, with the substitution of the detectors' monopole moments $m_\dd$ by the harmonic detectors' $Q_\dd$ position operator: $m_\dd\to Q_\dd$.

We will now show that if A is initially in a coherent state (that can be produced and described classically) there will be a signal transmitted to B at leading order  that can be read out just from the expectation value of its position operator $Q_\db$ (which is also classically accessible). Let us assume that A starts in a state $\rho_\da=\ket{\alpha}\!\bra{\alpha}$ where $a_\da\ket{\alpha}=\alpha\ket{\alpha}$, and that B starts in the ground state, $\rho_\db=\ket{0}\!\bra{0}$. Then, since
\begin{equation}
    \tr[Q_\da\ket{\alpha}\!\bra{\alpha}]=2\text{Re}(\alpha e^{-\ii\Omega_\da t}),
\end{equation}
and
\begin{equation}
    \big[Q_\db(t'),\rho_{\textsc{b}}\big]=e^{\ii\Omega_\db t'}\ket{1}\!\bra{0}-e^{-\ii\Omega_\db t'}\ket{0}\!\bra{1},
\end{equation}
the leading order signalling contribution becomes
\begin{align}\label{finalres}
\nonumber\rho_{\textsc{b},\text{sig}}^{(2)}&=2 \int_{-\infty}^{\infty}\!\!\!\text{d}t \int_{-\infty}^{\infty}\!\!\!\text{d}t' \, \eta_\da(t)\eta_\db(t')\text{Re}(\alpha e^{-\ii\Omega_\da t}) \mathcal{C}(t,t')\\
&\times\left[e^{\ii\Omega_\db t'}\ket{1}\!\bra{0}-e^{-\ii\Omega_\db t'}\ket{0}\!\bra{1}\right].
\end{align}

We can now compute the signalling contribution from the presence of Alice to the expectation value of a quadrature $Q'_\db=a_\db+a^\dagger_\db$  of B. First, we decompose this expectation value into a noise (local) contribution and a signalling contribution as
\begin{align}
    \langle{Q'_\db}\rangle_{\rho_{\textsc{b},T}}=&\underbrace{\lambda_\db\tr[Q'_\db\rho^{(1)}_{\textsc{b},\text{noise}}]}_{\langle{Q'_\db}\rangle_{\text{noise}}^{(1)}}+\underbrace{\lambda^2_\db\tr[Q'_\db\rho^{(2)}_{\textsc{b},\text{noise}}]}_{\langle{Q'_\db}\rangle_{\text{noise}}^{(2)}}\\
    &+\underbrace{\lambda_\da\lambda_\db\tr[Q'_\db\rho_{\textsc{b},\text{sig}}^{(2)}]}_{\langle{Q'_\db}\rangle_{\text{sig}}^{(2)}}+\mathcal{O}(\lambda_\dd^3).
\end{align}
We now see  that the expectation value of the $Q'_\db$ quadrature has a non-zero contribution from the presence of A that encodes information of: (a) whether A coupled to the field or not, (b) the coherent amplitude of A and (c) A's spatial and temporal localization. Indeed, using the fact that $\tr[Q'_\db\ket{0}\!\bra{1}]=1$, we obtain that 
\begin{align}\label{finalesp}
\nonumber\langle{Q'_\db}\rangle_{\text{sig}}^{(2)}=&4\ii \lambda_\da\lambda_\db \int_{-\infty}^{\infty}\!\!\!\text{d}t \int_{-\infty}^{\infty}\!\!\!\text{d}t' \, \eta_\da(t)\eta_\db(t') \mathcal{C}(t,t')\\
&\times \sin(\Omega_\db t')\text{Re}(\alpha e^{-\ii\Omega_\da t}),
\end{align}
which is, in general, non-vanishing. Therefore, we conclude that there is a leading order signal from a classical state of A (a coherent state), to a classical observable of B that is mediated through the (classical) radiation Green function. In summary, we showed in this section that even in a scenario where no genuinely quantum features of the field or the antennas play a role we have that the leading order signalling is $\mathcal{O}(\lambda_\da\lambda_\db)$, as described in the previous Sec.~\ref{sec:leadingordersignal}, and identified in \cite{jonsson_information_2015}.

\subsection{Time dilation effects  on signaling}
\label{sec:time_dilation_effects}

Time dilation has a direct impact on the signal strength between the two detectors, due to the exponential factor in the integrand of \eqref{eq:C2} and \eqref{eq:D2}.  
This raises the question of what is the effect of time dilation on communication:
Is it possible to enhance the transmission of information by exploiting dilation effects, or is time dilation always a hindrance to signaling?
In this section we discuss two contrary effects which tend to cancel each other out, in the context of our framework.

At first,  one could expect time dilation to potentially enhance communication.
For example, consider Alice and Bob at static positions outside  a black hole, at radial coordinates   $r_\da$ and $r_\db<r_\da$ respectively. In this setup, Bob  perceives Alice as blue-shifted because he is closer to the horizon.
This means that the total duration of the signal, as measured with respect to Bob's proper time, is decreased roughly\footnote{This statement is only approximate since the signal from Alice may be elongated differently on its way from Alice to different potential positions of Bob.}
by the relative blue-shift factor in comparison to the duration Bob would have observed if located at the same
 radius as Alice (i.e., at $r_\db=r_\da$). 
Say the relative blue-shift factor between $r_\db$ and $r_\da$  is two, then it seems as though Bob should be able to double the rate of signaling in the sense that he only needs half the  amount of proper time to receive Alice's signal 
than if he were located at the same radius as Alice.

However, the decrease of signal  duration with respect to Bob's proper time also leads to a lower leading order signal strength. In particular, Bob cannot overcome 
this lowering of the signal strength even if he adapts his detector frequency  
so as to be in resonance with Alice's blue-shifted signal: Decisive for the signal strength is the signal duration with respect to proper time, but not the frequency at which it is emitted or received.
This relation is known from flat spacetime. There, in 3+1 dimensions, for two identical detectors $(\Omega_\da=\Omega_\db=\Omega$) at rest at a distance of $L$ and for a time interval of length $T$ (using sharp switching functions), the leading order signal strength is \cite{jonsson_quantum_2017}
\begin{align}\label{eq:bound flat}
    \left|C_2\right|+ \left|D_2\right|=\frac{T}{4\pi L}+\frac{\left|\ee{\ii2\Omega T}-1\right|}{8\pi\Omega L}\leq\frac{T}{2\pi L},
\end{align}
showing that a change of detector frequency cannot increase the signal strength beyond the bound set by $T$.

In curved spacetime a similar bound exists, under certain assumptions, as discussed in detail in App.~\ref{app:propertime_bound}:
For every signal emitted by Alice, i.e., for every choice of worldline and detector parameters for Alice, one can give a constant
$C_{\mathbb{B}}<\infty$, such that the leading order signal strength is bounded by 
\begin{align}\label{eq:upperbound_body}
|C_2|+|D_2|\leq C_{\mathbb{B}} \,\Delta \tau_\db,
\end{align}
where $\Delta \tau_\db$ is the total amount of proper time  during which the receiver interacts with the signal. This bound applies to any 
(even time-varying) choice of  detector frequency $\Omega_\db$ for Bob, and the constant  $C_{\mathbb{B}}<\infty$ depends only on the region of spacetime inside of which Bob is allowed to couple his detector to the field. (However, there are no further restrictions on Bob's worldline.)

This behaviour of the signal strength is analogous to  the response of any resonantly-driven harmonic oscillators: Consider harmonic oscillators of different frequencies that are each resonantly driven by a driving force of the same amplitude, for the same number of oscillations. The oscillators with the lower frequencies get more excited. This is because the strength of the excitation is given by the Fourier integral of stationary phase (the resonance condition) during the driving and for lower frequency oscillators the driving period is longer.

In summary, the two effects above may exactly cancel each other out, so as to leave  the signal rate $(|C_2|+|D_2|)/\Delta \tau_\db$ unchanged. That is, whereas our first argument indicated that time dilation may be able to help Bob save proper time when receiving Alice's signal by a factor proportional to the dilation, the latter argument indicates a loss of signal strength also linear in the dilation.

Evaluating the integrals $|C_2|$ and $|D_2|$ requires handling the distributional nature of the scalar field Green function. Therefore, before studying signalling between detectors we need to study and analyze the Green function itself.

\section{The Green function}\label{sec:gf}

In this section we introduce the Green function and describe some its analytical features as well as the methods we shall use to efficiently evaluate it for our relevant scenarios.

A Green function 
is a distribution which depends on two spacetime points $\coord{x}$ and $\coord{x}'$.
It obeys the wave equation with a (invariant) Dirac $\delta$-distribution as a source: 
\begin{equation}\label{eq:GF}
\Box_{\coord{x}} G_{ret}(\coord{x},\coord{x}')=-4\pi \frac{\delta_4(\coord{x}-\coord{x}')}{\sqrt{-g(\coord{x})}},
\end{equation}
where $\Box_{\coord{x}}$ is the D'Alembertian with respect to $\coord{x}$.
Specifically, the {\it retarded} Green function $G_{ret}(\coord{x},\coord{x}')$ obeys this equation with the boundary condition that it is equal to zero if $\coord{x'}$ does not lie in the causal past of $\coord x$.
Heuristically, $G_{ret}(\coord{x},\coord{x}')$ may be thought of as the value of the field at $\coord{x}$ as created by an ``infinite" impulse at the base point $\coord{x}'$.

It is known that, in a general curved spacetime, $G_{ret}(\coord{x},\coord{x}')$ diverges whenever the points $\coord{x}$ and $\coord{x}'$ are connected via a null geodesic~\cite{Garabedian,Ikawa,casals2016global}.
As we shall see in later sections, these divergences play an important part in the behaviour
of $C_2$ and $D_2$ in the case of Schwarzschild spacetime.

In a general curved background spacetime, the retarded Green function can be quite  difficult  to calculate.  In the following we give an overview of general properties of the retarded Green function and state-of-the-art methods for calculating it, mostly specialized to Schwarzschild spacetime.
We split the overview into two subsections, one for  points  $\coord{x}$ and $\coord{x}'$   ``close" (Quasi-Local), and the other one for points ``far apart" (Distant Past). Each one of these regimes requires different methods for the calculation of the Green function, which we also describe.

\subsection{Quasi-Local}\label{sec:GF_quasilocal}

The divergence of the retarded Green function
for points connected via a null geodesic is manifest 
 in the so-called Hadamard form when the points are ``close". Specifically, the Hadamard form is only valid {\it locally} in a normal neighbourhood of $\coord{x}$, i.e., in a 
region containing $\coord{x}$ such
that every $\coord{x}'$ in that region is connected to $\coord{x}$ by a {\it unique} geodesic which lies within the region.
The Hadamard form (in ($3+1$)-dimensions) is~\cite{Hadamard,Friedlander,Poisson:2011nh}:
\begin{equation}\label{eq:hadamard}
G_{ret}(\coord{x},\coord{x}') =\left(U(\coord{x},\coord{x}')\delta(\sigma)- V(\coord{x},\coord{x}')\theta(-\sigma)\right)\theta(\Delta t),
\end{equation} 
where $U$ and $V$ are regular biscalars.
Here, $\sigma=\sigma(\coord{x},\coord{x}')$ is Synge's world-function~\cite{Synge}, which is equal to one-half of the squared distance along the (unique) geodesic connecting $\coord{x}$ and $\coord{x}'$. This means that $\sigma$
 is negative/zero/positive whenever that geodesic is, respectively, timelike/null/spacelike.
 
 Like the world-function, both  biscalars $U$ and $V$ depend only on the geometrical properties of the background spacetime.
Clearly, the term with $U$ only has support {\it on} the lightcone and is called the direct term, whereas the term with $V$ also has support {\it inside} the lightcone and is called the tail term.
If the tail term is zero, as is the case in flat spacetime (for massless fields in 3+1 dimensions), then the field only propagates at the speed-of-light, as per the direct term.
We then say that strong Huygens' principle holds. 
However, in most curved spacetimes (such as in Schwarzschild spacetime) the tail term is non-zero, indicating that the scalar (as well as the electromagnetic) field propagates at all speeds smaller than and equal to the speed-of-light.
In this case, strong Huygens' principle is not valid.

The biscalar $U$ is equal to the square root of the so-called van Vleck determinant~\cite{VanVleck:1928,Morette:1951,Visser:1993} and it may be calculated
by solving a transport equation along the geodesic joining $\coord{x}$ and $\coord{x}'$ -- see, e.g.,~\cite{Ottewill:2009uj} and App.~\ref{sec:Had} for details.
We used a  \textit{Mathematica}  version of the code in~\cite{Wardell-transport-Code} to solve the transport equation for $U$.

The biscalar $V$ can be calculated via the so-called Hadamard series:
\begin{equation}\label{eq:Had ser V}
V(\coord{x},\coord{x}')=\sum_{n=0}^{\infty}V_n(\coord{x},\coord{x}')\sigma^n(\coord{x},\coord{x}'),
\end{equation}
for some coefficients $V_n$.
The series in Eq.~\eqref{eq:Had ser V} is not a Taylor series and it converges uniformly in a subregion of the maximal normal neighbourhood of $\coord{x}$~\cite{Friedlander}.

For general points $\coord{x}$ and $\coord{x}'$ in a general spacetime, the coefficients $V_n$ cannot be obtained exactly but they can be expressed 
as covariant Taylor series expansions, whose coefficients can be calculated
exactly in terms of background geometrical tensors (e.g.,~\cite{Ottewill:Wardell:2008}).
In practice, however, it is hard to calculate the coefficients in these covariant Taylor series for $V_n$ to high order.
It is  more practical to instead calculate $V$ as a multiple power series expansion in the coordinate separations.
From now on and until the end of this section we specialize to the case of of Schwarzschild spacetime.
In this spacetime, the multiple power series for $V$ may be written as~\cite{Ottewill:Wardell:2008,CDOWb}:
\begin{equation}\label{eq:V power series}
V(\coord{x},\coord{x}')=\sum_{i,j,k=0}^{\infty}v_{ijk}(r)(t-t')^{2i}(1-\cos\gamma)^j(r-r')^k,
\end{equation}
for  some coefficients $v_{ijk}$, 
where $\gamma\in [0,\pi]$ is the angle separation between $\coord{x}$ and $\coord{x}'$.
Indeed, Eq.~\eqref{eq:V power series} is how in practice we calculated $V$ in this paper, except where otherwise explicitly indicated.
We used the code publicly available in~\cite{Hadamard-WKB-Code} to calculate the coefficients $v_{ijk}$
up to
$i=26-j$, $j=26$, $k=26-i-j$, thus in practise truncating the infinite series in Eq.~\eqref{eq:V power series}.

To summarize, within the maximal normal neighbourhood of $\coord{x}$, we calculate the retarded Green function in the following way:
We  use the Hadamard form Eq.~\eqref{eq:hadamard}, where we calculate $U$ numerically by solving a transport equation and $V$ via the power series Eq.~\eqref{eq:V power series} truncated up to 26.
Because of the truncation,  this series yields a value for the retarded Green function which is accurate ``enough" 
(i.e., within a certain desired accuracy)
only inside a {\it sub}region of the maximal normal neighbourhood of $\coord{x}$. 
We refer to such subregion as a \emph{``quasi-local" (QL) region}.

Finally, we note that in the particular case that the spacetime  points  are
connected via a radial null geodesic in Schwarzschild spacetime, 
it can be shown (see App.~\ref{sec:Had}) that $U(\coord{x},\coord{x}')= 1$.

\subsection{Distant Past}\label{sec:distantpast}

As just mentioned, the Hadamard form Eq.~\eqref{eq:hadamard} is only useful for calculating the Green function within a QL region. 
However, in  Schwarzschild spacetime, there are null geodesics which orbit around the black hole.
As we shall explain below, 
this implies that, given a certain  point $\coord{x}$, many points $\coord{x}'$  do not lie in a normal neighbourhood of $\coord{x}$, and so they lie outside a QL region. Therefore, the Hadamard form cannot be used for calculating the Green function between
$\coord{x}$ and these points $\coord{x}'$. 
We shall use a  method different from the Hamadard form for calculating the Green function between these points. 
We refer as the \emph{Distant Past (DP)} to the region where this different method yields values for the Green function which are within our required accuracy.  

{\it A priori}, there is no reason why there should be a  region of overlap between the DP and the QL region, i.e.,  a region where the calculations of the Green function using the two methods are both accurate enough. Indeed, the existence of such an overlap region depends on the specific methods used and on the accuracies obtained and required.
In practice, using our methods in Schwarzschild,     we find that there are regions of overlap for many 
choices of worldlines of Alice and Bob and switching functions in the calculations
of $C_2$ (and $D_2$), as we shall present
in later sections.

We next explain how orbiting null geodesics imply that not all points lie in a QL region and we describe
the singularities of the Green function  when 
$\coord{x}$ and $\coord{x}'$ are null-separated. For this purpose, we first give a name to the various types of null geodesics depending on how many orbits they have travelled around the black hole, i.e., depending on how many caustic points (these are points where null geodesics focus; in Schwarzschild spacetime, a caustic is thus a point  along a null geodesic \footnote{The coincidence point $\coord{x}=\coord{x}'$ is excluded from the definition of caustic.} at which $\gamma=0$ or $\gamma=\pi$) they have crossed.
We call a {\it direct} (or primary) null geodesic
a null geodesic which has travelled an angular distance equal to the angle separation $\gamma\in [0,\pi)$ (i.e., it has not crossed a caustic).
A secondary null geodesic is a null geodesic which has  travelled an angular distance equal to $2\pi-\gamma \in (\pi,2\pi)$ (i.e., it has  crossed one caustic).
A tertiary null geodesic  is 
a null geodesic which has  travelled an angular distance equal to $2\pi+\gamma\in(2\pi,3\pi)$ (i.e., it has  crossed two caustics).
A quaternary null geodesic is a null geodesic which has  travelled an angular distance equal to $4\pi-\gamma\in (3\pi,4\pi)$ (i.e., it has  crossed two caustics).
And similarly for null geodesics orbiting more times around the black hole.
The direction in which primary, secondary, etc null geodesics orbit around the black holes alternates. 

Consider now a given base point $\coord{x}'$, a given spatial position $\spacoord{x}$ ($\neq \spacoord{x}'$) and vary $t$.
For $t$ small enough, $\coord{x}$ and $\coord{x}'$ are not causally connected and so $G_{ret}(\coord{x},\coord{x}')=0$.
As $t$ increases, there will be a time when
$\coord{x}$ is connected to $\coord{x}'$ by a direct null geodesic; 
that marks the start of causal contact and so where $G_{ret}(\coord{x},\coord{x}')$ starts being non-zero.
As $t$ increases further, there will be a time when
$\coord{x}$ is connected to $\coord{x}'$ by a secondary null geodesic;
that marks the end of $\coord{x}'$ being in the maximal normal neighbourhood of $\coord{x}$.
Since QL is a subregion of the maximal normal neighbourhood, a QL region cannot include points connected by a null geodesic which has orbited around the black hole (i.e., which has crossed any caustics). If we wish to obtain the Green function for points arbitrarily separated and, in particular, if we wish to study the effect of orbiting null geodesics, we need to calculate the Green function outside a QL region.

As already mentioned, the Green function diverges when 
$\coord{x}$ and $\coord{x}'$ are null-separated and these divergences play an important part in $C_2$ and $D_2$.
The Hadamard form Eq.~\eqref{eq:hadamard} explicitly shows that the divergence
when $\coord{x}$ and $\coord{x}'$
are connected by a direct null geodesic  
is given by $\delta(\sigma)$.
Outside a maximal normal neighbourhood of $\coord{x}$, where Eq.~\eqref{eq:hadamard} is no longer valid, it has been shown, via a variety of methods, that the divergence displays a fourfold singularity structure~\cite{Ori1short,Dolan:2011fh,harte2012caustics,Zenginoglu:2012xe,casals2016global}. Its leading order is given by\footnote{This structure does not hold at caustic points. See~\cite{casals2016global} for the fourfold structure of the sub-leading order divergence away from caustics as well as for the structure at the divergences at caustics.} 
$\delta(\sigma)\to \text{PV}\left(1/\sigma\right)\to -\delta(\sigma)\to -\text{PV}\left(1/\sigma\right) \to \delta(\sigma)\dots$, where
$\text{PV}$ denotes the principal value distribution.
By ``$\sigma$" here we mean a well-defined extension of the world function outside normal neighbourhoods~\cite{casals2016global}.
Each change in the form of the singularity is basically due to the wavefront of the field passing through  a caustic point.
That is, the $\delta(\sigma)$ divergence arises from direct  null geodesics;
the $\text{PV}\left(1/\sigma\right)$  from secondary null geodesics;
the $-\delta(\sigma)$  from tertiary null geodesics;
the $-\text{PV}\left(1/\sigma\right)$  from quaternary null geodesics; and so on for  null geodesics orbiting the black hole more times.

In Fig.~\ref{fig:C2+D2dir_contour}(b) we exemplify the direct null geodesics;
in Fig.~\ref{fig:C2,contour_new} we exemplify the secondary %
null geodesics;
in Fig.~\ref{fig:GF_integrand_stepfeature} we show the singularity structure of the Green function at secondary and tertiary null geodesics -- we will discuss these figures in detail in their relevant places in the paper. 

We next describe the specific method that we used for calculating the Green function in the DP, which
is essentially the 
method described in App.~B of~\cite{mark2017recipe}.
First, we carry out a  decomposition in multipolar $\ell$-modes as:
\begin{align} \label{eq:Green}
&
G_{ret}(\coord{x},\coord{x}')=
\\ &
-\frac{\theta\left(u-u'\right)\theta\left(v-v'\right)}{r\, r'}
\sum_{\ell=0}^{\infty}
(2\ell+1)P_{\ell}(\cos\gamma)G_{\ell}(r,r'; \Delta t),
\nonumber
\end{align}
where $u:=t-r_*$ and $v:=t+r_*$ are null coordinates
and
\begin{equation}
r_*:= r+2M\ln\left(\frac{r}{M}-2\right)\in (-\infty,\infty)
\end{equation}
is the so-called tortoise coordinate.
The $\ell$-modes $G_{\ell}$ satisfy the following $(1+1)$-dimensional Green function equation:
\begin{equation}\label{eq:eq ell-mode}
\frac{\partial^2 G_{\ell}}{\partial v\partial u}+
\frac{f}{4}\left(\frac{\ell(\ell+1)}{r^2}+\frac{2M}{r^3}\right)G_{\ell}=-\frac{1}{2}\delta(u-u')\delta(v-v').
\end{equation}
These modes satisfy
the ``boundary" conditions that 
$G_{\ell}(v=v')=G_{\ell}(u=u')=-1/2$, and the ``initial" condition
that $G_{\ell}(v=v',u=u')=-1/2$.
Since these boundary conditions are along (radial) null geodesics, it is said to be a characteristic initial data (CID) problem.

We solved this CID problem in the following way.
We discretized the $u$-$v$ plane as a grid of stepsize $h$ so that 
the differential equation \eqref{eq:eq ell-mode} is turned into a difference equation.
This difference equation may be solved 
via the finite difference scheme described in~\cite{Lousto:1997wf}, which is of $\mathcal{O}\left(h^4\right)$. In our case, we extended this scheme to be of $\mathcal{O}\left(h^6\right)$.
We give the details of our extension of the scheme in App.~\ref{sec:CID}.

We now discuss some points about the $\ell$-mode in Eq.~\eqref{eq:Green} 
which are important for the practical evaluation of the Green function.
The divergences of
the Green function when the points $\coord{x}$ and $\coord{x}'$ are connected by a  null geodesic
arise from the infinite $\ell$-mode sum.
However, in practise, we must truncate the sum at some finite value $\ell_{\textrm{max}}$ of $\ell$, which we took to be $\ell_{\textrm{max}}=100$.
As explained in~\cite{CDOW13}, this truncation implies that some spurious oscillations may arise in the (approximated) Green function.
We removed these spurious oscillations by introducing a smoothing factor in the summand, which we took to be $\exp{\left(-\frac{\ell^2}{2\ell_{\textrm{cut}}^2}\right)}$ with $\ell_{\textrm{cut}}=\ell_{\textrm{max}}/5$ (see~\cite{CDOW13,Hardy}).
Due to this truncation and the smoothing factor, the (finite) $\ell$-sum does not yield an exact divergence in the Green function when the points are null-separated: the divergence is essentially `smoothed' out. 
In particular, this means that the finite $\ell$-mode sum will smooth out the Dirac $\delta$-divergence in \eqref{eq:hadamard} at the start of causal separation, i.e., when the points $\coord{x}$ and $\coord{x}'$ are connected by a direct null geodesic (i.e., such that $\sigma=0$).
As a consequence, the truncated $\ell$-sum will not approximate the Green function well near $\sigma=0$.
The region  where the  truncated $\ell$-sum does approximate the Green function within our required accuracy is the DP\footnote{Such definition would in principle exclude points ``close" to the null-crossing divergences. However, we still consider such points as part of the DP as long as the divergence does not correspond to the null geodesic and as long as the truncated $\ell$-sum yields a large ``enough" value.}, as mentioned above.

When evaluating the Green function in the DP and in the QL region as indicated so far, we were able to find an overlap between the two regions in the case that Alice is static at $r=6M$ and Bob is static at a radius down to around $r\approx 4M$. For Bob static at smaller radii, however, we found no overlap.
Essentially, the reason is that as Bob's radius diminishes (with Alice's radius fixed), the coordinate time interval for the start of causal separation increases and so the truncated series Eq.~\eqref{eq:V power series} struggles more to converge at the start of causal separation.
In order to extend the calculation to smaller radii, we applied the technique recently suggested in~\cite{ModesDirectPart}.
Basically, this technique consists of decomposing the direct part ``$U\delta(\sigma)$" of the Hadamard form \eqref{eq:hadamard} into $\ell$-modes.
These modes of the direct part are then subtracted from the $\ell$-modes $G_{\ell}$ of the Green function in Eq.~\eqref{eq:Green} and, only afterwards, the $\ell$-sum is performed.
We can do this subtraction since we calculate separately the contribution to the Green function from the direct and non-direct parts.
Such subtraction helps removing the spurious smoothing-out of the $\delta(\sigma)$ in the DP calculation, thus improving its region of validity. In practice, this allowed us to calculate the Green function accurately enough for Bob static down to a radius of $r\approx 2.26M$, which is a significant improvement with respect to the $r\approx 4M$ that we can achieve without such subtraction.

The analysis of  the Green function performed in this section allows us to next study some  general properties of communication between detectors in static spacetimes.

\section{Signaling in static spacetimes}\label{sec:staticspacetimes}

In this section we discuss some common characteristics of the signaling between particle detectors in any static spacetime, such as Schwarzschild spacetime.
Namely, we discuss a general time-mirror  symmetry of  communication scenarios in Sec.~\ref{sec:section_timemirrorsymmetry}, then   we show how the distributional nature of the Green function allows us to separate signals in two very different kinds of contributions (direct and non-direct) in Sec.~\ref{sec:direct_nondirect_contr}, and finally we present 
how Fourier analysis techniques can facilitate the evaluation of signaling scenarios in static spacetimes in Sec.~\ref{sec:Fouriertrick}.

Static spacetimes possess a global timelike Killing vector field. Therefore, its metric can be brought into the form
\begin{align}\label{eq:staticspacetimemetric}
\diff s^2%
=-\metricN(\spacoord{x})^2 \diff t^2+\sum_{i,j=1}^3
h_{ij}(\spacoord{x}) \diff x^i \diff x^j,
\end{align}
with a global timelike coordinate $t$, spatial coordinates $\spacoord{x}$ and metric components $\metricN$ and $h_{ij}$. For example, in Schwarzschild spacetime it is  $\metricN(\spacoord{x})=\sqrt{1-2M/|\spacoord{x}|}$.
Due to the time translational invariance of the metric, both the retarded and the advanced Green functions of the field equation are time translational invariant. This means that the Green functions, and thus the commutator function as well, 
depend on the coordinate times only via the
difference in coordinate time.
That is, they are of the general form
\begin{align}
G(\coord{x},\coord{x'})=G(t-t',\spacoord{x},\spacoord{x}').
\end{align}
This property allows for certain simplifications in the evaluation of the  $C_2$ and $D_2$ terms. In Section \ref{sec:Fouriertrick}, we discuss a method which evaluates these terms without any integration by viewing them as Fourier transforms and applying the convolution theorem.

In this section  we consider communication between static detectors, i.e., detectors located at fixed spatial coordinates.
The proper time of such a detector is linear with respect to the global coordinate time. We use this property to generally  choose Bob's proper time as
\begin{align}\label{eq:static tau_B}
    \tau_\db(t) &= \metricN\left(\bm x_\db\right) t.
\end{align}
If both Alice and Bob are static, then 
it is convenient to introduce a shift when defining Alice's proper time:
\begin{align}\label{eq:static tau_A}
    \tau_\da (t)= \metricN\left(\bm x_\da\right) (t+\Delta t_{\da\to\db}).
\end{align}
Here, $\Delta t_{\da\to\db}$ is the interval of coordinate time that it takes for a direct null geodesic to propagate from Alice at $\bm x_\da$ to Bob at $\bm x_\db$. With this choice we have that the direct null geodesic which reaches Bob at his proper time $\tau_\db$ emanates from Alice at her proper time
\begin{align}\label{eq:tildetau_and_nu}
\tilde\tau_\da(\tau_\db)=\nu \tau_\db,\text{ with } \nu:=\frac{\metricN\left(\bm x_\da\right)}{\metricN\left(\bm x_\db\right)}.
\end{align}
Analogously, we find $\tilde\tau_\db(\tau_\da)=\tau_\da/\nu$ for the proper time of Bob at which the direct null geodesic which emanates from Alice at her proper time $\tau_\da$ reaches Bob.

\subsection{Symmetry of signaling terms between time-mirrored scenarios}\label{sec:section_timemirrorsymmetry}

In curved spacetimes it is interesting to compare the signal strength from one region in spacetime to another to the signal strength in the reverse direction. For example, is it easier or harder to signal from a sender close to the black hole horizon to a more distant receiver than in a scenario where the sender is distant but the receiver is close to the horizon?

As we show in the following, the leading order signal strength in these two scenarios is identical if all other parameters except for the detector position are kept constant. This is  because the leading order signal strength $|C_2|+|D_2|$ is identical for pairs of signaling scenarios in static spacetimes which can be viewed as time-mirrored versions of each other.

By time-mirroring we mean the following procedure: 
Given one particular signaling scenario with worldlines $\coord{x}_\dd(t)$ and switching functions $\eta_\dd(t)$, the worldlines and switching functions of the time-mirrored scenario are obtained by inverting the sign of the argument, i.e., the worldlines become 
\begin{align}
   \coord{x}_\dd'(t)=\coord{x}_\dd(-t). 
\end{align}
for $\dd=\da,\db$, and the switching functions
\begin{align}
    \eta'_\dd(t)=\eta_\dd(-t).
\end{align}
(The wordlines and switching functions can always be assumed to be defined on an interval $t\in[-T,T]$.)

In the time-mirrored scenario, the roles of Alice and Bob are exchanged: We still assume the initial state of detectors and field to be a product state. However, now Bob acts as sender because he couples to the field first. Thus, Bob gets to encode a message for Alice into the initial state of his detector, and Alice will try to read out the message from the final partial state of her detector.

Note that the detector frequencies  are not changed in the mirrored scenario. For example,  Bob uses the same detector frequency $\Omega_\db$ in the mirrored scenario where he is the sender, as in the original scenario where he is the receiver.

As shown in Appendix \ref{app:timeinversion}, the signal terms that result  in the mirrored scenario relate to the original ones as
\begin{align}
    C_2'=C_2,\qquad D_2'=-D^*_2.
\end{align}
In this way, the leading order signal strength $|C_2'|+|D_2'|=|C_2|+|D_2|$ is the same for both scenarios.
This property of the leading order signal strength was shown to hold in Minkowski spacetime before \cite{jonsson_quantum_2017,jonsson_decoupling_2016}. However, because it only relies on the retarded Green function to fulfill 
\begin{align}\label{eq:timeconditiononG}
G_{ret}(t,\spacoord{x}, t',\spacoord{x}')= G_{ret}(-t',\spacoord{x}',-  t,\spacoord{x}),
\end{align}
it generalizes to all spacetimes with this property. This includes all static spacetimes and thus, in particular, also Schwarzschild spacetime.

\subsection{Direct and non-direct contributions}\label{sec:direct_nondirect_contr}

As discussed in Section \ref{sec:gf}, the Green function has support not only for null separated points, but generally also for timelike separated points. 
This means that the total signal strength is a combination of different contributions that propagate from the sender to the receiver along a continuum of different, null and timelike, paths.
In order to assess the individual contributions to the total signal strength which arise from different paths between sender and receiver, it is helpful to split up the signal terms accordingly, using that the terms  $C_2$ and $D_2$  are linear in the Green function  (see equations \eqref{eq:C2}, \eqref{eq:D2}, \eqref{eq:comm_as_GF}). 

In particular, it is helpful to split off the  part of the signal that propagates from Alice to Bob along a direct (shortest  possible) null geodesic. This part is often easy to evaluate because it arises from the  singular term in the Hadamard form for the Green function \eqref{eq:hadamard}. This \emph{direct contribution} reads (see \eqref{eq:C2Gret})
\begin{widetext}
\begin{align}\label{eq:C2d}
C_2^{\textrm{d}}&:= \frac{-\ii}{4\pi} \integral{\tau_\db}{-\infty}{\infty} \integral{\tau_\da}{-\infty}{\tau_\da(t(\tau_\db))}  \eta_\db\left(\tau_\db\right) \eta_\da\left(\tau_\da\right) \ee{\ii\left( \Omega_\db \tau_\db- \Omega_\da \tau_\da \right)}  U(\coord{x_B}(\tau_\db),\coord{x_A}(\tau_\da)) \delta(\sigma) \nn
&=\frac{-\ii}{4\pi}\integral{\tau_\db}{-\infty}{\infty} \eta_\db(\tau_\db)\eta_\da\left(\tilde\tau_\da(\tau_\db)\right) \ee{\ii(\Omega_\db\tau_\db-\Omega_\da\tilde\tau_\da(\tau_\db))}  \frac{U(\coord{x_B}(\tau_\db),\coord{x_A}(\tilde\tau_\da))}{\left|\Delta\lambda\,  t_\alpha(\lambda_0) u_\da^{\alpha}(\tilde\tau_\da) \right|}.
\end{align}
\end{widetext}
Here we used the fact that the partial derivative of the Synge world function with respect to $x'$ is \cite{Poisson:2011nh}
\begin{align}\label{eq:dsigma/dx}
     \sigma_{\alpha'}=\partialfrac{}{{x^{\alpha'} }}\sigma(x,x') = - \Delta\lambda{(x,x')} \, g_{\alpha'\beta'} t^{\beta'}
\end{align}
where $\Delta\lambda{(x,x')}=\lambda_1-\lambda_0>0$ is the difference in the affine parameter $\lambda\in [\lambda_0,\lambda_1]$ along the unique null geodesic $z(\lambda)$ such that
$x'=z\left(\lambda_0\right)$
and 
$x=z\left(\lambda_1\right)$, and $t^\mu=\difffrac{z^\mu}{\lambda}$ is a tangent  vector. 
This yields 
\begin{align}\label{eq:dsigma/dtau}
&\left.\difffrac{\sigma (\coord{x_B}(\tau_\db),\coord{x_A}(\tau_\da))}{\tau_\da}\right|_{\tau_\da=\tilde\tau_\da(\tau_\db)}
\nn &=
\left. \partialfrac{\sigma(\coord{x_B}(\tau_\db),\coord{x_A})}{\coord{x_A}^\alpha}  \difffrac{\coord{x_A}^\alpha}{\tau_\da} \right|_{\tau_\da=\tilde\tau_\da(\tau_\db)} 
=-\Delta\lambda_{\da,\db}\,  t_\alpha(\lambda_0) u_\da^{\alpha}(\tilde\tau_\da), 
\end{align}
with $u_\da^{\alpha}(\tau_\da):= \difffrac{}{\tau_\da}\coord{x_A}^\alpha(\tau_\da)$ being a tangent vector to Alice's wordline and $\Delta\lambda_{\da,\db}:= \Delta\lambda{(\coord{x_B}(\tau_\db),\coord{x_A}(\tau_\da))}$. Hence, in order to obtain the second expression
in  \eqref{eq:C2d}, 
we  replaced the  Dirac $\delta$-distribution factor  in the first expression  by
\begin{align}\label{eq:delta_of_synge_staticdetectors}
    \delta(\sigma(\coord{x_B}(\tau_\db),\coord{x_A}(\tau_\da)))
    =\frac{\delta\left(\tau_\da-\tilde\tau_\da(\tau_\db)\right)}{\left|\Delta\lambda_{\da,\db} \, t_\alpha(\lambda_0) u_\da^{\alpha}(\tilde\tau_\da) \right|} .
\end{align}

We can derive some general properties of the direct contribution  $C_2^{\textrm{d}}$ from  \eqref{eq:C2d}. 
E.g., we see directly that due to the factor $\eta_\db(\tau_\db)\eta_\da\left(\tilde\tau_\da(\tau_\db)\right)$ in the integrand, the direct contribution vanishes unless  Alice and Bob  interact with the field at points that are connected by a direct null geodesic. 

Furthermore, we see that  $|C_2^{\textrm{d}}|$ is maximal if  $\Omega_\db\tau_\db-\Omega_\da \tilde\tau_\da(\tau_\db)=0$, i.e., if Alice's and Bob's  detector frequencies are tuned such that they cancel the frequency shift arising between their wordlines due to motion and gravitation.
To see this, first note that all factors in the integrand, apart from the complex exponential, are non-negative: The switching functions take values in $\eta_\dd\in[0,1]$, the denominator is a non-negative real number, and  $U$ is equal to the square root of the van Vleck determinant, as discussed in Section \ref{sec:gf}.
Hence, in order to maximize  $|C_2^{\textrm{d}}|$, Alice and Bob need to choose their detector frequencies in such a way that the exponential term oscillates as little as possible.

However, while this choice is optimal for $|C_2^{\textrm{d}}|$, it may not always be the optimal choice with respect to $|C_2^{\textrm{d}}|+|D_2^{\textrm{d}}|$. From \eqref{eq:D2 from C2}, we see that the exponential factor in the integrand of $D_2$ is always oscillatory, except for detectors with a vanishing energy gap. Generally, this means that a non-resonant choice of detector frequencies, while leading to a smaller value of $|C_2^{\textrm{d}}|$, may achieve a larger value of $|C_2^{\textrm{d}}|+|D_2^{\textrm{d}}|$. This applies in particular to scenarios where the length of the detector-field coupling is comparable to the detector's period.

For example, we can see this effect in the case of stationary detectors in a static spacetime. Here, $\tilde\tau_\da(\tau_\db)= \nu \tau_\db + \tau_0$ is a linear function, so that 
equation \eqref{eq:C2d} simplifies  to
\begin{align}\label{eq:C2_direct_stationary}
C_2^{\textrm{d}} &= \frac{-\ii\, U(\spacoord{x}_\db,\spacoord{x}_\da)}{4\pi\left|\Delta\lambda\,  t_\alpha(\lambda_0) u_\da^{\alpha}(\tilde\tau_\da) \right|}
\nn* &\qquad\times 
\integral{\tau_\db}{-\infty}{\infty} \eta_\db(\tau_\db)\eta_\da\left(\tilde\tau_\da(\tau_\db)\right) \ee{\ii(\Omega_\db-\nu \Omega_\da) \tau_\db}.
\end{align}
Here we used that due to  the time-translational invariance we can rewrite $U(\coord{x_B}(\tau_\db),\coord{x_A}(\tilde\tau_\da))=U(\spacoord{x}_\db,\spacoord{x}_\da)$, and also $t_\alpha(\lambda_0) u_\da^{\alpha}(\tilde\tau_\da) $ is constant and does not depend on $\tilde\tau_\da$.
If we assume that the switching functions are sharp switching functions $\eta_\da(\tau_\da)=\eta_{[A_1,A_2]}(\tau_\da)$ and $\eta_\db(\tau_\db)=\eta_{[B_1,B_2]}(\tau_\db)$, as defined in \eqref{eq:sharpswitchingfn}, with switching times such that Bob receives all the direct null geodesics from Alice, i.e., $B_1\leq A_1/\nu$ and $B_2\geq A_2/\nu$, then
\begin{align}\label{eq:C2_direct_stationary_sharpswitch}
C_2^{\textrm{d}} 
    &=\frac{-\ii}{4\pi }\frac{U(\spacoord{x}_\db,\spacoord{x}_\da) } {\left|\Delta\lambda\,  t_\alpha(\lambda_0) u_\da^{\alpha}(\tilde\tau_\da)\right|}\ee{\ii\frac{(\Omega_\db-\nu\Omega_\da)(A_1+A_2)}{2\nu}} \nn*
    &\qquad\times\frac{A_2-A_1}{\nu}\sinc\left(\frac{(\Omega_\db-\nu\Omega_\da)(A_2-A_1)}{2\nu}\right).
\end{align}
For fixed switching times $A_1$ and $A_2$, the sinc-function explains  why $|C_2^{\textrm{d}}|+|D_2^{\textrm{d}}|$ is dominated by $|C_2^{\textrm{d}}|$ in the regime of large detector frequencies and thus maximized when $\Omega_\db=\nu\Omega_\da$, where
\begin{align}
\lim_{\Omega_\db\to \nu \Omega_\da} C_2^{\textrm{d}} &=\frac{-\ii\, U(\spacoord{x}_\db,\spacoord{x}_\da) (A_2-A_1)}
{4\pi \left|\Delta\lambda\,  t_\alpha(\lambda_0) u_\da^{\alpha}(\tilde\tau_\da)\right| \nu}.
\end{align}
Whereas for low detector frequencies choosing   one or both of the frequencies to vanish, can lead to a larger $|C_2^{\textrm{d}}|+|D_2^{\textrm{d}}|$ because the gain in $|D_2^{\textrm{d}}|$ overcomes the loss in $|C_2^{\textrm{d}}|$.

We also note that  $(A_2-A_1)/\nu$ corresponds to  the proper time for which Bob interacts with the signal, i.e., the signal strength obeys the linear bound \eqref{eq:upperbound_body} discussed in Sec.~\ref{sec:time_dilation_effects}.

In general it is not possible to avoid oscillations of the exponential term in all signaling scenarios. In fact it is possible that the oscillations limit  the magnitude of $C_2^{\textrm{d}}$  even if the detectors are allowed to interact with the field for arbitrary long times. This effect has been previously  studied between accelerated detectors in Minkowski spacetime in \cite{jonsson_quantum_2017,jonsson_decoupling_2016}.

In addition to the direct contribution $C_2^{\textrm{d}}$, which often dominates the signal, the timelike support of the Green function gives rise to further contributions to the signal, which we call the \emph{non-direct contribution}
\begin{align}
    C_2^{\textrm{nd}}=C_2^{\textrm{d}}-C_2.
\end{align}
(Note that the direct and non-direct contribution need to be added coherently, i.e.,  $|C_2|+|D_2|=\left|C_2^{\textrm{nd}}+C_2^{\textrm{d}}\right|+\left|D_2^{\textrm{nd}}+D_2^{\textrm{d}}\right|$ before taking the absolute value in the total signal strength.)
Since the specific properties of the timelike part of the Green function depend decisively on the spacetime geometry, it is difficult to derive general properties for the non-direct signaling contributions. 

Another challenge of the non-direct contribution is that it typically is more difficult to evaluate.
However, in static spacetimes for detectors at rest  at least the integral expression can be simplified:
By a change of integration variables one can  typically  perform one of the two integrations analytically. In that way, only one numerical integration is left. For this, see Appendix \ref{app:varchange}.
Another method, which is particularly helpful when the Green function is expressed as a series, is developed in the following section.

\subsection{Fourier method for non-direct contribution}\label{sec:Fouriertrick}

When the Green function is represented as a series, e.g., by the Hadamard series \eqref{eq:Had ser V}, then it is even possible to  avoid any integration in  evaluation  of the signal terms $C_2$ and $D_2$. For this we interpret them as a Fourier transformation, apply the convolution theorem and use that the Fourier transform of a series, e.g., $V(\coord{x},\coord{x'})$, yields a sum of derivatives of the Dirac $\delta$-distribution.
This results in a representation of the signal terms which 
highlights to what extent different modes of the field carry the signal, and which can be efficient for numerical evaluation.
In the scope of this work, for example, we used this method for consistency checks between different numerical methods and, depending on the particular setting, found it to be very efficient in numerical calculations. It could also prove useful in similar scenarios such as, e.g., those of \cite{Jorma,KeithResponse,Keith2018}.

For two detectors at rest in a static spacetime, the signal term $C_2$ in \eqref{eq:C2Gret}  can be cast into the form of the Fourier transform of the product of two functions:
\begin{align}
 C_2
    &=\frac{-\ii}{4\pi} \integral{t_1}{-\infty}{\infty} \integral{t_2}{-\infty}{t_1} \ee{\ii\left( N\left(\spacoord{x_B}\right) t_1- N\left(\spacoord{x_A}\right) (t_2+\Delta t_{\da\to\db}) \right)}\nn
    &\qquad\times N_\da N_\db \eta_\db\left(t_1\right) \eta_\da\left(t_2\right) G_{ret}\left(\coord{x_B}(t_1),\coord{x_A}(t_2)\right) \nn 
    &=  \frac{-\ii N_\da N_\db \ee{-\ii N_\da\Omega_\da \Delta t_{\da\to\db}}}{4\pi} \ft\left[\Theta \cdot G_{\da\db} \right]\left(-N_\db\Omega_\db,N_\da\Omega_\da\right)
\end{align}
where $N_\dd=N\left(\spacoord{x}_\da\right)=\difffrac{\tau_\dd(t)}{t}$ for $\dd=\da,\db$ is the metric component of \eqref{eq:staticspacetimemetric} at the detector locations, and the two-dimensional Fourier transform is denoted as
\begin{align}
    \ft[f](k,k')=\integral{t}{-\infty}\infty\integral{t'}{-\infty}\infty f(t,t')\ee{-\ii (kt+k't')}.
\end{align}
The two functions being transformed are, firstly,
\begin{align}\label{eq:thetaswitchingfuncprod}
    \Theta(t,t')=\eta_\db(t)\eta_\da(t') \theta\left(t-t'-\Delta t_{\da\to\db}\right)
\end{align}
which, in particular, contains 
a Heaviside-function 
which restricts the integral to the support of the retarded Green function\footnote{Strictly, we should use $ \theta(t-t'-\Delta t_{\da\to\db}+\epsilon)$, for some infinitesimally small and positive $\epsilon$,  to ensure that the Green function's direct $\delta(\sigma)$-singularity   contributes to the integral.}, which in turn lies inside the integration boundaries $t_1\geq t_2$ of the original expression.
Secondly, the function
\begin{align}
G_{\da\db}(t_1,t_2)%
=G_{ret}\left(\coord{x_\db}(t_1),\coord{x_\da}(t_2)\right)
\end{align}
gives the retarded Green function between Bob's and Alice's location which, in a static spacetime, is a function of the difference in coordinate time only $G_{\da\db}(t_1,t_2)=G_{\da\db}(t_1-t_2)$.

Applying the convolution theorem
\begin{align}
    &\ft[f\cdot g](k,k')=\frac1{4\pi^2}\left(\ft[f]*\ft[g]\right)(k,k')\nn 
    &=\frac1{4\pi^2} \integral{l}{-\infty}\infty\integral{l'}{-\infty}{\infty} \ft[f](l,l')\ft[g](k-l,k'-l'),
\end{align}
the signal term can be written  as
\begin{align}\label{eq:C2asFTconvol}
 C_2&=  \frac{-\ii N_\da N_\db \ee{-\ii N_\da \Omega_\da\Delta t_{\da\to\db}}}{16\pi^3}\nn*
 &\qquad\times \ft\left[\Theta\right]*\ft\left[G_{\da\db}\right]\left(-N_\db\Omega_\db,N_\da\Omega_\da\right)
\end{align}
The  Fourier transform of  $G_{\da\db}(t_1,t_2)$  takes the following form, with $s=t_1-t_2$ $p=\frac12(t_2+t_2)$:
\begin{align}
    \mathcal{F}\left[G_{\da\db}\right](k_1,k_2) &= \integral{p}{-\infty}\infty \ee{-\ii(k_1+k_2)p}\nn &\quad\times 
    \integral{s}{-\infty}{\infty} G_{\da\db}(s)  \ee{-\ii \frac{k_1-k_2}2 s} \nn 
    &= 2\pi \delta(k_1+k_2) \mathcal{F}\left[ G_{\da\db}\right]\left(\frac{k_1-k_2}2\right).
\end{align}
This makes it possible to interpret $C_2$ as resulting from the Fourier transform of the product $\mathcal{F}\left[\Theta\right]$ of switching functions, which has been convoluted along the diagonal $k_1=-k_2$ by the Fourier transform of the Green function.

This general expression is of particular interest when the Green function can be expressed in terms of a power function.
For example, for the calculation of the non-direct contribution to $C_2$ from this tail term, we would replace $G_{\da\db}$ by $-V_{\da\db}$ above (see \eqref{eq:hadamard}). As discussed in Sec.~\ref{sec:GF_quasilocal}, the  tail term in the Hadamard form which can be expanded as a series (see \eqref{eq:Had ser V})
\begin{align}\label{eq:t-exp V}
    V_{\da\db}(t_1,t_2)=V(\coord{x_\da}(t_1),\coord{x_\db}(t_2))&= \sum_{n=0}^\infty c_n\,(t_1-t_2)^n,
\end{align}
for some coefficients $c_n\in\mathbb{R}$.
Thus, its Fourier transform is given by  $\delta$-distributions and their derivatives,
\begin{align}
    \mathcal{F}\left[V_{\da\db}\right](k_1,k_2) &= 2\pi \delta(k_1+k_2) \integral{s}{-\infty}{\infty} \sum_n c_n s^n  \ee{-\ii \frac{k_1-k_2}2 s} \nn
    &=4\pi^2 \delta(k_1+k_2) \sum_n c_n \delta^{(n)}\left(\frac{k_1-k_2}2\right).
\end{align}
The derivatives of the $\delta$-distribution are defined by $\integral{x}{}{} f(x)\delta^{(n)}(x)=(-1)^nf^{(n)}(0)$, such that
\begin{align}
    \left[f *\delta^{(n)}\right](x)=\integral{y}{}{} f(x-y)\delta^{(n)}(y)=f^{(n)}(x).
\end{align}
Thus, replacing $G_{\da\db}$ by $-V_{\da\db}$ in \eqref{eq:C2asFTconvol} and calculating the convolution, we can rewrite the signal term as
\begin{align}
 C_2
 &=   \frac{\ii N_\da N_\db \ee{-\ii N_\da \Omega_\da \Delta t_{\da\to\db}}}{4\pi}\nn* 
 &\quad\times\sum_n c_n \left. \frac{\diff^n}{\diff l^n} \ft[\Theta](-N_\db\Omega_\db+l,n_\da\Omega_\da-l)\right|_{l=0}.
\end{align}

With this expression, it is possible to replace the integration by differentation, in the evaluation of the signal term $C_2$ provided that the Green function can be expressed as a power series and that the Fourier transform $\ft[\Theta]$ (and its derivatives) are known. In App.~\ref{app:FTofswitching} we give $\ft[\Theta]$ for the sharp switching functions considered here.

\section{Static observers near a Schwarzschild black hole}\label{sec:schwarzschild_static}

In this section we calculate the signal strength between static observers in the vicinity of a Schwarzschild black hole.
The total signal strength, presented here at the beginning of the section, comprises direct and non-direct contributions, which we study separately and in detail in Sections \ref{eq:direct_contribution_static_schwarzschild} and \ref{sec:schwarzschild_static_nondirect}, respectively.

\begin{figure}[tb]
\subfloat[$\Omega_B=1/M$%
]{
  \includegraphics[width=0.47\textwidth]{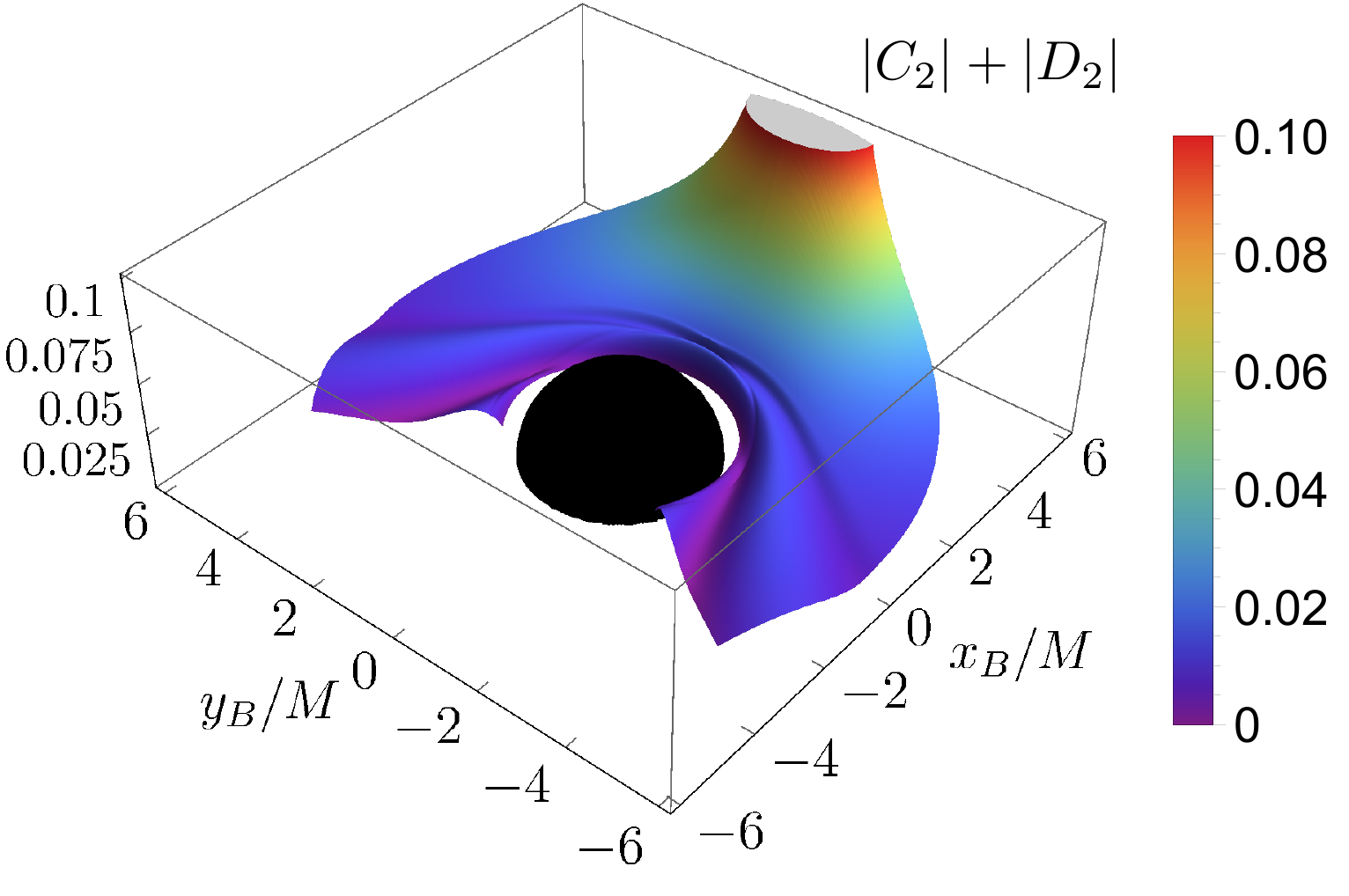}
  }
  
  \subfloat[$\Omega_B=1/(2M)$
  \label{fig:totalsignal_static_bottom}]{
    \includegraphics[width=0.47\textwidth]{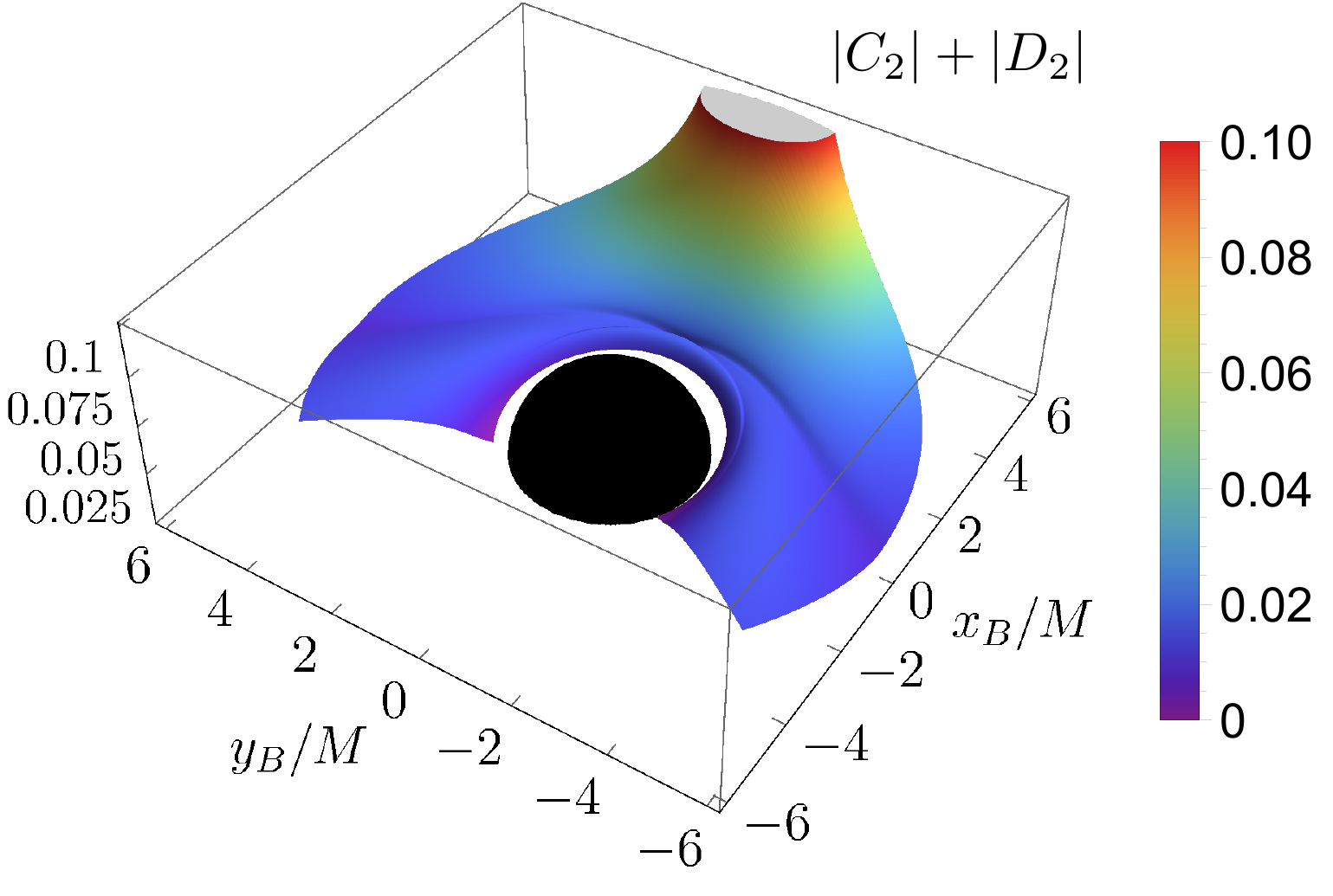}
    }
\caption{
Leading order signal strength between Alice, spatially fixed at 
radius $r_\da=6M$ and Bob,
spatially fixed at varying radii $r_\db$ and angular separation $\gamma$ from Alice. The plot labels Bob's position by $x_\db=r_\db \cos\gamma$ and $y_\db=r_\db\sin\gamma$.
The detectors couple to the field through sharp switching functions for $0\leq\tau_\da\leq M$ and $0\leq\tau_\db\leq 15M$, respectively.
The top of the plots is capped close to Alice's position  because the direct contribution diverges at exactly that point.
The plot covers the radial coordinate position of Bob down to $r_\db=2.26M$. 
The plot only covers positions up to a certain angular separation between Alice and Bob because the numerical evaluation of   the signal strength is infeasible beyond this region.
}
\label{fig:totalsignal_static} 
\end{figure}

To achieve a high signal strength when signaling via massless fields it is important for the receiver to catch the lightrays, i.e., the null geodesics, emanating from the sender.
In flat spacetime the situation is very simple: Because there is a unique null geodesic which connects the sender's interaction with the field to the receiver's worldline, the receiver  should be switched on when the lightrays emanating from the sender arrive at the receiver. (In fact, in 3+1 dimensional flat spacetime there are no other signals apart from that, whereas, e.g., in lower dimensions, where the strong Huygens principle does not hold, signals can propagate slower than the speed of light \cite{jonsson_information_2016,jonsson_information_2015,jonsson_quantum_2017}.)

When spacetime is curved, e.g., by a black hole, a much more complex and interesting picture emerges:
Part of the signal can now propagate inside the future light cone, slower than the speed of light, along a multitude of paths. In particular, these paths can include not only timelike paths but also non-direct null geodesics which, e.g., orbit the black hole on their way from the sender to the receiver.
If the receiver interacts with the field for long enough, then contributions from  all of these  timelike and null paths combine to yield the total signal strength.

We will see below that the part of the signal which propagates along null geodesics tends to carry the largest contribution to the signal strength. This is because the Green function is singular for points that are connected by null geodesics (see  Sec.~\ref{sec:GF_quasilocal}).
As discussed in Sec.~\ref{sec:distantpast}, after the $\delta( \sigma)$-singularity for the direct null geodesic, there follow $\pv(1/\sigma)$, $- \delta(\sigma)$,$-\pv(1/\sigma)$, ...,  singularities corresponding to secondary, tertiary, quarternary, etc null geodesics.

Depending on the position, the motion and the coupling parameters of the sender and receiver, the different contributions to the signal may combine constructively or destructively, thus potentially creating bright or dark spots for communication.
In this article we study these effects for signaling between detectors in the vicinity of a Schwarzschild black hole. In particular, in this section we address signaling between static detectors, before addressing infalling detectors in the next section.

Fig.~\ref{fig:totalsignal_static} shows the leading order signal strength $|C_2|+|D_2|$ between  a static sender (Alice) and a static receiver (Bob),  as a function of Bob's position. 
More specifically, Alice is kept at a fixed spatial position and switched on for a fixed  interval $A_1\leq\tau_\da\leq A_2$ of her proper time, using a sharp switching function.  Bob is placed at the locations plotted on the $x$- and $y$-axes and switched on sharply for $B_1\leq\tau_\db\leq B_2$.
As detailed in App.~\ref{app:varchange}, in this scenario the expression for $C_2$ in \eqref{eq:C2Gret}  can be brought into the form
\begin{widetext}
\begin{align}
    C_2  
    &=\integral{s}{\max\left[\nu B_1-A_2,0\right]}{\nu B_2-A_1} \frac{\ee{\ii\, \Omega_\da s}     
     G_{ret}(s/v(\spacoord{x}_\da)+ \Delta t_{\da\to\db},\spacoord{x}_\db,\spacoord{x_A})}{4\pi(\Omega_\db-\nu\Omega_\da)} 
     \left( \ee{\ii\, (\Omega_\db-\nu\Omega_\da) {\max\left[B_1, (s+A_1)/\nu\right]}}-\ee{\ii\, (\Omega_\db-\nu\Omega_\da) {\min\left[ B_2, (s+A_2)/\nu\right]}} \right).\label{eq:C2_static_aftervchange}
\end{align}
\end{widetext}
We used Eq.~\eqref{eq:C2_static_aftervchange}, together with the corresponding one for $D_2$, in the numerical evaluations of the signal strength presented in this section.

In Fig.~\ref{fig:totalsignal_static}, Alice  is always placed at a radial coordinate $r_\da=6M$. 
There, her detector is coupled to the field by a sharp switching function (see \eqref{eq:sharpswitchingfn}) during the interval $A_1=0\leq \tau_\da\leq A_2=M$ of her proper time.
Bob's spatial location varies. %
At all different locations his detector is coupled to the field during an interval of his proper time given by $B_1=0\leq\tau_\db\leq B_2= 15M$.
Due to the relations \eqref{eq:static tau_B} and \eqref{eq:static tau_A} between  detector proper times and  time coordinate, this means that at all different positions Bob switches on his detector exactly when  his spatial position is reached by the direct null geodesic which  emanates from the spacetime event at which Alices switched on her detector.

It then depends on Bob's position whether he receives more 
null geodesics 
than just the direct (primary) null geodesics from Alice while his detector is coupled to the field:
At some spatial positions, Bob's detector receives secondary null geodesics, tertiary null geodesics (which have fully orbited the black hole before connecting Alice and Bob), or even quarternary null geodesics.
(Note that, due to the varying gravitational redshift factor, the total amount of {\it coordinate} time during which Bob couples to the field  depends on his radial coordinate $r_\db$.)%

Fig.~\ref{fig:totalsignal_static} only covers positions for Bob down to a minimal coordinate of $r_\db \approx 2.26M$ and it also excludes a region around the line of caustics, which is at angular separation $\gamma=\pi$ between Alice and Bob. The reason for this being that the numerical evaluation of the Green function, and thus of the signal strength $|C_2|+|D_2|$ grows increasingly difficult
as Bob's position approaches a caustic.

For each given position of Bob, the different contributions to the signal combine to give the leading order signal strength $|C_2|+|D_2|$ plotted in Fig.~\ref{fig:totalsignal_static}. 
It shows that the most important factor is the distance between Alice and Bob. In fact, as Bob's location approaches Alice the signal strength diverges. Below we show that this is due to the direct contributions $C_2^{\textrm{d}}$ and $D_2^{\textrm{d}}$ which dominate the leading order signal strength.
As Bob's location is moved away from Alice, the signal strength generally drops off. However, the smooth decay is modulated by ripple-like features at certain distances from Alice. These features are caused exactly by null geodesics that orbit around the black hole before arriving at Bob's detector, as we show in Sec.~\ref{sec:schwarzschild_static_nondirect}.

Before analyzing the different contributions  to the total signal strength, let us briefly discuss  the units used here. 
In Schwarzschild spacetime, the mass $M$ of the black hole sets a length scale, which is half of the Schwarzschild radius $r_{BH}=2M$. We use this to  measure distances in units of $M$. To measure the frequency $\Omega$ of a detector, we relate it to the wavelength $\mu$ of radiation associated to the frequency via $\Omega=2\pi/\mu$. Hence, we use $M^{-1}$ as units for detector frequency.

For example, for a black hole with the mass of the Sun, the Schwarzschild radius is $r_{BH}=2M\approx2.9\times10^3$m. Hence, the frequency $\Omega=1/M$ corresponds to radiation with a wavelength of about $\mu=2\pi/\Omega=2\pi M\approx 9.1\times 10^3$m.
Conversely, for a detector in the microwave regime, with a wavelength of $\mu=10^{-2}\text{m}\approx 6.9\times 10^{-6}\text{M}$, the  detector frequency expressed in units  of $M^{-1}$ reads $\Omega=2\pi/\mu\approx 9.1\times 10^5 M^{-1}$.

\subsection{Direct contribution}\label{eq:direct_contribution_static_schwarzschild}

\begin{figure}[tb]
\subfloat[\label{fig:C2+D2dir_3d}
Direct contribution  to the signal strength of Fig.~\ref{fig:totalsignal_static_bottom}.]{
  \includegraphics[width=0.45\textwidth]{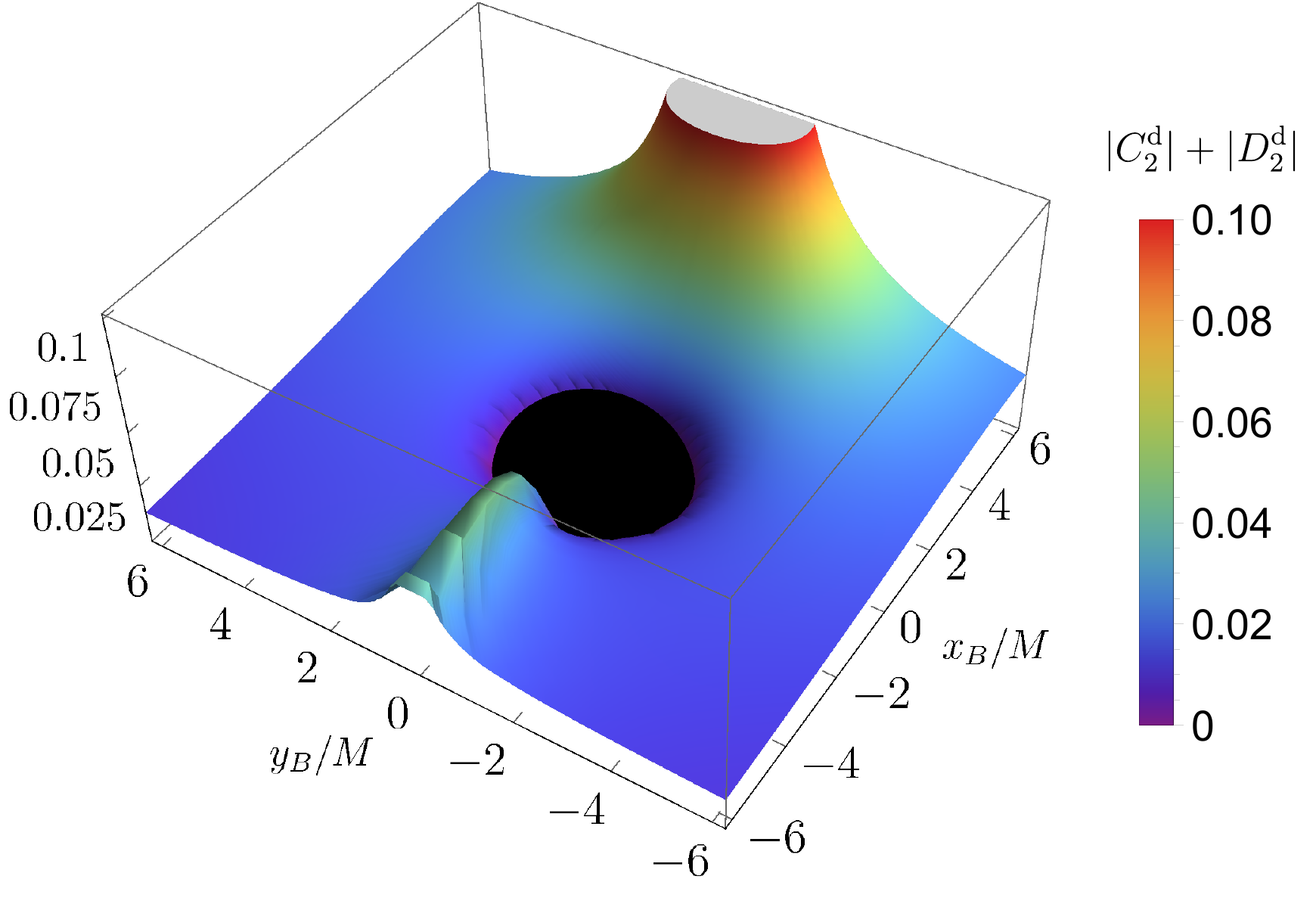}
  }
  
\subfloat[\label{fig:C2+D2dir_contour}
  Contour plot version of Fig.~\ref{fig:C2+D2dir_3d}, also showing  the direct null geodesic between the  marked locations. 
  ]{
    \includegraphics[width=0.45\textwidth]{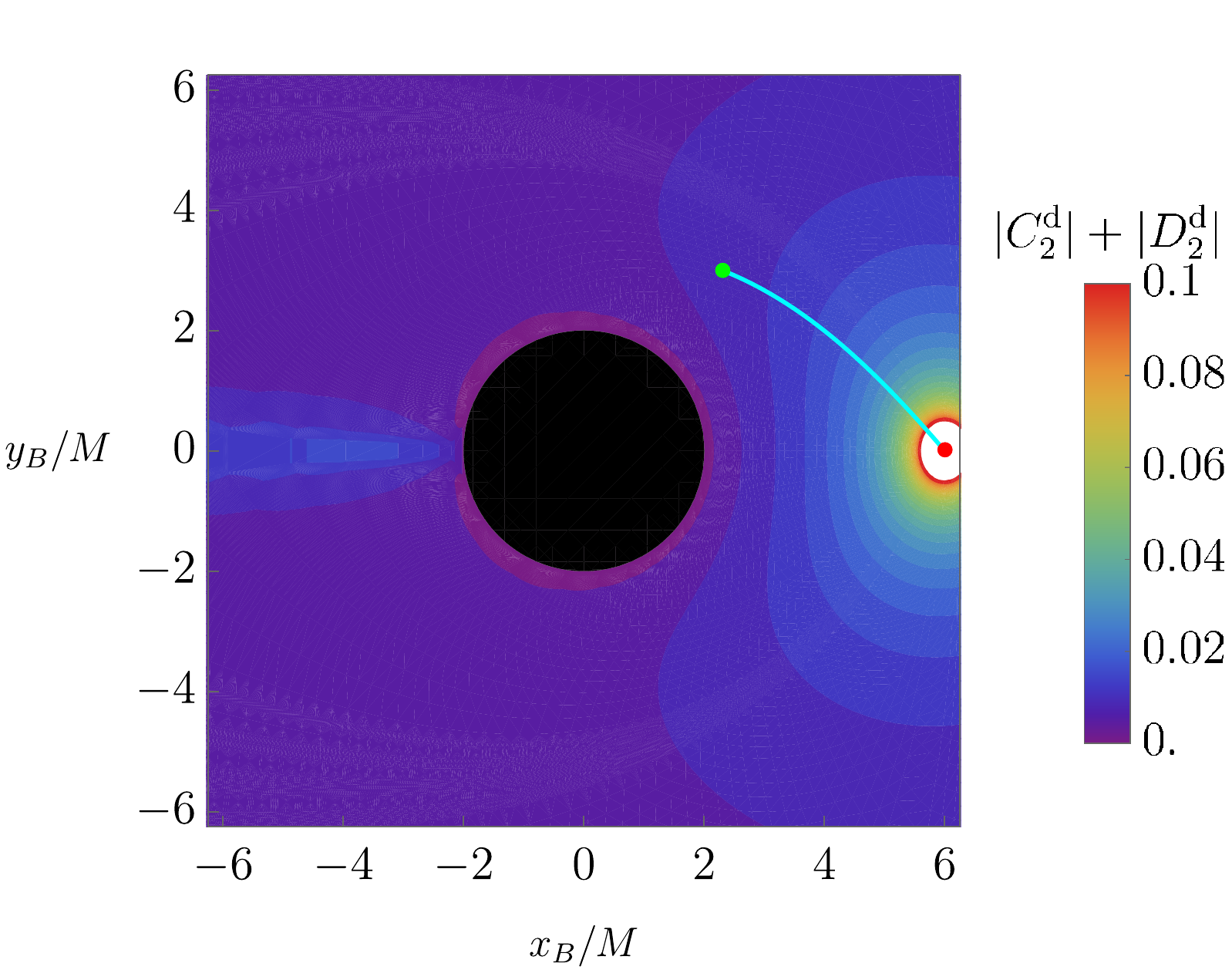}
    }    
\caption{
Direct contributions $|C_2^\textrm{d}|+|D_2^\textrm{d}|$ to the signal strength in the scenario of Fig.~\ref{fig:totalsignal_static_bottom}:
The detectors are static: Alice is placed at a fixed radial coordinate $r_\da=6M$, whereas Bob's radial coordinate $r_\db$ and angular separation $\gamma$ vary, as labelled by $x_\db=r_\db\cos\gamma$ and $y_\db=r_\db\sin\gamma$. 
The detector gaps are $\Omega_\da=1/M$ and $\Omega_\db=1/(2M)$, and Alice couples to the field for her proper time interval $0\leq\tau_\da\leq M$.
A red dot in Fig.~\ref{fig:C2+D2dir_contour} indicates Alice's position, and the  plot shows the direct null geodesic emanating from there to the green dot, as one example for Bob's location.
Note that the direct contribution is not defined  at the exact angular separation $\gamma=\pi$; the plot shows a numerical interpolation.
}
\label{fig:C2+D2dir} 
\end{figure}

The leading order signal strength plotted in Fig.~\ref{fig:totalsignal_static} is generally dominated by the contribution from the part of the signal which propagates from Alice to Bob along direct null geodesics.
This direct contribution, as we defined in Sec.~\ref{sec:direct_nondirect_contr}, is plotted in Fig.~\ref{fig:C2+D2dir} which shows $|C_2^{\textrm{d}}|+|D_2^{\textrm{d}}|$.
In the scenario we are considering here, the switching functions are such that Bob always receives all  direct null geodesics from Alice. 
Hence,   $C_2^{\textrm{d}}$ and $D_2^{\textrm{d}}$ are obtained from  Eq.~\eqref{eq:C2_direct_stationary_sharpswitch}.
The squared root $U$ of the van Vleck determinant 
and the affine parameter interval $\Delta \lambda$
appearing there were calculated numerically by solving a transport equation, as discussed in Sec.~\ref{sec:GF_quasilocal} and detailed in App.\ref{sec:Had}. This allows for the evaluation of the direct contribution for arbitrary angular separations between Alice and Bob.

However, the direct contribution is actually not  defined at a caustic: two null-separated points with angular separation $\gamma=\pi$ are in fact connected by a continuum of null geodesics rather than a single geodesic, thus causing the Hadamard form \eqref{eq:hadamard} to break down. (The values plotted in Fig.~\ref{fig:C2+D2dir} at $\gamma=\pi$ are numerical interpolations.)

Away from $\gamma=\pi$ the direct contribution to $C_2^{\textrm{d}}$ or $D_2^{\textrm{d}}$ is well-defined.
However, we see that it diverges both when Bob's location approaches the caustic at $\gamma=\pi$ and when  Bob's position approaches Alice's position.
The latter divergence arises because $\Delta \lambda\to0$ (see definition below \eqref{eq:C2d}) in the denominator vanishes at the coincidence limit.

The former divergence arises, at $\gamma\to\pi$, because the van Vleck determinant diverges there.
However,  this divergence needs to be interpreted carefully and does not necessarily mean that the total signal strength grows unbounded. 
The reason is that, if Bob is too close to a caustic, then any direct null geodesic will be followed soon by its secondary counterpart, causing the non-direct contribution to the signal to be of comparable magnitude to the direct contribution.
Thus, the direct contribution alone is not indicative of the total signal strength in this region, because it may be counteracted by an equally large non-direct contribution.

\begin{figure}[tb]
\begin{center}
  \includegraphics[width=0.45\textwidth]{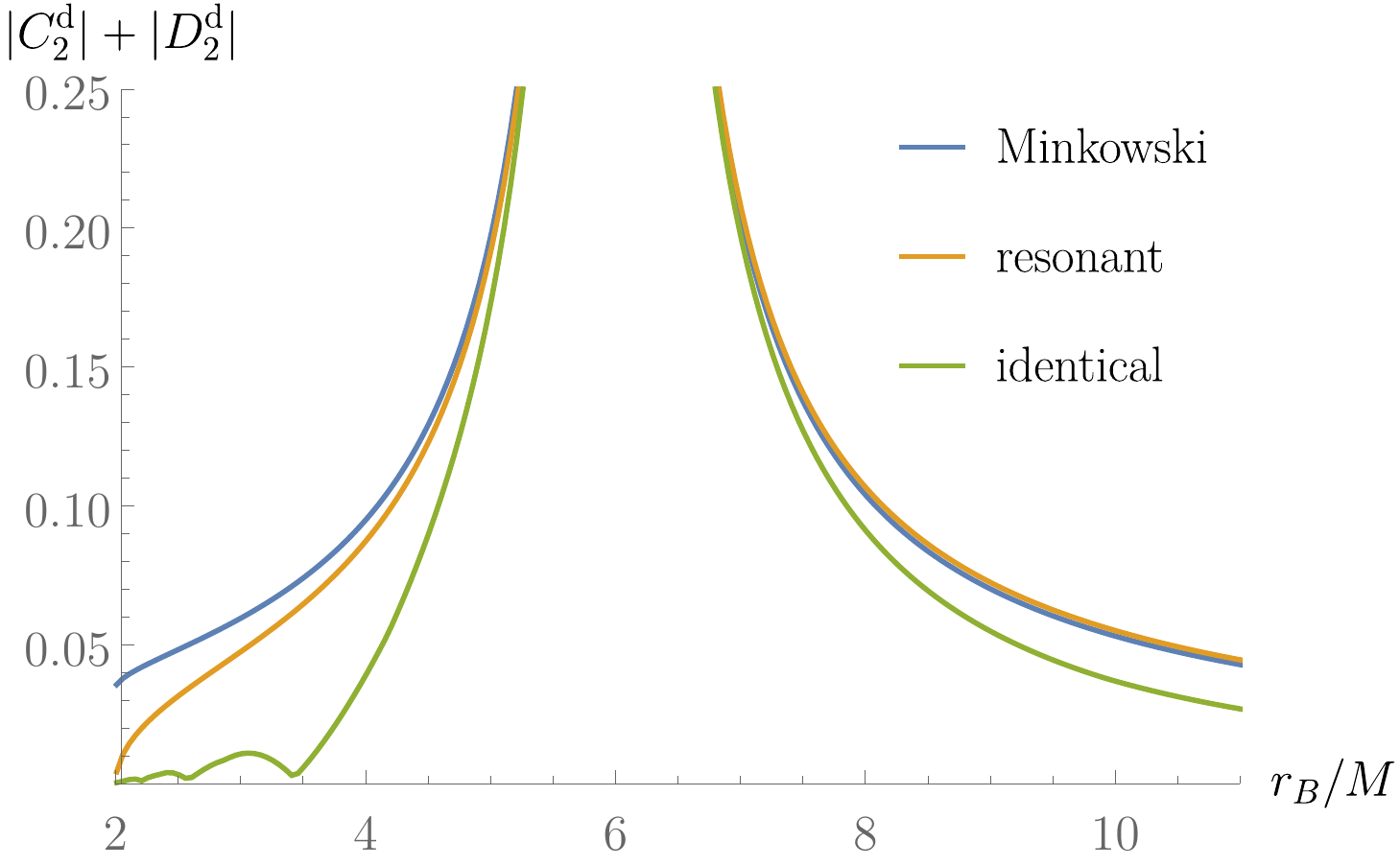}
 \end{center}  
\caption{
Logarithmic plot comparing the direct signal strength $|C_2^{\textrm{d}}|+|D_2^{\textrm{d}}|$ for radially separated static detectors in Schwarzschild and in Minkowski spacetime. 
In Schwarzschild spacetime Alice is located at $r_\da=6M$, her detector gap is $\Omega_\da=10/M$	and she couples to the field for a proper time duration of $A_2-A_1=3M$. Bob's   radial coordinate is $r_\db$. He couples to the field in such a way that he receives all of Alice's direct null geodesics.
The resulting signal strengths in Schwarzschild spacetime are shown for Bob using an identical detector with $\Omega_\db=10/M$ (green), and a resonant detector with $\Omega_\db=\nu \Omega_\da$ (yellow).
In Minkowski spacetime (blue), Alice and Bob use identical detectors ($\Omega_\da=\Omega_\db=10/M$) which are placed so that their  their static distance in Minkowski spacetime  $d(r_\da,r_\db)$ is the same as in the Schwarzschild scenario (see Eq.~\eqref{eq:static_dist}). 
}
\label{fig:direct_signal_strength} 
\end{figure}

In order to further analyze the characteristics of the direct contribution, in the following we focus  on a special case where the direct contribution can be solved analytically. This is the case where sender and receiver have identical angular variables, i.e., zero angular separation $\gamma=0 $, which we refer to as radially separated detectors.
The radial null geodesic connecting Alice at radial coordinate $r_\da$ to Bob at $r_\db$ is of the form
\begin{align}\label{eq:radial geod}
    t(\lambda)&= \lambda\mp 2M\ln\frac{r_\da\mp \lambda-2M}{r_\da-2M},\quad
    r(\lambda) =  r_\da\mp \lambda,
\end{align}
where the negative sign applies if   $r_\db<r_\da$ and  the positive sign applies if $r_\da<r_\db$.
We choose the  affine parameter so that $r(\lambda=0)=r_\da$
and so that the geodesic reaches Bob at affine parameter value  $\lambda = |r_\da-r_\db|$.
Furthermore,  the van Vleck determinant appearing in \eqref{eq:C2_direct_stationary_sharpswitch} is equal to 1 for radially separated detectors, because it is equal to 1 between points connected by a radial null geodesic (see App.~\ref{sec:Had-rad null}).
Altogether, for radially separated, static detectors, the direct contribution to $C_2$ in \eqref{eq:C2_direct_stationary_sharpswitch} thus reads
\begin{align}\label{eq:C2d static}
C_2^{\textrm{d}}&=\frac{-\ii }{2\pi}\ee{\ii\frac{(\Omega_\db-\nu \Omega_\da)(A_1+A_2)}{2\nu}}\frac{ \sqrt{1-\frac{2M}{r_\da}} }{r_A-r_B} \nn 
&\qquad\times\frac1{\Omega_\db-\nu\Omega_\da} \sin\left(\frac{(\Omega_\db-\nu \Omega_\da)(A_2-A_1)}{2\nu}\right),
\end{align}
where $\nu=\sqrt{(1-(2M/r_\da)}/\sqrt{(1-(2M/r_\db)}$ is the red-shift factor between Alice and Bob, as defined in \eqref{eq:tildetau_and_nu}.
Fig.~\ref{fig:direct_signal_strength} shows the direct contribution $|C^{\textrm{d}}_2|+|D_2^{\textrm{d}}|$ to the signal strength  for identical  detectors ($\Omega_\db=\Omega_\da$) and for resonant detectors ($\Omega_\db=\Omega_\da\nu$) with different radial separations. 

The gravitational red-shift caused by the spacetime curvature, impacts  on the value of  $C_2^{\textrm{d}}$ in two different ways.
The first effect  is that the red-shift impacts on the resonance between the detectors. The second effect is that the proper time during which the receiver gets to interact with the direct contribution is affected by the red-shift.

If the detectors are  off-resonant, i.e., $\Omega_\da \nu\neq\Omega_\db$, there is a bound on the magnitude of $C_2^{\textrm{d}}$ which is independent of the duration $A_2-A_1$ of Alice's signal. This is because 
of the last sine-factor in $C_2^{\textrm{d}}$ above, which yields
\begin{align}
    |C_2^{\textrm{d}}|\leq \frac{\sqrt{1-2M/r_\da}}{2\pi |r_\da-r_\db|\,|\Omega_B-\nu\Omega_A|}.
\end{align}
Analogously, $|D_2^{\text{d}}|$ is then bounded by 
\begin{align}
|D_2^{\textrm{d}}|\leq \frac{\sqrt{1-2M/r_\da}}{2\pi |r_\da-r_\db|\,|\Omega_B+\nu\Omega_A|}.
\end{align}

A linear growth of signal strength with the duration of the signal requires resonance, i.e.,  Bob needs to account for the red-shift and tune his detector to the frequency $\Omega_B=\nu\Omega_A$.
In this case, the direct contribution to $C_2$ simplifies to
\begin{align}\label{eq:C2d resonant}
    C_2^{\textrm{d}}= \frac{-\ii \sqrt{1-\frac{2M}{r_\da}} }{4\pi(r_A-r_B)} \frac{A_2-A_1}{\nu}%
\end{align}
Hence, for resonant detectors the direct contribution grows linearly with the duration of the signal. 
It is interesting to note that the specific value of $\Omega_\da$ and $\Omega_\db$ has no impact on $C_2^{\textrm{d}}$ as long as the detectors are resonant. Instead, we see that the determining factor for the magnitude of the direct contribution between resonant detectors is the duration of the signal as measured in terms of Bob's proper time, which is $(A_2-A_1)/\nu$. 
Thus, a linear bound  of the form \eqref{eq:upperbound_body} also applies to this direct contribution here in the case of sharp switching functions (whereas the arguments given in its derivation in App.~\ref{app:propertime_bound} assumed smooth switching functions).
In particular, as Bob approaches the horizon (i.e., $r_\db\to2M$), the red-shift factor  diverges (i.e., $\nu\to\infty$), and so the duration of the signal with respect to Bob's proper time goes to zero and $C_2^{\textrm{d}}\to0$: Bob becomes increasingly transparent for {\it incoming} signals as Bob is placed increasing closer to the horizon.

An interesting question is how the signal strength between static observers in curved Schwarzschild spacetime compares to  flat Minkowski spacetime as a function of the distance between sender and receiver. However, a priori, it is not clear which notion of distance between the observers is appropriate for this comparison. Various notions could be thought of that coincide in Minkowski spacetime, but give different results in Schwarzschild spacetime, as we  illustrate in the following.

A distance measure between static observers which we find to result in similar signal strengths in Schwarzschild and Minkowski distance,
we will refer to as \emph{static distance} (for the purpose of this subsection).
It is most easily obtained by picking a slice of constant coordinate time, using Schwarzschild coordinates in Schwarzschild spacetime and standard coordinates in Minkowski spacetime. 
(The spatial coordinates of sender and receiver are independent of the choice of time slice, because sender and receiver are static.)
The static distance is then given by the proper distance along the shortest (spacelike) geodesic connecting the sender to the receiver on the slice of constant time.
For radially separated detectors in Schwarzschild spacetime, located at radial coordinates $r_\da$ and $r_\db$, this static distance is
\begin{align}\label{eq:static_dist}
d(r_\da,r_\db)&= \left| r_\da \sqrt{1-\frac{2 M}{r_\da}} -r_\db \sqrt{1-\frac{2 M}{r_\db}} \right. \nn* 
&\qquad \left.+M \log \left(\frac{r_\da \sqrt{1-\frac{2 M}{r_\da}}-M+r_\da}{r_\db \sqrt{1-\frac{2
   M}{r_\db}}-M+r_\db }\right)\right|,
\end{align}
while in flat Minkowski spacetime it just corresponds to   $d(\spacoord{x_\da},\spacoord{x_\db})=|\spacoord{x_\da}-\spacoord{x_\db}|$. 
In a coordinate-independent fashion, the static distance can be defined as the  proper distance along the shortest spacelike geodesic connecting the  static sender and static receiver, orthogonal to the timelike Killing vector field of the static spacetime.
Note that, as Bob approaches the horizon in Schwarzschild, the static distance approaches a finite limit
\begin{align}\label{eq:limit_staticdist}
    \lim_{r_\db\to2M} d(r_\da,r_\db)&= \left| r_\da \sqrt{1-\frac{2 M}{r_\da}} \right.\nn* 
    &\qquad\left.+M \log \left( \frac{r_\da \sqrt{1-\frac{2 M}{r_\da}}-M+r_\da}{M} \right)\right|.
\end{align}

As seen in Fig.~\ref{fig:direct_signal_strength}, resonant detectors in Schwarzschild spacetime achieve a direct signal strength which resembles the signal strength between detectors at equal static distance in Minkowski spacetime. (Where in Minkowski spacetime identical and thus resonant detectors are chosen, which generally maximizes the signal strength for long enough coupling times.) 
In fact, if $r_\db>r_\da$ the signal strength in Schwarzschild spacetime is slightly larger than the signal strength in Minkowski spacetime.
In the other direction, where Bob is closer to the horizon and $r_\db<r_\da$, we find the opposite: The signal strength in Schwarzschild  spacetime is smaller; in particular, it drops down to zero as Bob approaches the horizon. The behaviour in both directions arises because, in Schwarzschild spacetime, Bob  has, respectively, more or less proper time at hand to interact with Alice's signal, as explained above.

\begin{figure}[tb]
\begin{center}
  \includegraphics[width=0.45\textwidth]{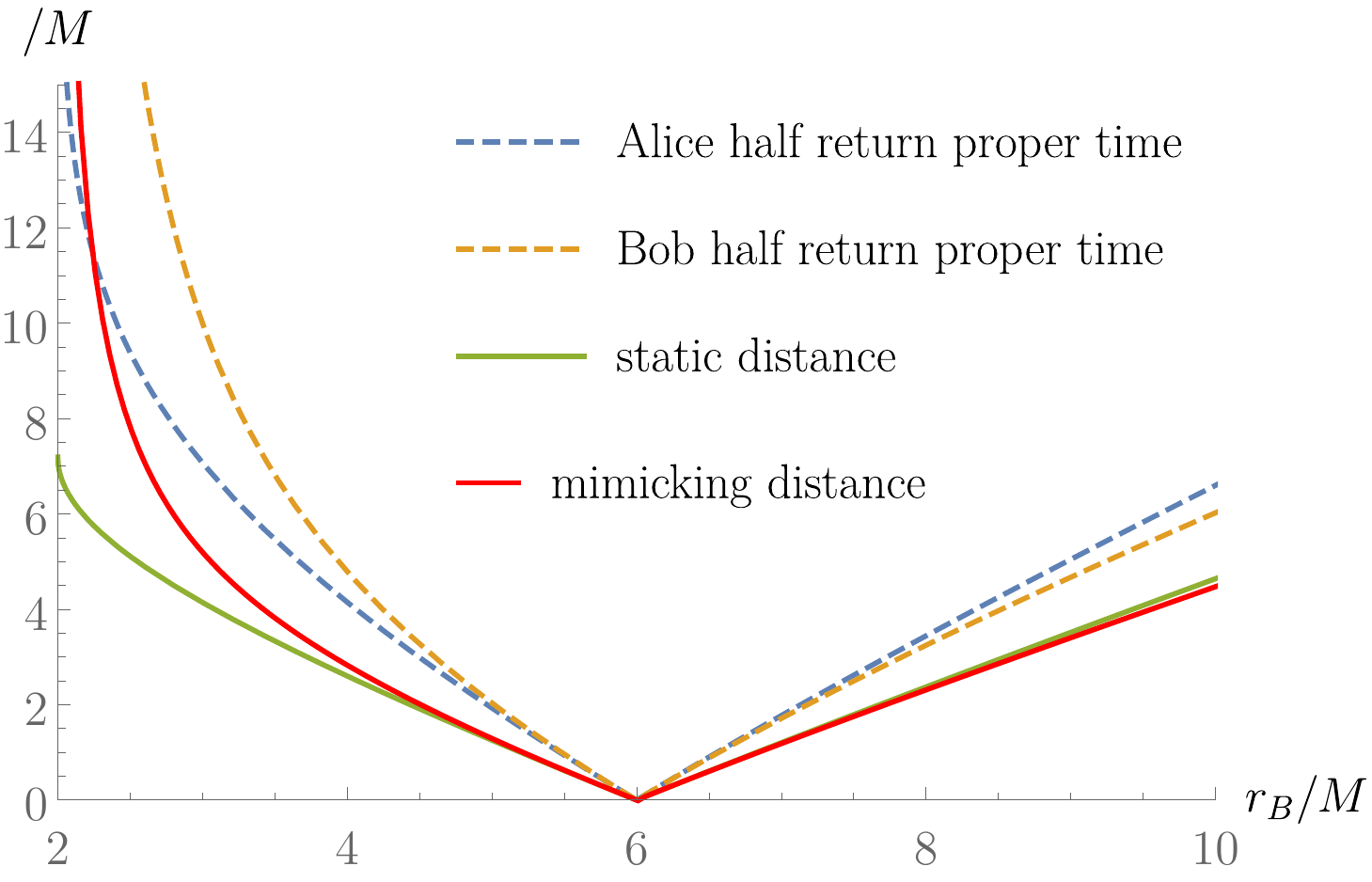}
 \end{center}  
\caption{
The figure compares various measures of distance between two radially separated static observers, Alice (sender) at $r_\da=6M$ and Bob at varying $r_\db$, in Schwarzschild spacetime.
The plot shows  half of the signal return time for Alice (blue, dashed) and Bob (yellow, dashed), i.e., the time it takes for a signal to propagate from Alice to Bob measured in terms of the respective proper times of Alice and Bob, and the static distance (green, solid) in Eq.~\eqref{eq:static_dist}.
The plot also shows the ``mimicking distance'' (red, solid), i.e., the distance in Minkowski spacetime at which two identical detectors ($\Omega_\db=\Omega_\da$) achieve the same direct signal strength as the two detectors in Schwarzschild spacetime at $r_\da$ and $r_\db$ when they are  resonantly-tuned  ($\Omega_\db=\nu\Omega_\da$). (The return times and the mimicking distance diverge, as $r_\db\to2M$, whereas the static distance remains finite, see \eqref{eq:limit_staticdist}.)
}
\label{fig:mimick_distance} 
\end{figure}

The use of the static distance  for the comparison of signal strength between Schwarzschild and Minkowski spacetimes above may appear rather ad hoc. 
One could think of other ways to measure the distance between two given static detectors  which, arguably, could even be more physical or operational.

For example, a very operational approach would be for Alice and Bob to  measure the distance in terms of the  proper time that they observe it takes for a signal to propagate along  direct null geodesics from the sender to the receiver, and back again to the sender. 
From this perspective, we would compare a given scenario in Schwarzschild spacetime with scenarios in Minkowsi spacetime that have the same signal return  time.
One caveat with this approach is that it is asymmetric. In curved spacetime Alice and Bob will measure the signal-return time in terms of their respective proper times and thus assess the distance between them differently.

In flat Minkowski spacetime all these notions coincide: Alice and Bob both measure the same signal return time, and the signal return time coincides with two times the static distance (due to $c=1$). Of course, all of these notions just correspond to the one natural notion of distance between two static observers in flat spacetime.

In curved spacetime, all of these notions of distance differ, as  Fig.~\ref{fig:mimick_distance} illustrates for Schwarzschild spacetime.
There, two static, radially separated observers are placed at radial coordinates $r_\da=6M$ and $r_\db$. The plot shows the static distance between them (green) as well as half of the signal-return time as measured in Alice's proper time (dashed blue) and in Bob's proper time (dashed yellow).

In addition, Fig.~\ref{fig:mimick_distance} plots a ``mimicking distance''  (red) which is the distance  in Minkowski spacetime for which the direct signal strength between two identical detectors ($\Omega_\db=\Omega_\da$) in Minkowski spacetime is the same as between the two radially separated detectors in Schwarzschild spacetime at $r_\da$ and $r_\db$ which are tuned into resonance ($\Omega_\db=\nu\Omega_\da$). (Note that this distance is independent of  Alice's detector frequency $\Omega_\da$.)

Fig.~\ref{fig:mimick_distance} motivates our previous use of the static distance to compare Schwarzschild and Minkowski spacetime because for small distances it resembles the mimicking distance more closely than the signal-return times.
The differences between the different distance measures actually may open up for an interesting way of measuring spacetime curvature. Because, as noted above,  in regions without spacetime curvature all four notions of distance  would coincide, Alice and Bob may be able to detect and quantify spacetime curvature by measuring and comparing signal strength and signal-return times.

\subsection{Non-direct contribution}\label{sec:schwarzschild_static_nondirect}
After the direct part of the signal has passed by, propagating from Alice to Bob along the shortest  (i.e., direct) null geodesic, Bob still continues to receive signals from Alice: 
We call this part of the signal %
the non-direct contribution.
In this subsection we  analyze its physical features and show that they account for the ripple-like features observed in Fig.~\ref{fig:totalsignal_static}. In principle,  all timelike separations between Alice's and Bob's detector couplings contribute to the signal. 
However, we find that the most distinct features of the signal strength, like the mentioned ripples, can be understood as arising from the part of the signal propagating close to secondary and higher-orbiting null geodesics.
These are null geodesics which propagate around the black hole on their way from Alice's position to Bob's position (which throughout this Section \ref{sec:schwarzschild_static} continue to be static) as, e.g., seen in Fig.~\ref{fig:C2,contour_new} for secondary null geodesics.

More precisely, in order to obtain the non-direct contribution $C_2^{\textrm{nd}}$ to $C_2$ we first subtract  the singular direct part from the Green function (compare \eqref{eq:hadamard}) as
\begin{align}
    G^{\textrm{nd}}_{ret}\left(\coord{x},\coord{x'}\right)=G_{ret}\left(\coord{x},\coord{x'}\right)-U\left(\coord{x},\coord{x'}\right) \delta\left( \sigma %
    \right),
\end{align}
when $x'$ is in a normal neighbourhood of $x$;
outside a normal neighbourhood, we define $G^{\textrm{nd}}_{ret}$ to be simply equal to $G_{ret}$.
We then use this non-direct part $G^{\textrm{nd}}_{ret}$ of the Green function instead of $G_{ret}$ in the expressions  \eqref{eq:C2_static_aftervchange} or \eqref{eq:C2Gret} (and  analogously for $D_2$). In this way, the full coefficient splits up into a direct and a non-direct contribution, $C_2=C_2^{\textrm{d}}+C_2^{\textrm{nd}}$.

As discussed in Sec.~\ref{sec:gf}, the non-direct part $G^{\textrm{nd}}_{ret}$ of the Green function exhibits singularities between points which are connected by secondary and other higher-orbiting null geodesics.
In view of these singularities, it is conceivable that the use of discontinuous sharp switching functions could render the integral expressions for $C_2$ and $D_2$ divergent and ill-defined. However, in the previous section we analytically saw that the $\delta(\sigma)$-singularity in the direct contribution
(or similar $\delta(\sigma)$-singularities in null geodesics orbiting around the black hole, such as in tertiary null geodesics)
yields a finite and well-defined contribution. And in App.~\ref{app:secondary_geodesics_signal} we show analytically that  the $\pv\frac1\sigma$-singularity arising, e.g., for secondary null geodesics, also yields finite and well-defined contributions to $C_2$ and $D_2$. Thus the full (and exact) solutions $C_2^{\textrm{nd}}$ and $D_2^{\textrm{nd}}$ are finite everywhere, apart from potentially the caustics of the spacetime, even for sharp switching functions.
We note that, regardless of that, the numerical results for $C_2^{\textrm{nd}}$ and $D_2^{\textrm{nd}}$  are necessarily finite everywhere anyway, since we effectively approximate the exact non-direct part of the Green function by a smooth function (for which the Green function singularities are smeared).

\subsubsection{Shifting Bob's coupling}\label{sec:shfitBobcouple}

An immediate question to ask about the non-direct contribution to $C_2$ (or $D_2$) is  how its  magnitude compares to the direct contribution. To this end,  we study  a signaling scenario where Bob couples to the field for a time interval whose length is synchronized to the duration of Alice's signal but the time when Bob switches on his detector is delayed more and more. 
In this way, as the switching time of Bob's detector becomes later, Bob's detector soon no more interacts  with any of the direct null geodesics emanating from Alice's detector. Instead, e.g., for some late switching times, it interacts with the part of the signal that propagates along secondary and higher-orbiting null geodesics from Alice to him.

\begin{figure}[tb]
\begin{center}
  \includegraphics[width=0.45\textwidth]{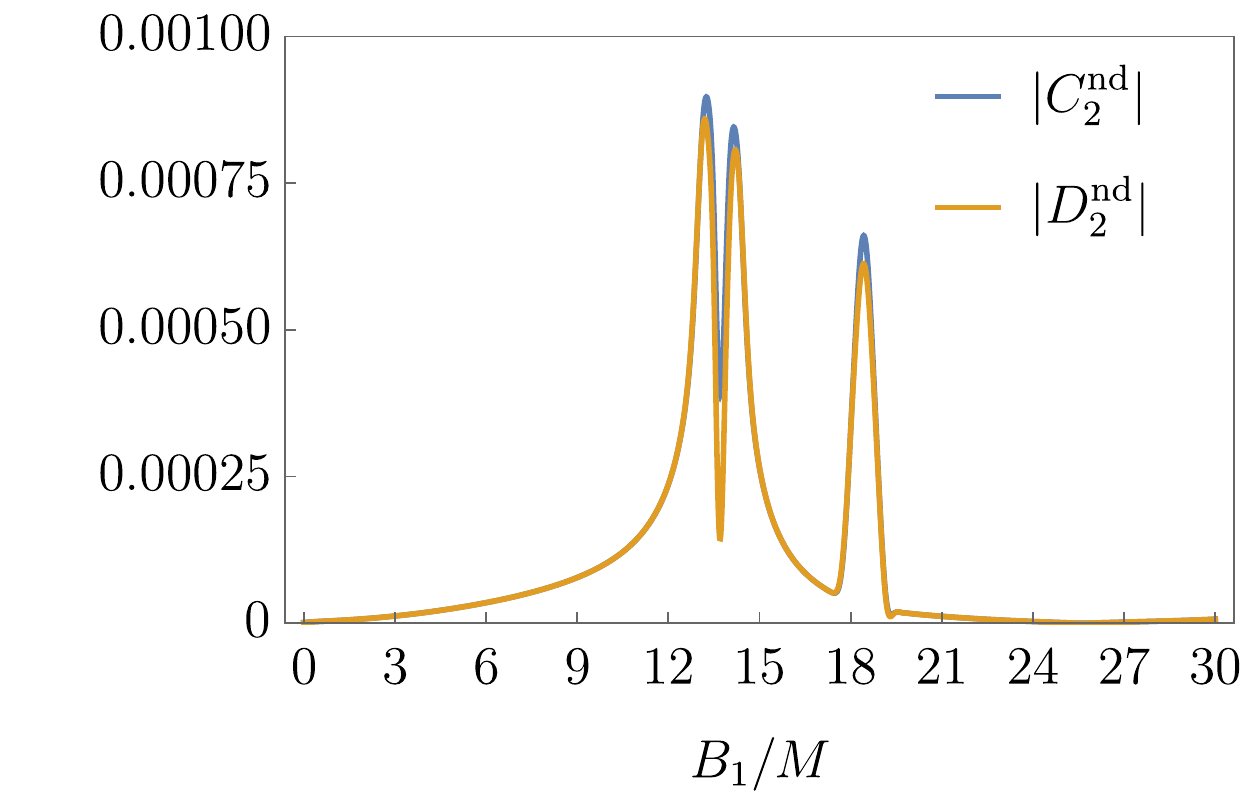}
  \end{center}
\caption{The non-direct contribution $\left|C_2^{\textrm{nd}}\right|$ and $\left|D_2^{\textrm{nd}}\right|$ as a function of $B_1$  for static detectors 
at $r_\da=6M$ and $r_\db \approx 3.01M$  and  $\gamma=\pi/4$, with
 $\Omega_A=\Omega_B=1/M$.
The switching on/off proper times are
$A_1=0$, $A_2=M$ and
$B_2=B_1+\tilde\tau_\db(A_2)$, with $\tilde\tau_\db(A_2)\approx 0.71M$. 
Note that the corresponding direct contribution at $B_1=0$ would be $|C_2^{\textrm{d}}|\approx 0.0121551$ and $|D_2^\textrm{d}|\approx0.0107647$ which then drops down to zero as soon as $B_1>\tilde\tau_\db(A_2)$.
}
\label{fig:C2NonDirectB1ShiftPlot} 
\end{figure}

The resulting non-direct contribution is plotted in Fig.~\ref{fig:C2NonDirectB1ShiftPlot}. Specifically, the plot shows the scenario where Alice  
is located at radial coordinate $r_\da=6M$ and Bob at radial coordinate $r_\db \approx 3.01M$  with a total angular separation of  $\gamma=\pi/4$. Both detectors have identical energy gaps $\Omega_\da=\Omega_\db=1/M$.
Alice switches on her detector over a proper time  interval from $A_1=0$ to $A_2=M$.
Since Bob is closer to the horizon than Alice is, the length of the signal is shorter in terms of his proper time, and is given by $\tilde\tau_\db(A_2)-\tilde\tau_\db(A_1)=\tilde\tau_\db(A_2)\approx 0.71 M$.
While in Fig.~\ref{fig:C2NonDirectB1ShiftPlot} Bob always couples to the field for an interval of this duration, $B_2-B_1=\tilde\tau_\db(A_2)$, we vary the point in time at which Bob switches on his detector, i.e., for a given switch-on proper time $B_1$ we have $B_2=B_1+\tilde\tau_B(A_2)$.

This means that, for $B_1=0$,  Bob switches on his detector when the first direct light signal from Alice (i.e., the one emanating at her proper time equal to $A_1$) arrives at his location, and switches it off when the last direct light signal from Alice (i.e., the one emanating at her proper time equal to $A_2$) arrives. In this case and for the given parameters, we have a direct contribution of magnitude $|C_2^{\textrm{d}}|\approx 0.0121551$ and $|D_2^{\textrm{d}}|\approx 0.0107647$.
For $B_1>0$, Bob switches on his detector {\it after} the first direct null geodesic has passed through his spatial location, hence the direct contribution decreases. Once $B_1>\tilde\tau_\db(A_2)$, Bob's detector does not interact with any direct null geodesics, and the direct contribution $C_2^{\textrm{d}}=0$ vanishes.
However, we see that for later switching times, in the interval $12M\leq B_1\leq 20M$, several spikes arise in the non-direct contribution which reach up to one fifth of the magnitude of the maximal direct contribution.
These peaks are due to the part of the signal which propagates along secondary and tertiary null geodesics from Alice to Bob, and they arise due to the singular behaviour of the Green function along null geodesics.

For example, as discussed in Sec.~\ref{sec:distantpast}, 
a $-\delta(\sigma)$-singularity appears in the Green function between points which are connected by a tertiary null geodesic, in similarity to points connected by direct primary geodesics.
Accordingly, we find a peak in the non-direct contribution when Bob's coupling interval is such that it exactly covers the arrival of all the tertiary null geodesics emanating from Alice. This is the case when Bob switches on exactly when the first of the tertiary null geodesics arrives at his location at $B_1=B_{ter}\approx 18M$.

Particularly interesting is the contribution from the secondary null geodesics: They cause the double peak structure in Fig.~\ref{fig:C2NonDirectB1ShiftPlot} centered around $B_1=B_{sec}\approx 14M$, which is the proper time of Bob at which the first (i.e., emitted from Alice at $\tau_A=A_1$) secondary null geodesic arrives at Bob's location.
As discussed in Sec.~\ref{sec:distantpast}, the Green function diverges like a $\pv\frac{1}{\sigma(\coord{x},\coord{x}')}$-distribution in the neighbourhood of points $\coord{x}$ which are connected to $\coord{x}'$ by secondary null geodesics. We can therefore qualitatively understand the properties of the part of the signal propagating along secondary null geodesics from the analytic solutions of App.~\ref{app:secondary_geodesics_signal}. There, we approximate the behaviour of the Green function near the secondary null geodesic as $\pv\frac{1}{\sigma}$, thus ignoring its regular coefficient:
The most significant difference in comparison to the direct contribution (from $\delta(\sigma)$-distribution) is that the contributions from the secondary null geodesics have tails which extend beyond points that are connected by secondary null geodesics. That is, even if Bob switches off his detector (close to, but still) before the first secondary null geodesic arives, or after the last one has passed by, there is still a contribution to the signal strength from the $\pv\frac1\sigma$-distribution.

The results of App.~\ref{app:secondary_geodesics_signal} show that the signal features arising from secondary null geodesics are always  roughly symmetric about Bob's switch-on time of $B_1=B_{sec}$, i.e., when Bob's switching is aligned so that he exactly interacts with all secondary null geodesics emanating from Alice. For the parameters of Fig.~\ref{fig:C2NonDirectB1ShiftPlot} this point happens to be local minimum of the signal strength.
For other parameters, in particular for longer interaction duration, richer features than the double-peak structure can arise, as seen in Fig.~\ref{fig:PV_Signal} of App.~\ref{app:secondary_geodesics_signal}. In particular, if the detectors are also resonant ($\Omega_\db=\nu\Omega_\da$) then the signal strength exhibits a peak (with overlaid oscillatory features) around $B_1=B_{sec}$.

\subsubsection{Long time-like coupling of Bob}\label{sec:static,nd}
\begin{figure*}%

\subfloat[$B_2=15M$ and $\Omega_B=1/M$\label{fig:rainbow_a}]{ \includegraphics[width=.45\textwidth]{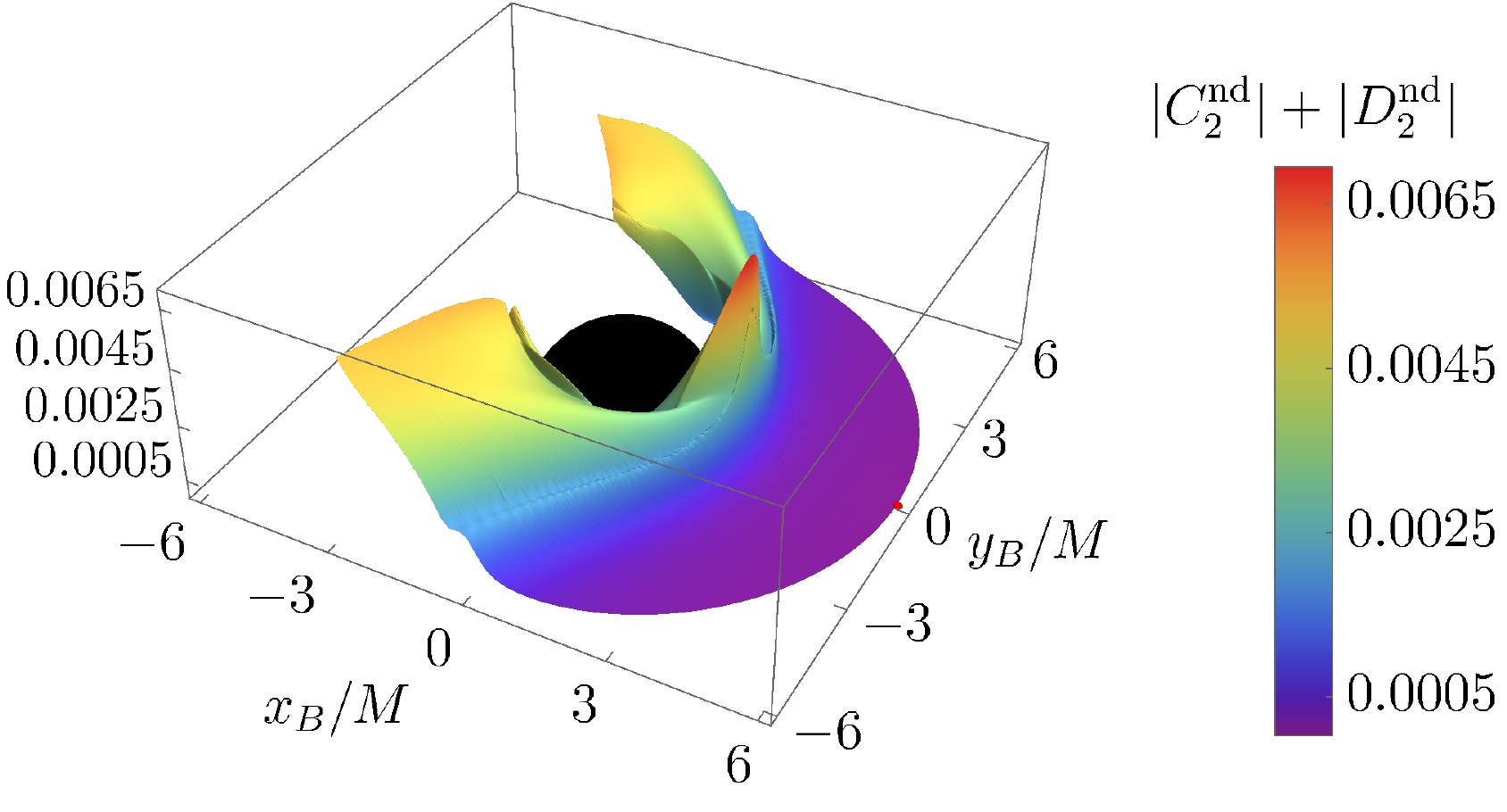}
}
\subfloat[$B_2=15M$ and $\Omega_B=1/(2M)$ \label{fig:middlerainbow}\label{fig:rainbow_b}]{ \includegraphics[width=.45\textwidth]{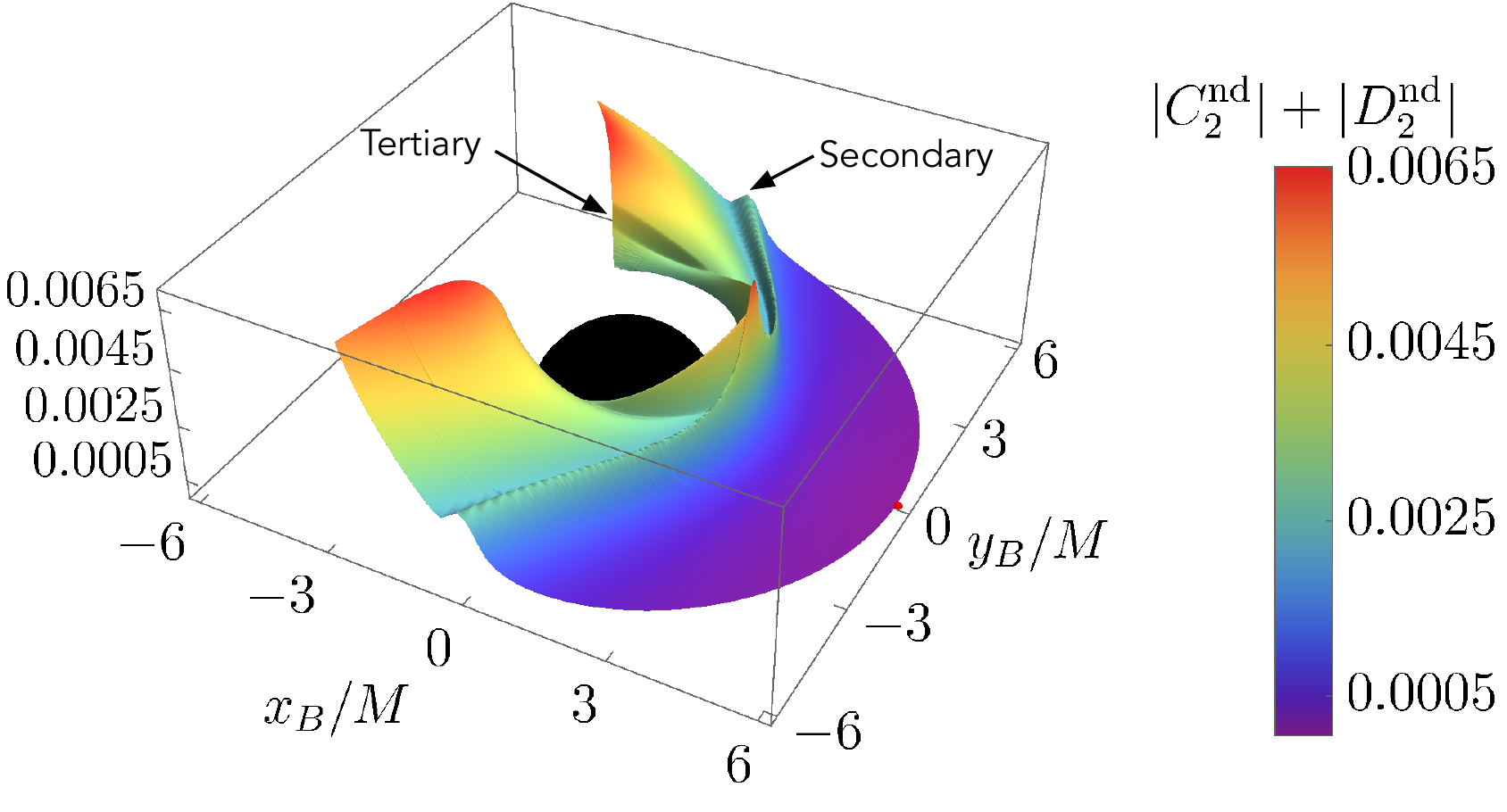}}

\subfloat[$B_2=25M$ and $\Omega_B=1/M$\label{fig:rainbow_c}]{ \includegraphics[width=.45\textwidth]{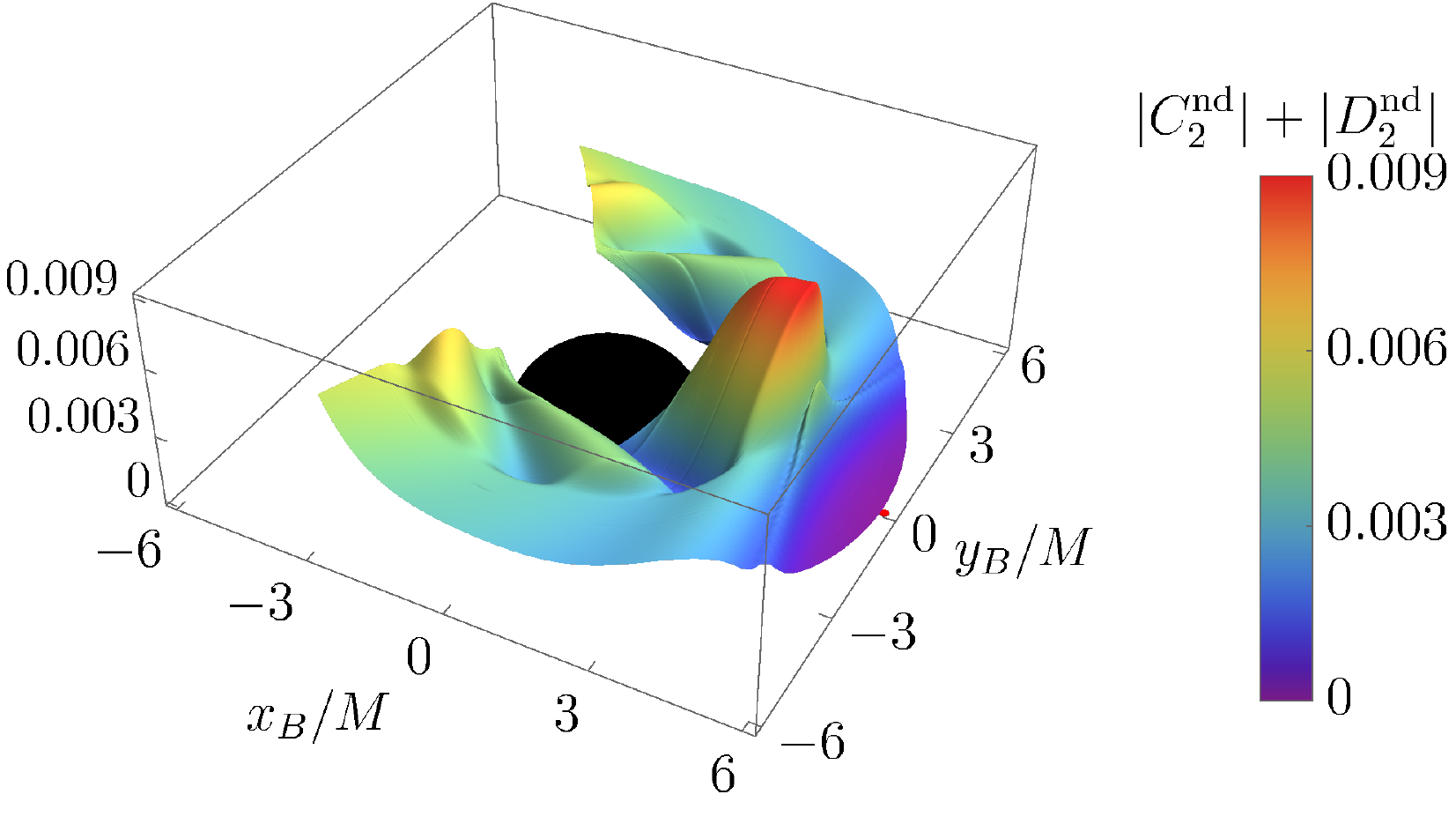}}
\subfloat[$B_2=25M$ and $\Omega_B=1/(2M)$
\label{fig:rainbowbottom}\label{fig:rainbow_d}]{ \includegraphics[width=.45\textwidth]{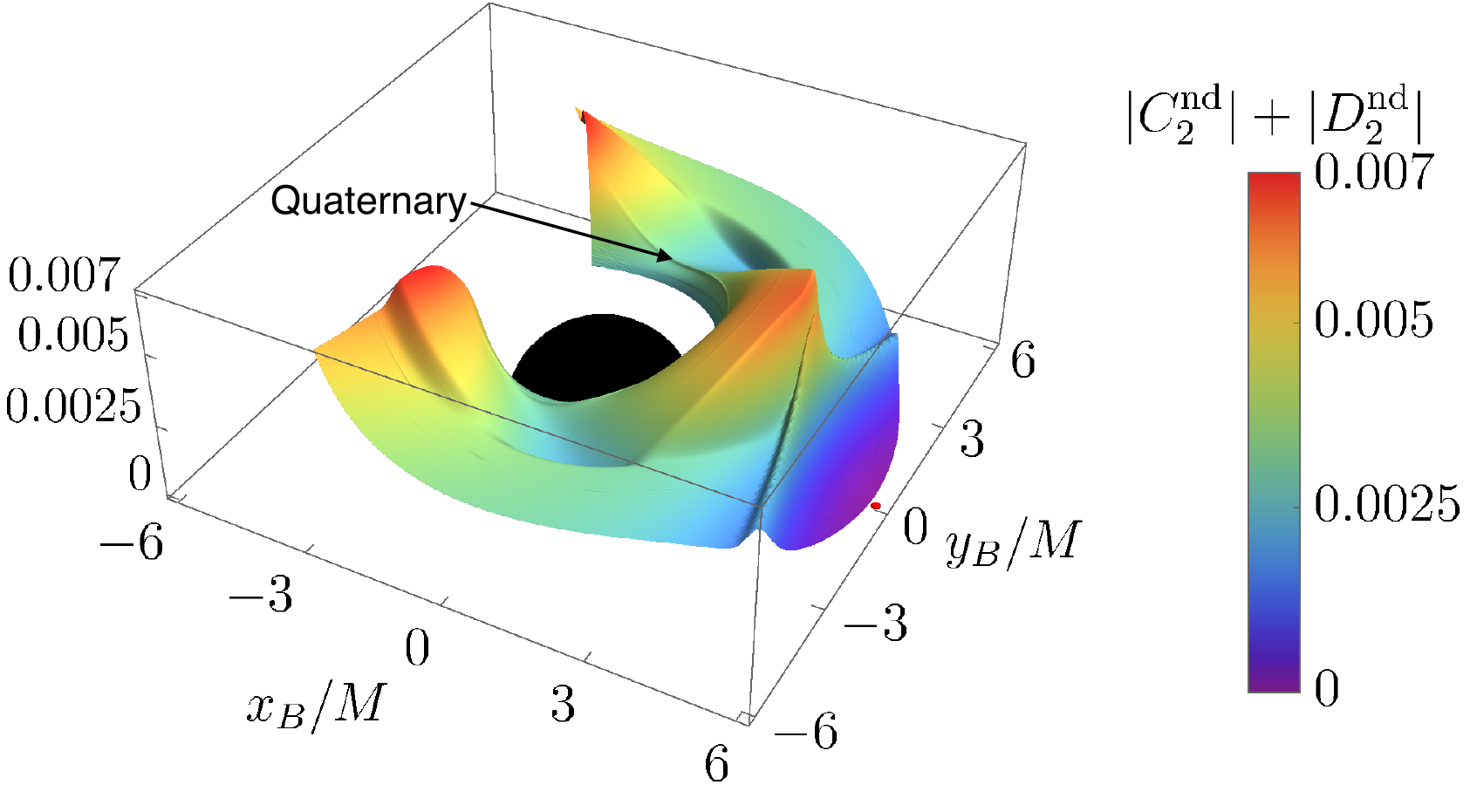}}
\caption{
Non-direct contributions $\left|C_2^{\textrm{nd}}\right|+\left|D_2^{\textrm{nd}}\right|$ for static detectors as a function of Bob's location for 
$A_1=B_1=0$, $A_2=M$, $\Omega_A=1/M$ and with further parameters as specified, comparing
two different values of $B_2$ and $\Omega_\db$.
A red dot indicates  Alice's location at $r_\da=6M$. Figs.~\ref{fig:rainbow_a} and \ref{fig:rainbow_b} show the non-direct contribution to the total signal strength of Fig.~\ref{fig:totalsignal_static}.
}
\label{fig:C2D2_NonDirect_RainbowPlots} 
\end{figure*}

The signaling scenario of Fig.~\ref{fig:totalsignal_static} is slightly different from the scenario that we just considered  in Fig.~\ref{fig:C2NonDirectB1ShiftPlot}.
Whereas in Fig.~\ref{fig:C2NonDirectB1ShiftPlot} the proper time window during which Bob couples to the field varies, while Bob remains at the same position,  Fig.~\ref{fig:totalsignal_static} compares the signal strength for different  (static) positions of Bob, while his proper time window is fixed.
At all the different positions, Bob switches on his detector at his proper time $\tau_\db=0$ which is when the first primary null geodesic from Alice reaches him, which emanated from Alice at her proper time $\tau_\da=0$. Thus, depending on Bob's spatial position, his switch-on happens at different coordinate times.
At all positions, Bob is switched off after a fixed amount of his proper time has passed. Hence, Bob's position determines to what extent his detector gets to interact with non-direct contributions to the signal.
Thus, while the 2-D plot of Fig.~\ref{fig:C2NonDirectB1ShiftPlot} plotted the signal strength as a function of the coupling times, Fig.~\ref{fig:totalsignal_static} and Fig.~\ref{fig:C2D2_NonDirect_RainbowPlots} are functions of the detector position.

Fig.~\ref{fig:C2D2_NonDirect_RainbowPlots} shows non-direct contributions for different detector and switching parameters, in particular, including the scenarios of Fig.~\ref{fig:totalsignal_static}. Some features which are due to secondary, tertiary and quarternary non-direct null geodesics are highlighted by labels.
Their characteristics are somewhat different from the characteristics observed in Fig.~\ref{fig:C2NonDirectB1ShiftPlot}, 
because now  all contributions are integrated up over a long interaction time of Bob whereas previously, in Fig.~\ref{fig:C2NonDirectB1ShiftPlot}, a short interaction time window of Bob was shifted over various switch-on times.
In particular, we find that the secondary null geodesics now create a single ripple in the 3D plots of Fig.~\ref{fig:C2D2_NonDirect_RainbowPlots} rather than a double-peak as observed above. Also, the tertiary null geodesics create a step-like feature rather than a peak as above.

Let us first focus on the ``outermost" distinct feature (i.e., the distinct feature at the largest radius for a fixed angle), which is a ripple, in the plots in Fig.~\ref{fig:C2D2_NonDirect_RainbowPlots}.
This ripple is a consequence of secondary null geodesics.
The  plots in Figs.~\ref{fig:rainbow_c} and \ref{fig:rainbow_d} show how the ripple moves to smaller  angular separation $\gamma$ than in Figs.~\ref{fig:rainbow_a} and \ref{fig:rainbow_b}.
That is, this outermost ripple moves closer to Alice for larger switch-off times $B_2$ of Bob. This is  expected since increasing $B_2$ means that the secondary null geodesics have more time to propagate around the black hole to reach Bob before he switches off his detector.
Whereas the position of the ripple  only depends on $B_2$, thus is identical for  Figs.~\ref{fig:rainbow_a} and \ref{fig:rainbow_b}, and Figs.~\ref{fig:rainbow_c} and \ref{fig:rainbow_d}, the shape of the ripple also depends on the energy gaps of the detectors. This becomes clear by comparing Fig.~\ref{fig:rainbow_a} to Fig.~\ref{fig:rainbow_b}, and  Fig.~\ref{fig:rainbow_c} to Fig.~\ref{fig:rainbow_d}, which only differ in Bob's detector energy changing from being identical to Alice's, $\Omega_\db=\Omega_\da=1/M$, to being half of Alice's, $\Omega_\db=1/(2M)$.

\begin{figure}[tb]
\begin{center}
 \includegraphics[width=0.45\textwidth]{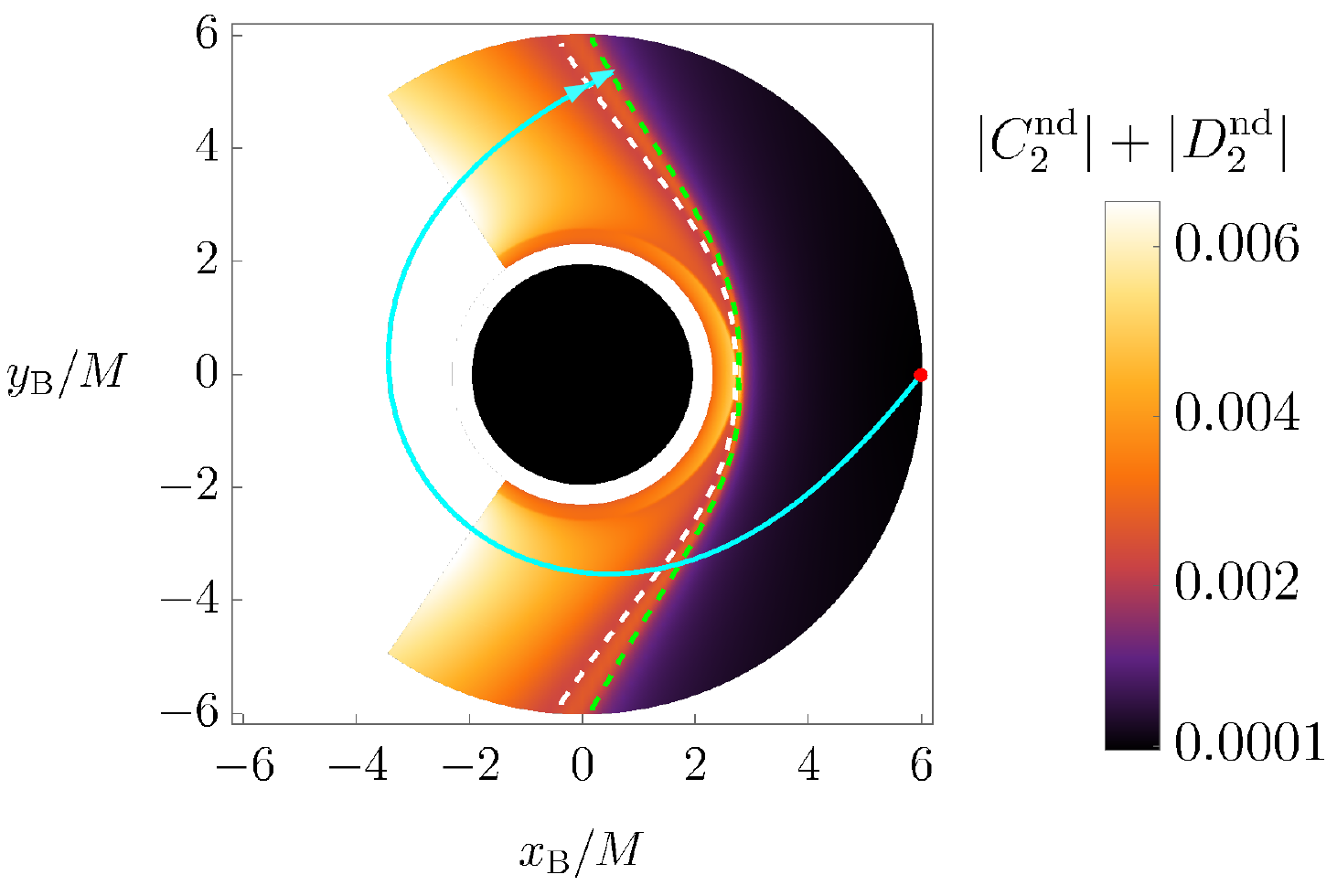}
 \end{center}
\caption{Non-direct contribution  $\left|C_2^{\textrm{nd}}\right|+\left|D_2^{\textrm{nd}}\right|$
for 
$A_1=B_1=0$, $A_2=M$, $\Omega_A=1/M$, 
$B_2=15M$ and $\Omega_B=1/(2M)$.
A red dot indicates  Alice's location.
This is a contourplot version of  Fig.~\ref{fig:rainbow_b}.
The dashed lines indicate how far the secondary null rays emitted by Alice have propagated at the time when Bob switches off his detector:
the green line, with smaller angular separation $\gamma$, shows the earliest  secondary null rays from Alice, and the white line, with  larger $\gamma$, shows the last secondary null rays from Alice.
}
\label{fig:C2,contour_new}
\end{figure}

The ripple has the form we expect it to have  based on Fig.~\ref{fig:PV_Cumu_short} of App.~\ref{app:secondary_geodesics_signal}, which shows the signal contribution from a $\pv\frac1\sigma$-distribution for the detector frequencies and sender switching times corresponding to the 3D plots of Fig.~\ref{fig:C2D2_NonDirect_RainbowPlots}.
Again in Fig.~\ref{fig:PV_Cumu_short}, the $\pv\frac1\sigma$-distribution is used to approximate the Green function along secondary null geodesics up to an overall prefactor. Thus, it yields the qualitative behaviour of the contribution from secondary null geodesics to the non-direct contribution of Fig.~\ref{fig:C2D2_NonDirect_RainbowPlots}.

In fact, Fig.~\ref{fig:C2,contour_new}, which is a contourplot of Fig.~\ref{fig:middlerainbow}, shows that  a (local) peak appears between the arrival of the first and of the last secondary null geodesic. This matches the behaviour of the $\pv\frac1\sigma$-signal  observed for $\Omega_\da=2 \Omega_\db=1/M$ in Fig.~\ref{fig:PV_Cumu_short}.
In Fig.~\ref{fig:C2,contour_new}, the  dashed green line  (i.e., the dashed line corresponding to the smaller separation angles $\gamma$ for a given radius) corresponds to points where the first  secondary null geodesics from Alice (i.e., emitted at $\tau_\da=A_1=0$) arrive at Bob's position exactly when Bob switches off the detector.
The dashed white line (i.e., the dashed line corresponding to the larger angles $\gamma$ for a given radius) corresponds to points where the last secondary null geodesic (i.e., emitted at $\tau_\da=A_2=M$) arrives at Bob's location when Bob switches off the detector.
Therefore, for the points with a value of $\gamma$ larger than that of a point on the larger-angle (white) dashed  line at the same radius, i.e., further away from Alice, {\it all} secondary null geodesics arrive while Bob's detector is switched on.
Whereas for points with a value of $\gamma$ smaller than that of a point on the smaller-angle (green) dashed line at the same radius, i.e., closer to Alice, {\it none} of the secondary null rays arrive before Bob's switch-off.
In-between the two dashed lines, i.e., where Bob switches the detector off roughly when Alice's ``middle" secondary null geodesic reaches him, lies the crest of the ripple and the magnitude of the non-direct contribution achieves a local maximum.

The appearance of the distinct ripple at the position that we just discussed, is a consequence of the coupling parameters that we have chosen for the numerical evaluation of the full non-direct contribution in Fig.~\ref{fig:C2D2_NonDirect_RainbowPlots}. 
The analytical solutions of App.~\ref{app:secondary_geodesics_signal}, which approximate the Green function near divergences,
show that when Alice emits longer signals, the signal strength from secondary null geodesics depends on Bob's total coupling duration in an oscillatory fashion up to about the time when Alice's last secondary null geodesic arrives at Bob, as seen in Fig.~\ref{fig:PV_Cumu_long}. In particular, the maximal magnitude of signal strength does not increase just because the duration of Alice's signal is increased.
However, if Bob tunes his detector resonant, i.e., $\Omega_\da=\nu \Omega_\da$, then the signal contribution from the secondary null geodesics  increases  roughly linearly with the duration of the signal.

\begin{figure}[tb]
    
    \subfloat[$r_\db\approx 2.583M$\label{fig:GF_integrand_stepfeature_top}]{\includegraphics[width=0.45\textwidth]{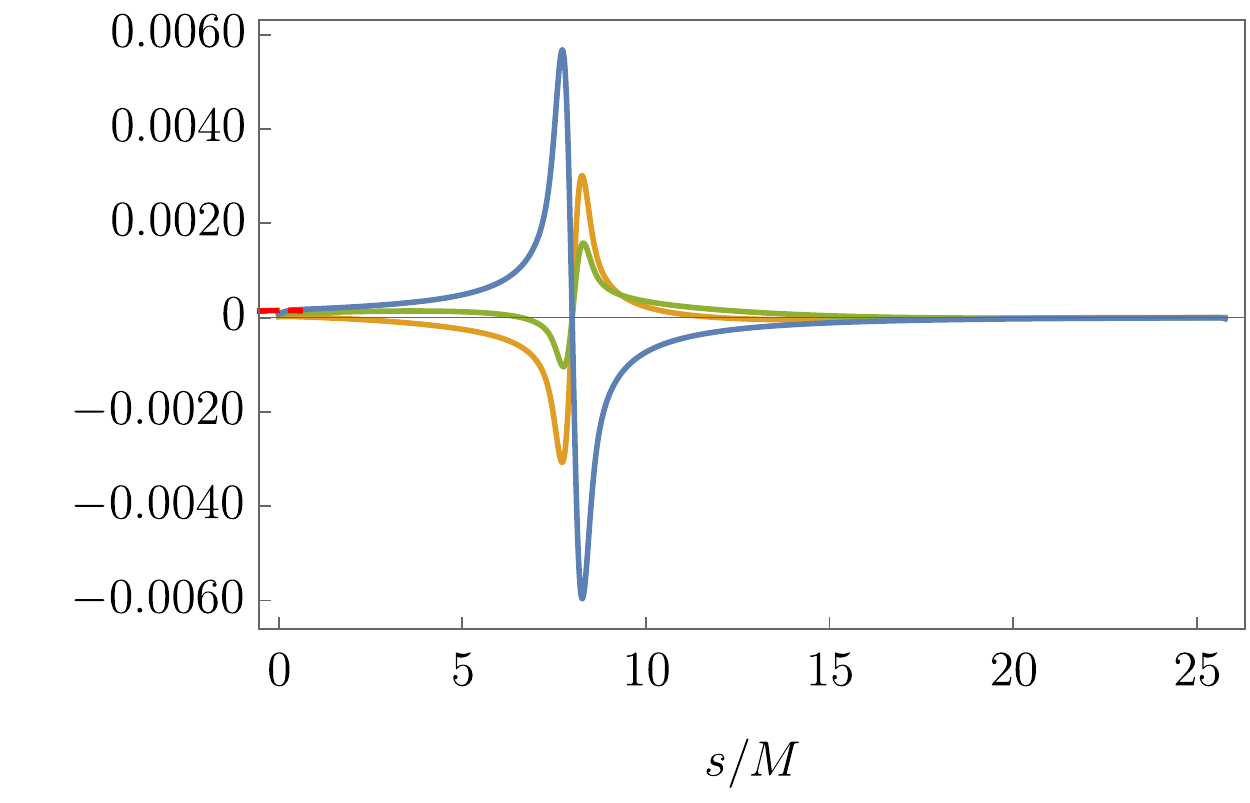}}
    
    \subfloat[$r_\db\approx 2.479M$\label{fig:GF_integrand_stepfeature_bottom}]{\includegraphics[width=0.45\textwidth]{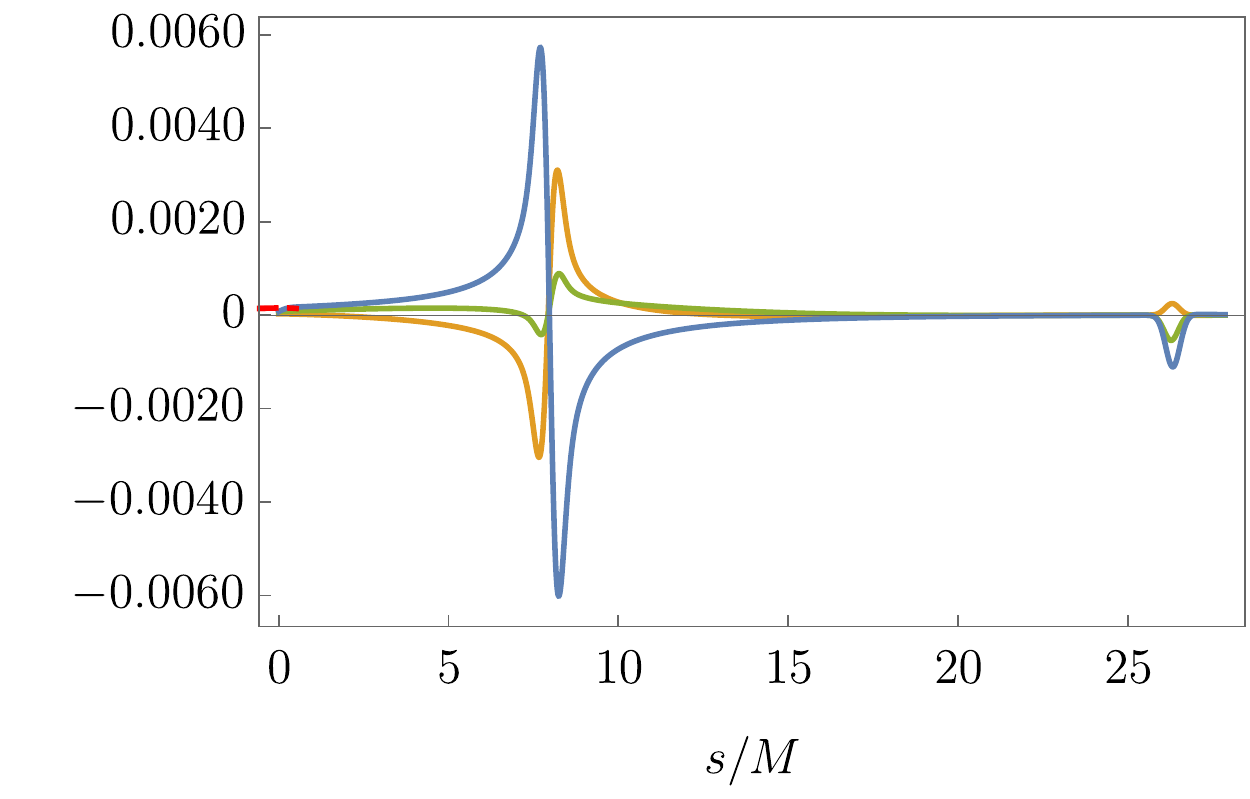}}    
\caption{Comparison of the integrand in $C_2^{\textrm{nd}}$ in Eq.~\eqref{eq:C2_static_aftervchange} for two different positions of Bob in the scenario of Fig.~\ref{fig:middlerainbow}:
The top/bottom plots correspond to positions of Bob located at the top/bottom of the steplike feature due to tertiary light rays in Fig.~\ref{fig:middlerainbow}. Both positions have angular separation $\gamma=69\pi/100$ from Alice, but 
with different radial coordinates.
The  orange and  green curves correspond to, respectively, the real and imaginary parts of the integrand (times $M$) in  Eq.~\eqref{eq:C2_static_aftervchange}.
The solid blue and dashed red curves correspond to the Green function $M^2G_{ret}/4\pi$   (which is a factor in that integrand) in, respectively, the DP and QL regions (extended to slightly negative values of $s/M$ for ease of visualization).
The horizontal axes contain the integration variable $s$ in Eq.~\eqref{eq:C2_static_aftervchange} divided by $M$.
(Recall that the switching parameters are  $A_1=B_1=0$, $A_2=M$, $B_2=15M$, $\Omega_A=1/M$ and $\Omega_B=1/(2M)$.)
}
\label{fig:GF_integrand_stepfeature} 
\end{figure}

In  Figs.~\ref{fig:rainbow_b} and \ref{fig:rainbow_d}  we can also see the effect of  {\it tertiary} light rays on the non-direct contribution. These rays are the main cause of the second outermost distinct feature in the plots, which is labelled ``tertiary" in Fig.~\ref{fig:middlerainbow}. 
This feature is not quite a ripple, like the  outermost feature  was, but it is more steplike, at least for certain angles away from $\gamma=0$.
The tertiary feature is also more localized than the secondary one, because the Green function has a $-\delta(\sigma)$-singularity for tertiary null geodesics. Hence, this singularity does not contribute to the integral in $C_2^{\textrm{nd}}$ if Bob is located on one side of the steplike feature in Fig.~\ref{fig:middlerainbow}, but it does contribute to $C_2^{\textrm{nd}}$ if Bob is located on the other side.
Perhaps less intuitive is the fact that the contribution from the tertiary lightrays can decrease the magnitude of the non-direct contribution. This occurs when the sign of the tertiary contribution to the signal is opposite to the earlier contributions, as illustrated in Fig.~\ref{fig:GF_integrand_stepfeature}.

Fig.~\ref{fig:GF_integrand_stepfeature} compares the integrand of expression \eqref{eq:C2_static_aftervchange} for $C_2^{\textrm{nd}}$,
as well as the Green function in that integrand,
for two different locations of Bob:  Fig.~\ref{fig:GF_integrand_stepfeature_top} corresponds to a location of Bob where  all tertiary null geodesics arrive only after Bob has switched off the detector, i.e., a location above the step-like feature in Fig.~\ref{fig:middlerainbow}.  On the other hand, Fig.~\ref{fig:GF_integrand_stepfeature_bottom} corresponds to a location where all tertiary null geodesics arrive while Bob is still coupled to the field.

The main features of the integrand are due to the $\pv(1/\sigma)$- singularity of the Green function around secondary null geodesics, which appears in both figures. In addition to that, the peak at the right end of Fig.~\ref{fig:GF_integrand_stepfeature_bottom} is due to the $-\delta(\sigma)$-singularity from tertiary null geodesics.
Whereas the exact expression for the integrand is singular at these places, the figure only plots the numerical approximation to the Green function of Sec.~\ref{sec:distantpast}.
This is why the singularities in the exact integrand appear smeared in these plots.

The full integration of the integrand plotted in Fig.~\ref{fig:GF_integrand_stepfeature_top} results in a  real (imaginary) part of $C_2^{\textrm{nd}}$ which is negative (positive).
However, the extra contribution due to the tertiary light rays in Fig.~\ref{fig:GF_integrand_stepfeature_bottom}  is  negative (positive) for the real (imaginary) part. Thus, the contribution from the tertiary rays  reduces the   real and imaginary parts of $C_2^{\textrm{nd}}$  and so also its absolute value. 
Hence, for the   parameters  at hand, the effect of the signal propagating along the tertiary null geodesics is to  reduce the non-direct contribution to the signal strength.

Additionally, Fig.~\ref{fig:rainbowbottom} even displays a feature due to quaternary lightrays, which is labelled ``quaternary".
This feature is, similarly to the secondary effect, like a ripple, as one would expect from the fact that the singularity of the Green function is of similar type (i.e., $\text{PV}\left(1/\sigma\right)$) at secondary and quaternary light-crossings.
We have checked that this feature is indeed due to quaternary rays by an analysis of the integrand similar to that described above for the feature due to tertiary rays.

This subsection explained the origin and nature of the modulation of the total leading-order signal strength $|C_2|+|D_2|$ observed in Fig.~\ref{fig:totalsignal_static},
which results from the combination of all possible null and timelike separations between Alice and Bob while they couple to the field.
(Note that the direct and non-direct contributions always had to be added coherently, before taking their absolute values, i.e., $|C_2|=|C_2^{\textrm{d}}+C_2^{\textrm{nd}}|$, and accordingly  $|D_2|=|D_2^{\textrm{d}}+D_2^{\textrm{nd}}|$.)
We found the  signal strength to  be generally  dominated by the direct contribution. 
Furthermore, the most distinct modulation of the total signal strength can be explained  by the parts of the signal which propagate along, or close to, secondary and higher-orbiting null geodesics.

\section{Radial infall towards a Schwarzschild black hole}\label{sec:infall}

Up to now we considered  scenarios where both detectors were static. In this section we instead consider the scenario where Bob continues to be static  but Alice is on a radially-infalling geodesic  in Schwarzschild spacetime.

\begin{figure}[tb!]
\begin{center}
  \includegraphics[width=0.45\textwidth]{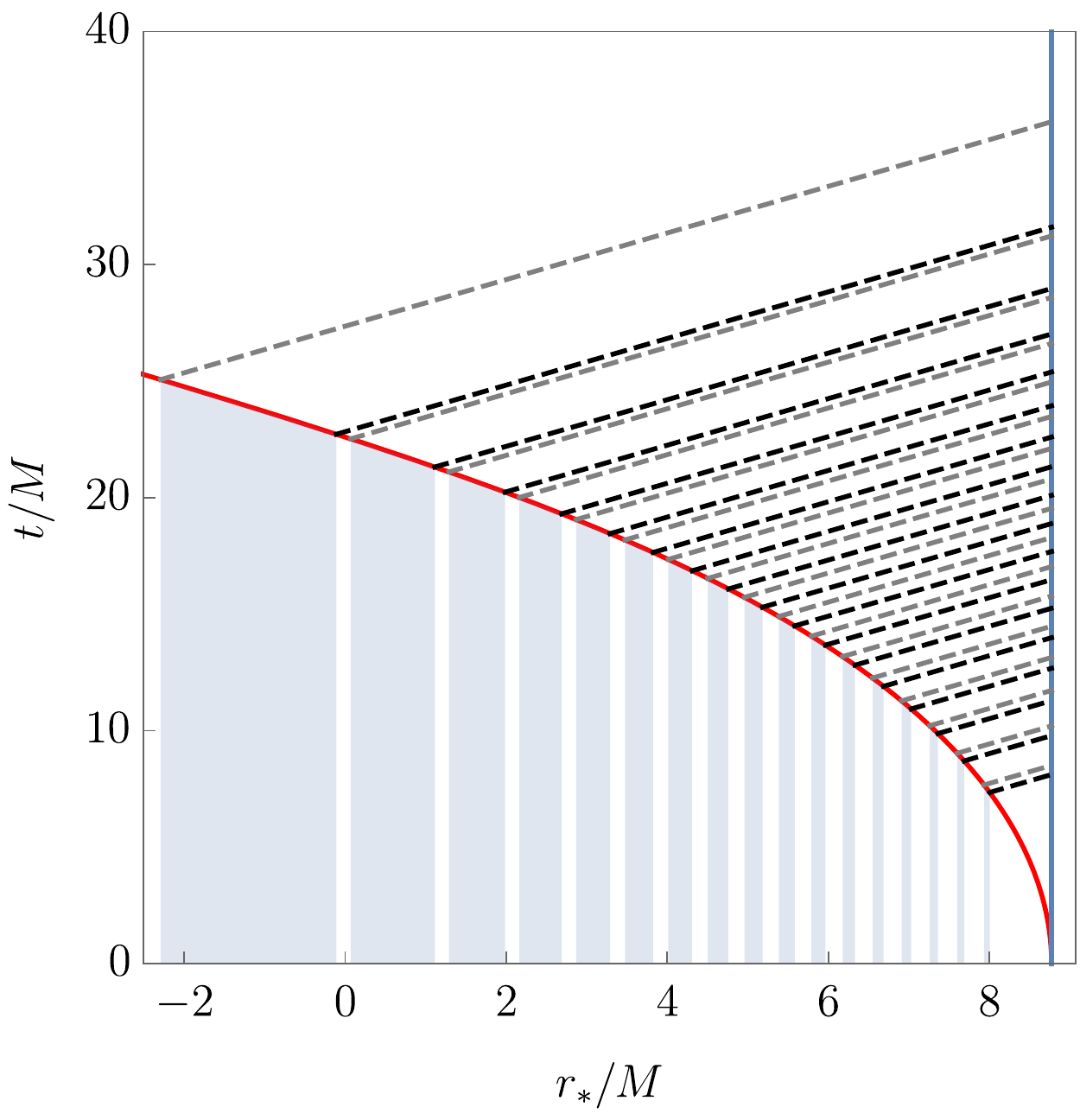}
  \end{center}
\caption{
Scenario where Alice follows a radially-infalling timelike geodesic starting from rest at $r=6M$, and Bob is static at $r=6M$ (which corresponds to $r_*=6M+2M\ln{4}$). Alice's worldline
$t=t(r_*(r_A))$
is given by the red line, Bob's
$t=t(r_*(r_B))$
by the blue vertical line. 
The  shaded regions indicate various coupling intervals during which Alice couples her detector to the field. The black-dashed lines represent the first radially-outgoing null geodesics emanating from Alice for  each interval, the grey-dashed lines represent the last ones.
All intervals last for the same amount of Alice's proper time  $\Delta\tau_\da=M/4$. 
Time windows starting later on Alice's worldline, i.e., closer to the horizon, extend over larger intervals of coordinate time (as well as of tortoise radial coordinate $r_*$).
}
\label{fig:radial infall}
\end{figure}

Specifically, Bob remains static at $r=6M$ and Alice's radially-infalling geodesic starts from rest at $r=6M$.
We compare the signal strength that arises when Alice switches on her detector at different points along her trajectory, but always for the same duration of her proper time: $\Delta\tau_\da=M/4$.
Bob correspondingly switches on/off his detector at the instant when he receives the direct radially-outgoing null ray emitted by Alice whenever she switches on/off her detector.
This  allows us to investigate how the channel capacity varies as Alice falls in.
We choose the detectors to be equal, i.e., $\Omega_\da=\Omega_\db$.
Fig.~\ref{fig:radial infall} illustrates how a coupling interval that starts when Alice is closer to the horizon thus extends over a longer interval in coordinate time (as well as in Bob's proper time).

\begin{figure}[tb!]
\begin{center}
    \includegraphics[width=0.45\textwidth]{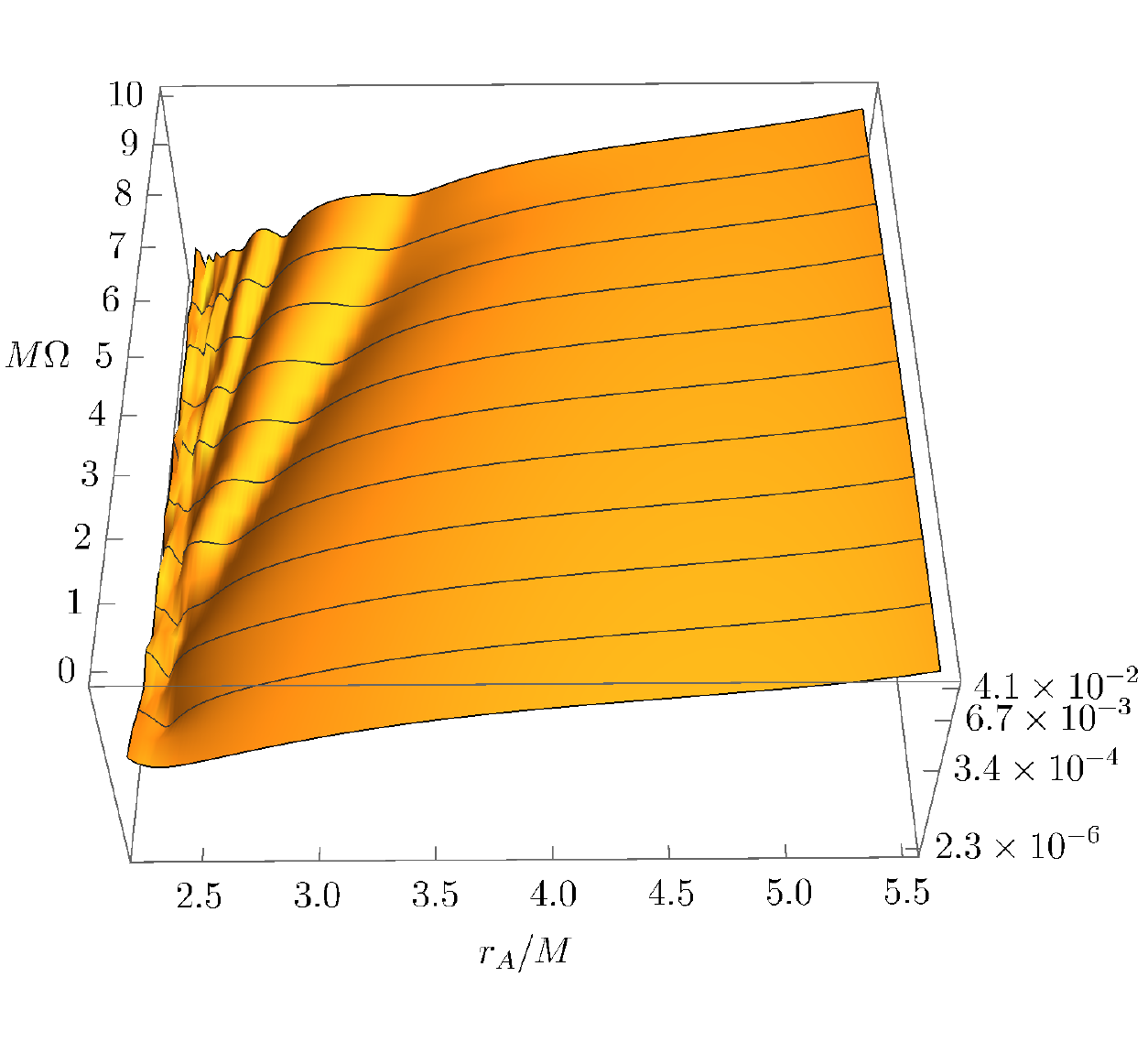}
    \\
    \includegraphics[width=0.45\textwidth]{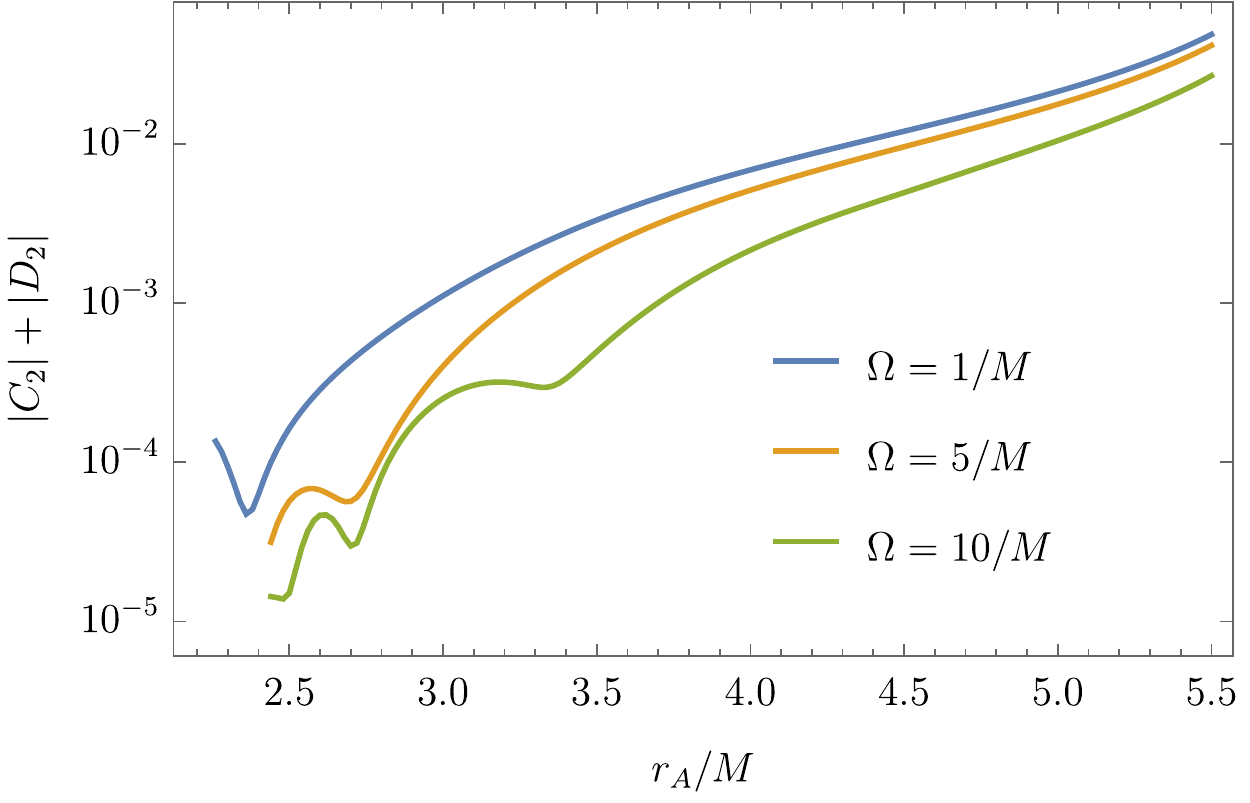}
  \end{center}
\caption{Plots of $|C_2|+|D_2|$ in the radial-infall scenario as a function of the radial coordinate $r_A$ (when Alice switches on her detector) and detectors' frequency $\Omega=\Omega_A=\Omega_B$. Independently of the value of $r_A$, Alice switches off her detector after a fixed amount $\Delta\tau_\da=M/4$ of her proper time. Top: 3D  plot (the red line corresponds to the case where $\Omega=1/M$).
Bottom: 2D plot as a function of $r_A$  for a sample of values of $\Omega$ (so these curves are just cross-sections of the 3D plot at the top).
}
\label{fig:CC2AndD2FullContributionRadialInfall}
\end{figure}

\begin{figure}[tb!]
\begin{center}
   \includegraphics[width=0.45\textwidth]{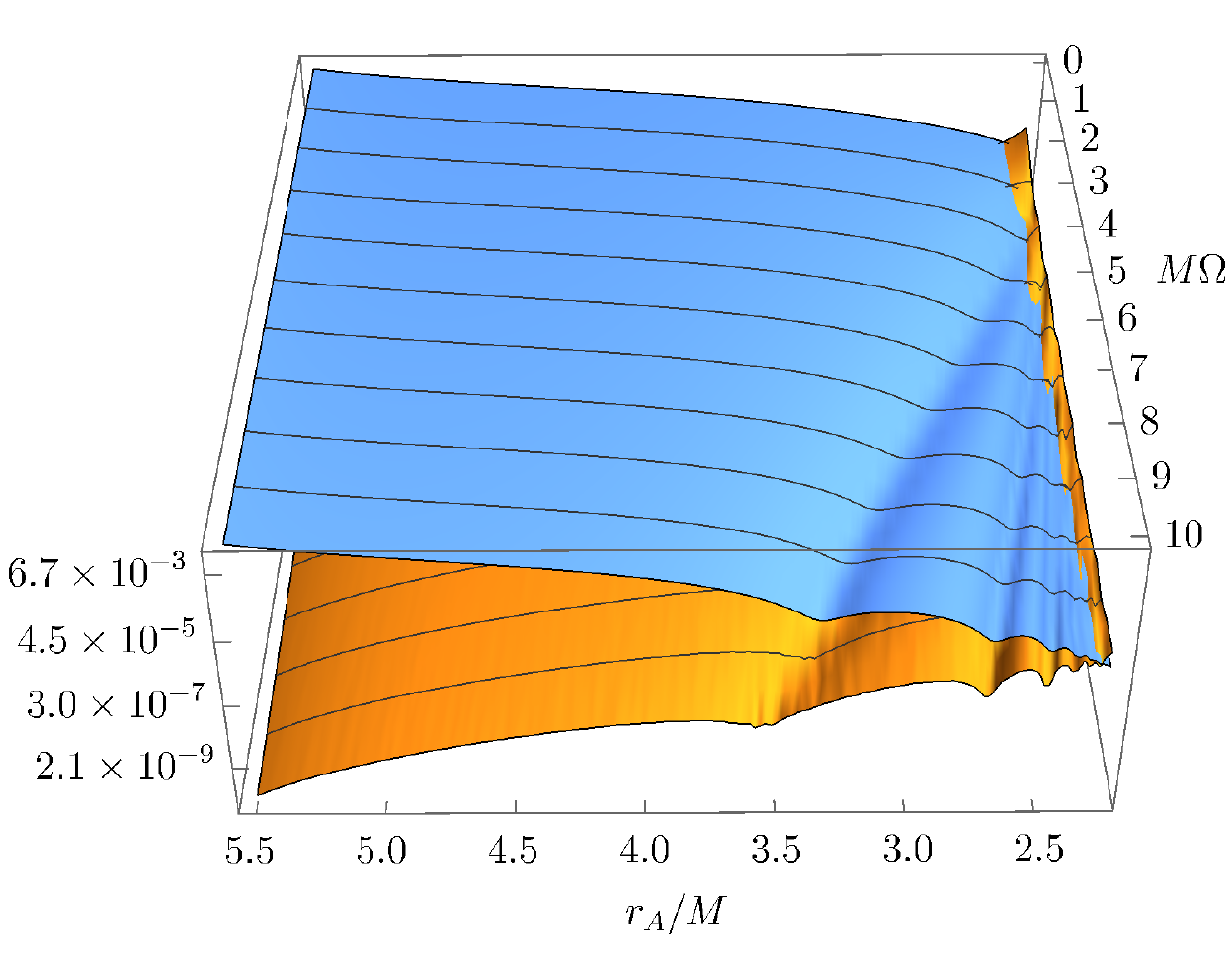}
   \\
   \includegraphics[width=0.45\textwidth]{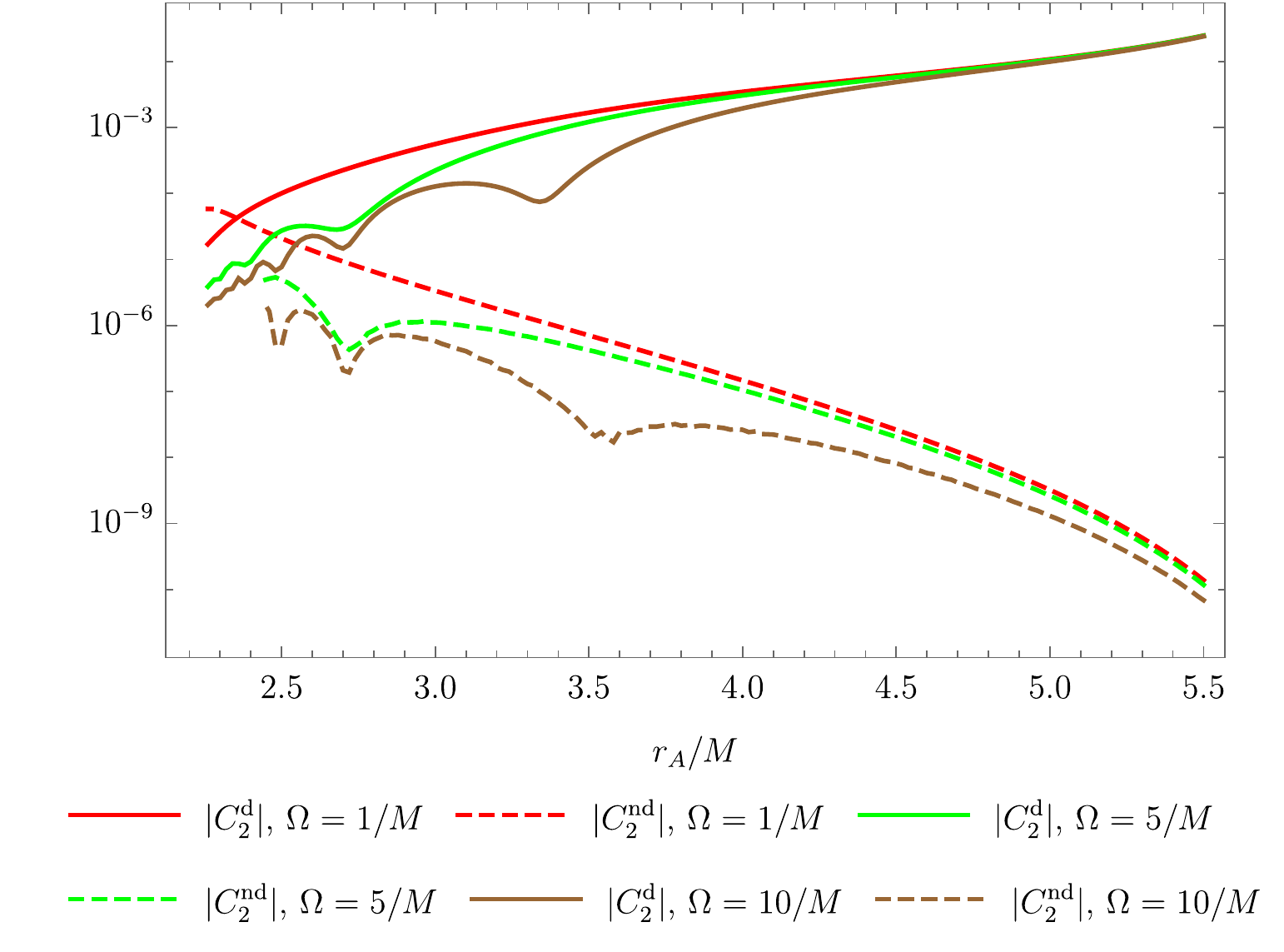}
  \end{center}
\caption{Plots of direct $|C_2^{\textrm{d}}|$ and non-direct $|C_2^\textrm{nd}|$ contributions to $C_2$ in the radial-infall scenario as functions of $r_A$ (when
Alice switches on her detector) and detectors' frequency $\Omega$.
Top: 3D plot where the blue and orange surfaces respectively correspond to
$|C_2^{\textrm{d}}|$ and $|C_2^\textrm{nd}|$.
Bottom: 2D plot as a function of $r_A$  for a sample of values of $\Omega$ (so these curves are just cross-sections of the 3D plot at the top).
}
\label{fig:C2DirectAndNonDirectRadialInfall}
\end{figure}

We used Eq.~\eqref{eq:C2d} to calculate the direct contribution $C_2^\textrm{d}$  to $C_2$,
where we used the property that $U\equiv 1$ along a radial null geodesic in Schwarzschild spacetime, as shown in App.~\ref{sec:Had}.
In its turn, for the non-direct contribution $C_2^{\textrm{nd}}$, we used Eq.~\eqref{eq:C2Gret}, where we did not include the direct term in $G_{ret}$. We obtained the equivalent expressions for $D_2$ by using Eq.~\eqref{eq:D2 from C2}.

In Fig.~\ref{fig:CC2AndD2FullContributionRadialInfall} we show   the total signal strength $|C_2|+|D_2|$ as a function of
two quantities:
Alice's radius $r_\da$ at which she switches on her detector and the frequency $\Omega:= \Omega_A=\Omega_B$ of both detectors.
The plot shows that, apart from some dips, the overall magnitude decreases as Alice approaches the horizon for fixed $\Omega$.
Also,  the signal strength decreases if $\Omega$ is increased while $r_\da$ is kept fixed, for most values of $r_\da$. However, closer to the horizon, where the signal strength is oscillatory, this ordering is broken.

In order to understand better this behaviour of the total signal strength $|C_2|+|D_2|$ seen in Fig.~\ref{fig:CC2AndD2FullContributionRadialInfall}, 
it is helpful to consider the direct and non-direct contributions to the signal strength separately.
In Fig.~\ref{fig:C2DirectAndNonDirectRadialInfall} we thus plot the direct and non-direct contributions to $C_2$ (the contributions to $D_2$ behave similarily).
The figure shows that the direct part is the dominant contribution to $C_2$ throughout most of the phase space $(r_A,\Omega)$ that we have covered, except for when Alice is near the horizon, in which case the non-direct contribution dominates:
As a consequence of the fact that, as Alice switches on her detector  closer to the horizon, the non-direct contribution $C_2^{\textrm{nd}}$ increases whereas the direct contribution $C_2^{\textrm{d}}$ decreases,  two distinct regions arise.
For $2.3M\lessapprox r_A$\footnote{This region varies slightly with $\Omega$: it is $2.3M\lessapprox r_A$ for $\Omega=10/M$ and 
$2.35M\lessapprox r_A$ for $\Omega=1/M$.}, the non-direct contribution is up to many orders of magnitude smaller than the direct contribution. (This also matches our general observations for static detectors in Sec.~\ref{sec:schwarzschild_static_nondirect} and, e.g., Fig.~\ref{fig:C2NonDirectB1ShiftPlot}.)
On the other hand, for $r_A\lessapprox 2.3M$, the non-direct contribution to $C_2$ becomes larger than the direct one.
We also note that Alice's switch-on radius where the non-direct and the direct contributions are of the same size  is approximately  $r_A\approx 2.3M$ for all $\Omega$. %

What are the reasons for  the two contributions to behave differently as Alice approaches the horizon?
The decrease in $|C_2^{\textrm{d}}|$ can be explained since this contribution essentially diminishes inversely-proportionally to the affine parameter distance along a radial null geodesic between Alice and Bob.

In its turn, the increase in $C_2^{\textrm{nd}}$ is probably related to the singular structure of the Green function and the arrival of secondary null geodesics.
The singular structure of secondary null rays (i.e., $-\textrm{PV}\frac{1}{\sigma}$\footnote{We note, however, that, in this setting where Alice and Bob are radially separated (i.e., $\gamma=0$), the secondary (and any higher-orbiting) null geodesic meet, as a whole one-dimensional envelope, at a caustic point. The divergence of the Green function at caustics is ``enhanced" and its precise form is given in~\cite{casals2016global}.}) means that the retarded Green function increases monotonically near the arrival of the secondary null rays. 
The closer to the horizon Alice is when she couples to the field, the smaller is the coordinate time interval between the time when the first secondary null geodesics from Alice's coupling reach Bob's position and the (earlier) time when Bob switches off the detector. 
The reasons for that are two-fold: Firstly, when Alice couples closer to the horizon then Bob is switched on for a longer interval of coordinate time (as seen in Fig.~\ref{fig:radial infall}). Secondly, it takes secondary null geodesics less coordinate time to propagate from Alice to Bob, when they emanate from Alice closer to the horizon.

Essentially, the non-direct contribution grows as Alice's switch-on radius approaches the horizon, since it  is ``anticipating" the arrival of the secondary null rays, which arrive closer to Bob's switch-off.
In fact, if first secondary null geodesics from Alice were reaching Bob before Bob's switch-off, then the non-direct contribution could decrease from that point again, due to the $-\pv\frac1\sigma$ structure of the Green function (as discussed in Sec.~\ref{sec:schwarzschild_static_nondirect} and App.~\ref{app:secondary_geodesics_signal}). However, we have  checked that, for our range of parameters, no secondary null geodesics have  time to reach Bob before he switches off his detector  (as opposed to the static case in Sec.~\ref{sec:schwarzschild_static_nondirect}). Thus the non-direct contribution grows monotonically as Alice's radius decreases over the range of radial coordinates considered here.

Fig.~\ref{fig:C2DirectAndNonDirectRadialInfall} also illuminates the origin of the dips in the total signal strength observed in Fig.~\ref{fig:CC2AndD2FullContributionRadialInfall}:
The dips (which are dips because we plot absolute values of quantities, but would be oscillations if we did not take the absolute value)  appear separately in both $C_2$, as seen in Fig.~\ref{fig:C2DirectAndNonDirectRadialInfall}, and in $D_2$, as we have checked separately.
Furthermore, we see that dips also appear separately in the direct contribution $C_2^{\textrm{d}}$ and the non-direct contribution $C_2^{\textrm{nd}}$.

The appearance of the dips is similar to the case of static observers where we observed them, e.g., in the direct contribution in Fig.~\ref{fig:direct_signal_strength}, which is based on Eq.\eqref{eq:C2d static}.
Just as there, also here in the case of an infalling sender, it is plausible that the dips are due to the relative detuning between sender and receiver which is caused both by the motion of the sender and their gravitational red-shift.
In fact, the motion of the sender seems to give rise to a certain difference between the static and the infalling scenario. Whereas in the infalling scenario, as shown in Fig.~\ref{fig:direct_signal_strength}, the integrals in the direct contribution vanish completely at certain points, the same does not happen for the case of the infalling sender considered here.

{We note that, around $\Omega=1/M$, the first dip in $|C_2|+|D_2|$ as a function of $r_A$ (see Fig.~\ref{fig:CC2AndD2FullContributionRadialInfall}) happens around the same radius as where the main contribution to $|C_2|$ swops between the non-direct and the direct contributions (see Fig.~\ref{fig:C2DirectAndNonDirectRadialInfall}).
However, this is a coincidence around $\Omega=1/M$, as can be seen by comparing the two figures at other values of $\Omega$.}

To conclude, the   infalling sender scenario of this section 
demonstrates that the non-direct contribution to the signal is essential for the calculation of the leading order signal strength close to the horizon. As Alice falls towards the black hole, it is increasingly difficult for her to send a signal back to Bob. However, the decrease in signal strength is not reduced as one expects if only considering direct null geodesics, but the non-direct contribution can counteract the loss of signal strength to some extent in the proximity to the horizon.

\section{Discussion}\label{sec:discussion}

We analyzed the quantum channel between two first-quantized qubit quantum systems that communicate via a quantum field.
To this end, we applied tools for the study of wave propagation in curved spacetimes and self-forces (see Apps.~\ref{sec:Had} and \ref{sec:CID} as well as~\cite{CDOWb,mark2017recipe,ModesDirectPart,Ottewill:2009uj,Ottewill:Wardell:2008}) to the study of particle detectors in quantum field theory in curved spacetimes.  

We then specialized to the case of a Schwarzschild black hole and identified three different contributions to the information exchanged between two particle detectors: the information carried by direct null geodesics, the information carried by black hole-orbiting (secondary, tertiary, etc) null geodesics, and  the information carried by timelike communication which arises due to the violation of the strong Huygens principle.

In summary, while usually in studies of communication in the presence of black holes the mechanism of communication is not described from first principles, we here worked out the communication between  emitters and receivers that are quantum and that exchange signals through a quantized field. In doing so, we have found several effects that were not anticipated by previous studies in flat spacetime. For example, we have determined the regimes in which the contributions of timelike and non-direct null-signals are relevant, and we have found an emergent transparency of particle detectors near the event horizon.

Concretely, we considered two distinct setups: one with two static detectors and another one with a static receiver and a radially-infalling emitter.
In the case of static detectors, we found that black hole-orbiting  null  geodesics  create, due to the corresponding
singularity structure of the Green function, ripples  or  steplike 
features in the total signal strength,
as a function of the  receiver location,
that depend  on  whether  the  number  of  orbits  is  even  or  odd, respectively (Sec.\ref{sec:static,nd}). Also, generally, the contribution to the total signal strength from direct null geodesics dominates over the non-direct contribution. 

However, we also found that
in the case of a radially-infalling emitter and a static receiver, the non-direct contribution (which consists  both of black-hole orbiting null geodesics and, due to the violation of the strong Huygens principle, to timelike signals, neither of which  possessing an analog in flat spacetime), dominates over the direct contribution when the emitter is near the horizon (Sec.\ref{sec:infall}). 

Further, 
in the case of radially-separated and non-resonant detectors (whether both static or with an infalling emitter), the total signal strength
has dips as a function of the radius of one of the detectors due to the relative
detuning  between  the detectors  as  caused
by the combination of their motion and their gravitational
red-shift
(Eq.\eqref{eq:C2d static} and Sec.\ref{sec:infall}). All these features are specific to the way waves propagate on curved (and, particularly,  Schwarzschild) spacetime.

Also, a particularly interesting and perhaps unintuitive result is that, as a stationary receiver, Bob, is placed closer and closer to the horizon, the amount of information that Bob receives from a stationary emitter (that is fixed further away from the horizon) diminishes in the sense that Bob becomes increasingly transparent for incoming signals. 
This is the case independently of the tuning of the resonance frequency of the receiver, including the case where it is chosen resonant with the blueshifted signal arriving from the sender. 
Technically, this phenomenon is related to the fact that the duration  in proper time  of the receipt of a message from the sender diminishes as the receiver is placed closer to the horizon and the signal is blueshifted. A similar phenomenon arises in the resonant driving of simple classical and quantum harmonic oscillators. Their amplitude response decreases if the oscillator frequency is increased but the driving force is kept constant and kept driving until the same number of oscillations is reached.

This is because the amplitude of the driven oscillator is the Fourier integral of the driving force evaluated for the duration of the driving. At resonance, the integrand is at stationary phase, i.e., in this case, the integrand is constant. Therefore, during resonance, the amplitude ramps up proportionally to the driving time. Similarly in our case here, the less time the receiver spends in contact with the sender's signal, the less its amplitude can build up. As the frequency of the signal is increased by blueshifting, the same number of oscillations is reached earlier leading to less resonant amplitude build-up. 

Finally, the methods and results presented here should generally be useful for further investigations in the field of relativistic quantum information. A key example would be the study of not only the classical but also the quantum channel capacity of the quantum channel between first-quantized systems that communicate via a quantum field in curved spacetime. For example, in our second scenario, where Alice is at a fixed radius, and Bob is moved closer and closer to the horizon, we found that Bob loses classical channel capacity. It would be interesting to track also the quantum channel capacity in that case, i.e., to track the ability of the quantum channel from Alice to Bob to transmit entanglement with an ancilla.

\section{Acknowledgements}

RHJ  acknowledges support from the Knut and Alice Wallenberg Foundation, ERC Advanced grant 321029, the VILLUM FONDEN via the QMATH center of excellence (grant no. 10059), and by the Wenner-Gren Foundations.
MC is thankful to Abraham Harte for useful discussions on the Hadamard bitensors.
MC acknowledges partial financial support by CNPq (Brazil), process number 310200/2017-2. EMM and AK acknowledge support through the NSERC Discovery program. EMM acknowledges support through an Ontario Early Researcher Award and AK acknowledges support through a Google Faculty Research Award. DQA acknowledges support from FAPERJ (process number 200.804/2019) and CNPq (process number 140951/2017-2).

\appendix

\begin{widetext}
\section{Upper bound on leading order signal strength}\label{app:propertime_bound}

In this appendix we wish to show how for arbitrary wordlines of Alice and Bob the upper bound \eqref{eq:upperbound_body} on the leading order signal strength $|C_2|+|D_2|$ arises  from the amount of proper time during which the receiver interacts with the field.
First, using Eq.~\eqref{eq:C2Gret} we see that
\begin{align}
    |C_2|&=\left| \frac{-\ii}{4\pi}\integral{\tau_\db}{-\infty}{\infty} \eta_\db(\tau_\db) \ee{\ii \Omega_\db \tau_\db} \integral{\tau_\da}{-\infty}{\tau_\da(t(\tau_\db))}  \eta_\da(\tau_\da)\ee{-\ii\Omega_\da\tau_\da} G_{ret}(\coord{x_\db}(\tau_\db),\coord{x_\da}(\tau_\da))
    \right|\\
&    \leq\frac1{4\pi}\integral{\tau_\db}{B_1}{B_2}\left|  \integral{\tau_\da}{-\infty}{\tau_\da(t(\tau_\db))}  \eta_\da(\tau_\da)\ee{-\ii\Omega_\da\tau_\da} G_{ret}(\coord{x_\db}(\tau_\db),\coord{x_\da}(\tau_\da))
    \right|,
\end{align}
\end{widetext}
where we use the fact that the support of the receiver's switching function is the interval $[B_1,B_2]$. 

For a given signal emitted by Alice, i.e., for every choice of switching parameters, detector frequency and wordline of Alice, we can consider
\begin{align}
    F(\coord{x_\db}):=\left|  \integral{\tau_\da}{-\infty}{\tau_\da(t(\coord{x_\db}))}  \eta_\da(\tau_\da)\ee{-\ii\Omega_\da\tau_\da} G_{ret}(\coord{x_\db},\coord{x_\da}(\tau_\da))
    \right|
\end{align}
as a function of Bob's position in spacetime.
Let $\mathbb B $ denote a region of spacetime
containing the part of Bob's worldline where he couples his detector to the field.
The bound \eqref{eq:upperbound_body} now follows immediately, if $F$ is bounded in $\mathbb B$:
with
\begin{align}\label{eq:CB}
    C_{\mathbb B}:=\frac1{2\pi}\sup_{\coord{x_\db}\in\mathbb B} F(\coord{x_\db})
\end{align}
and
using $D_2(\Omega_\da,\Omega_\db)=-C_2(\Omega_\da,-\Omega_\db)$,
we obtain that
\begin{align}\label{eq:bound C2+D2}
    |C_2|+|D_2|\leq C_{\mathbb B}\, (B_2-B_1).
\end{align}
Hence the question is under what conditions is $F$  bounded in $\mathbb B$.

A necessary condition on $\mathbb B$ is that it does not contain the part of Alice's wordline where Alice couples to the field.
In fact, there needs to exist a neighbourhood of the points on Alice's worldline at which her detector is coupled to the field which does not intersect with $\mathbb B$.
Otherwise, $F(\coord{x_\db})$ will diverge as $\coord{x_\db}$ approaches such a point of Alice's worldline, due to the contribution from the Dirac $\delta$-distribution part in the Hadamard form for $G_{ret}$ (see Sec.~\ref{sec:GF_quasilocal} and, in particular, Eqs.~\eqref{eq:hadamard} and \eqref{eq:delta_of_synge_staticdetectors}).

Apart from this restriction on $\mathbb B$, we expect on physical grounds that the function $\mathbb B$ is always bounded: 
This is because $F(\coord{x_\db})$ corresponds to  the amplitude of a solution to the Klein-Gordon equation with a source term given by that of a point (unit) scalar charge at Alice's worldline evaluated at Bob's worldline (multiplied by $\eta_\da(\tau_\da)\ee{-\ii\Omega_\da\tau_\da}$).
Hence, we expect $F(\coord{x_\db})$ to  be  bounded in all of spacetime outside of the neighbourhood of Alice's wordline.
A detailed proof of boundedness needs to consider the geometry of the given spacetime and resulting properties of $G_{ret}$, paying particular attention  to its singular contributions.

In Schwarzschild spacetime, we observe that the singular contributions from the $\delta(\sigma)$-distribution and $\pv\frac1\sigma$-distribution to the Green function, which appear for points connected by null geodesics, whether direct or black-hole--orbiting ones, always result in a finite value of $F(\coord{x_\db})$ if the switching function $\eta_\da$ is sufficiently differentiable.
(The behaviour of $F$ at caustic points of the spacetime requires further detailed analysis.)

For discontinuous sudden switching functions the argument above does not apply, because $F(\coord{x_B})$ generally would not be bounded, e.g., due to the $\pv\frac1\sigma$-singularities.
Nevertheless, we find that even for discontinuous sudden switching functions for both detectors  the null singularities of $G_{ret}$ give a contribution to the signal strength $|C_2|+|D_2|$ which also obeys a linear bound in  $B_2-B_1$, as above.
That is, we find analytical solutions to the contributions to $|C_2|+|D_2|$ resulting from the $\delta(\sigma)$-singularity in Sec.~\ref{sec:schwarzschild_static}, and from the $\pv\frac1\sigma$-singularity in  App.~\ref{app:secondary_geodesics_signal}.
These solutions are finite and bounded linearly in $B_2-B_1$.

\section{Hadamard bitensors}\label{sec:Had}

In this appendix we present
the calculation of the van Vleck determinant
$\Delta$ and the derivatives of the world function.
In Sec.\ref{sec:transp eqs} we 
give the system of coupled transport equations satisfied by
these quantities and
in Sec.\ref{sec:Had-rad null}
 we give their analytical values  along a radial null geodesic in Schwarzschild spacetime.

\subsection{Transport equations}\label{sec:transp eqs}

The van Vleck determinant $\Delta(x,x')$ between two points $x$ and $x'$ (in a normal neighbourhood of  $x$) obeys the following transport equation along the unique geodesic connecting the two points~\cite{Poisson:2011nh,Ottewill:2009uj}:
\begin{equation} \label{eq:VV}
\difffrac{\Delta^{1/2}}{\lambda}=\frac{1}{2 \lambda }\left(4-\sigma ^{\alpha}{}_{\alpha}\right)\Delta^{1/2},
\end{equation}
where $\lambda$ is an affine parameter along the geodesic and $\sigma ^{\alpha}{}_{\beta}:= \nabla^{\alpha}\nabla_{\beta}\sigma(x,x')$.
The initial condition for Eq.~\eqref{eq:VV} is $\Delta(x,x)=1$.

In their turn, the covariant derivatives of the world function can be obtained by solving the following transport equation~\cite{Ottewill:2009uj,PhysRevD.92.104030}:
\begin{align} \label{eq:transp eq Q}
&
\difffrac{Q^{\alpha}{}_{\beta}}{\lambda}=
u^{\delta} Q^{\alpha}{}_{\gamma} \Gamma^{\gamma}{}_{\beta \delta}-
u^{\delta} \Gamma^{\alpha}{}_{\gamma \delta} Q^{\gamma}{}_{\beta}-
\nonumber \\ &
\frac{1}{\lambda }(Q^{\alpha}{}_{\gamma} Q^{\gamma}{}_{\beta}+
Q^{\alpha}{}_{\beta})-
\lambda  R^{\alpha}{}_{\gamma\beta\delta} u^{\gamma}u^{\delta},
\end{align}
 where $Q^{\alpha}{}_{\beta}:= 
\sigma ^{\alpha}{}_{\beta}
 -\delta^{\alpha}{}_{\beta}$, $u^{\alpha}=\mathrm{d}x^{\alpha}/\mathrm{d}\lambda$ is a tangent vector to the geodesic between $x$ and $x'$, $\Gamma^{\gamma}{}_{\beta \delta}$ 
 are the Christoffel symbols
 and $R^{\alpha}{}_{\gamma\beta\delta}$ are the components of the Riemman tensor.
 The initial condition for Eq.~\eqref{eq:transp eq Q} is $Q^{\alpha}{}_{\beta}(x,x)=0$.
 Given the symmetry $\sigma_{\alpha\beta}=\sigma_{\beta\alpha}$, Eqs.\eqref{eq:transp eq Q} form a set of 10 coupled, nonlinear, first-order ordinary differential equations.

We numerically solved Eqs.(\ref{eq:VV}) and (\ref{eq:transp eq Q}) simultaneously using the code in~\cite{Wardell-transport-Code}.
In the particular case of radial null geodesics we can make analytical progress, as we show in the next subsection.

\subsection{Van Vleck determinant and $\sigma ^{\alpha}{}_{\beta}$ along radial null geodesics in Schwarzschild}\label{sec:Had-rad null}

The statement
in Sec.~\ref{sec:GF_quasilocal} that
$\Delta(\coord{x},\coord{x}')= 1$  along
a radial null geodesic in Schwarzschild spacetime
simply follows from the
fact that 
the Penrose limit 
(essentially, a limit of 
the geometry near a null geodesic in an arbitrary spacetime, which yields
a plane wave
spacetime that encodes various properties of the original spacetime)
is flat for radial null geodesics in Schwarzschild spacetime~\cite{hollowood2009refractive}.

We have furthermore numerically observed that
$\Delta(\coord{x},\coord{x}')= 1$  along
a radial null geodesic in Schwarzschild spacetime by solving the transport equation given above.
We note that this result   has been independently derived, and extended to any null geodesic
tangent to any principal null direction in any vacuum spacetime (including Kerr),  in~\cite{Harte}.

We can in fact go further and analytically obtain
$\sigma ^{\alpha}{}_{\beta}$ along a radial null geodesic.
Denoting by an overdot the derivative with respect $\lambda$, the 4-velocity of a radial null geodesic with energy $E$ can be written as
$$u^\mu=(\dot{t},\dot{r},\dot{\theta},\dot{\phi})=E(f^{-1},\epsilon,0,0)=:Et^\mu$$
where  $\epsilon$ is equal to -1 (1) for ingoing (outgoing) geodesics. Since $\dot{r}$ is constant, we are able to use $r$ as an affine parameter.
From now on all expressions will be valid for radial null geodesics.

Eqs.\eqref{eq:VV} and \eqref{eq:transp eq Q} then take on the forms
\begin{align}
    (r-r')\difffrac{\Delta^{1/2}}{r}&=-\frac{1}{2}{Q^\alpha}_\alpha\Delta^{1/2},\\
    (r-r')\difffrac{{Q^\mu}_\nu}{r}&=\,
    \epsilon(r-r')\lP{Q^\mu}_\alpha\Gamma^\alpha_{\nu\beta}t^\beta-{Q^\alpha}_\nu\Gamma^\mu_{\alpha\beta}t^\beta\rP-
\nonumber    \\&
    {Q^\mu}_\alpha{Q^\alpha}_\nu-{Q^\mu}_\nu
    -(r-r')^2{R^\mu}_{\alpha\nu\beta}t^\alpha t^\beta.
    \label{eq:transp eq Q rad}
\end{align}
From $\sigma_{\alpha\beta}=\sigma_{\beta\alpha}$ it follows that 
\begin{align}\label{eq:Qrt-Qtr}
    {Q^r}_t=-f^2{Q^t}_r,
\end{align}
and from the symmetries of the physical setup here it follows that
\begin{align}\label{eqn:QNSolution}
    {Q^A}_B(x,x')=&\,0,
    \end{align}
for all $A,B\in \{\theta,\phi\}$.

Using \eqref{eqn:QNSolution} in Eqs.\eqref{eq:transp eq Q rad} for the nonzero components ${Q^t}_t$, 
${Q^r}_r$ and
${Q^t}_r=-f^{-2}{Q^r}_t$,
 we obtain
\begin{align}\label{eqn:firstTE}
   & (r-r')\difffrac{{Q^t}_t}{r}=\,-\frac{2M(r-r')^2}{r^3f}-{Q^t}_t-({Q^t}_t)^2
   +
     \\&
   \lP f {Q^t}_r\rP^2
      +\frac{\epsilon 2M(r-r')}{r^2}{Q^t}_r,
\nonumber\\
 &   (r-r')\difffrac{{Q^r}_r}{r}=\,\frac{2M(r-r')^2}{r^3f}-{Q^r}_r-({Q^r}_r)^2
    +
     \\&
   \lP f {Q^t}_r\rP^2
      -\frac{\epsilon 2M(r-r')}{r^2}{Q^t}_r,
    \nonumber \label{eqn:secondTE}\\
  &  (r-r')\difffrac{{Q^t}_r}{r}=\,\frac{\epsilon2M(r-r')^2}{r^3f^2}-{Q^t}_r-
    \\&
  {Q^t}_r({Q^t}_t+{Q^r}_r)
    -
    \frac{M(r-r')}{r^2f^2}\lP2 f{Q^t}_r-\epsilon{Q^t}_t+\epsilon{Q^r}_r\rP.  \nonumber
\end{align}

Now, from $\Delta(\coord{x},\coord{x}')= 1$ together with Eqs.\eqref{eq:transp eq Q rad} and  (\ref{eqn:QNSolution}) we can conclude that
\begin{align}\label{eq:cond TrQ}
    {Q^\alpha}_\alpha=\,&{Q^t}_t+{Q^r}_r=0 \implies   {Q^r}_r=-{Q^t}_t.
\end{align}
Adding \eqref{eqn:firstTE} and \eqref{eqn:secondTE} and using $Q^{\alpha}{}_\alpha=0=const$ together with
\eqref{eq:Qrt-Qtr} and \eqref{eq:cond TrQ}, we obtain
\begin{align}
    &(r-r')\difffrac{{Q^\alpha}_\alpha}{r}=
        -2({Q^t}_t)^2+2\lP f{Q^t}_r\rP^2=0
   \nonumber     \\ &
        \implies {Q^t}_t=\pm f{Q^t}_r.
     \label{eq:cond dTrQ}
\end{align}

Substituting   Eqs.\eqref{eq:cond TrQ} and \eqref{eq:cond dTrQ} back into the transport equations (\ref{eqn:firstTE}) and (\ref{eqn:secondTE}) we see that, for these equations to be consistent, the $+/-$ sign in \eqref{eq:cond dTrQ} must be chosen so that it corresponds to ingoing/outgoing geodesics respectively (i.e., so that it is equal to ``$-\epsilon$").
The resulting first-order, linear ordinary differential equation for $Q^t{}_t$ is:
\begin{align}\label{eqn:QttIngoingRadislTE}
    (r-r')\difffrac{{Q^t}_t}{r}=\,&-\frac{2M(r-r')^2}{r^3f}-\frac{r^2-2Mr'}{r^2f}{Q^t}_t.
\end{align}
Integrating this equation, and using Eqs.\eqref{eq:cond TrQ}, \eqref{eq:cond dTrQ} and \eqref{eq:Qrt-Qtr},
 we finally obtain
\begin{align}\label{eq:Qtt,Qrr,Qtr,Qrt}
    {Q^t}_t=\,&
    -{Q^r}_r=
        \frac{M}{f}\left(\frac{3r-r'}{r^2}-\frac{2\ln\lP\frac{r}{r'}\rP}{r-r'}\right),
    \\
    {Q^t}_r=\,&-\frac{{Q^r}_t}{f^2}=\pm\frac{{Q^t}_t}{f}=
   \pm \frac{M}{f^2}\left(\frac{3r-r'}{r^2}-\frac{2\ln\lP\frac{r}{r'}\rP}{r-r'}\right),
    \nonumber
\end{align}
where the $+/-$ sign corresponds to ingoing/outgoing geodesics respectively. 

Eqs.\eqref{eqn:QNSolution} and \eqref{eq:Qtt,Qrr,Qtr,Qrt}
together provide
analytical expressions for all the components of ${Q^\alpha}_\beta$ along a
radial null 
geodesic in Schwarzschild spacetime.
We have analytically verified that these expressions for ${Q^\alpha}_\beta$ are indeed a solution of the system \eqref{eq:transp eq Q}, and thus that $\Delta(x,x')=1$ along radial null geodesics in Schwarzschild.

\section{Solving the Characteristic Initial Data problem}\label{sec:CID}

In this Appendix we provide a brief explanation of how we solved our CID problem in Eq.\eqref{eq:eq ell-mode}. 
We followed the finite difference method in \cite{mark2017recipe} but extended the order of the method from being  order $h^2$ to order $h^4$, where $h$ is the stepsize of the grid.
Essentially, \cite{mark2017recipe} write,
omitting $u'$ and $v'$ as arguments of $G_\ell$,
\begin{equation}
G_\ell(v,u)=\GlCID(v,u)\theta(u-u')\theta(v-v')
\end{equation}
and it can be shown that the modes $\GlCID$ satisfy
\begin{align}\label{eqn:CIDP}
    \frac{\partial^2\GlCID}{\partial v\partial u}+Q(r)\GlCID=0,\\
    \begin{array}{l}\label{eqn:initialData}
        \GlCID(v=v',u)=-\frac{1}{2},\\
        \GlCID(v,u=u')=-\frac{1}{2},
    \end{array}
\end{align}
where
$$
    Q(r):=\frac{1}{4}\left(1-\frac{2M}{r}\right)\left(\frac{\ell(\ell+1)}{r^2}+\frac{2M}{r^3}\right).
$$
We note that $G_\ell$ and $\GlCID$ also depend on
$v'$ and $u'$ but, for the sake of simplicity, we  omit these  arguments  in this appendix since we keep them fixed.
\begin{figure}
    \centering
    \includegraphics[width=0.33\textwidth]{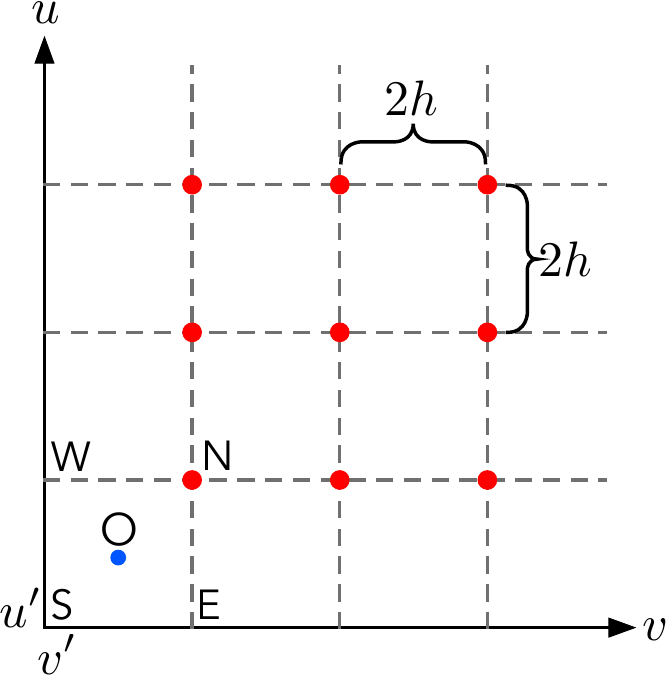}
    \caption{Grid distribution on the $(v,u)$-plane.}
    \label{fig:vuGrid}
\end{figure}
 Ref.~\cite{mark2017recipe} solved this CID problem by using the scheme proposed in
Lousto and Price~\cite{Lousto:1997wf}.
We next describe this scheme including our extension to order $h^4$.

This CID problem can be solved by constructing an equally-spaced grid in the $(v,u)$-plane. Let $2h$ be the stepsize between the nodes of the grid along either the $u$-direction or the $v$-direction. In Fig.~\ref{fig:vuGrid} we show the grid arrangement in the $(v,u)$-plane.

The value of $G_{\ell}(v,u)$ at each node is then calculated by integrating Eq.~\eqref{eqn:CIDP} over each square formed by four nodes (e.g., $S$, $E$, $N$ and $W$ in Fig.~\eqref{fig:vuGrid}) of the grid. For instance, in Fig. \ref{fig:vuGrid} the integration of Eq.~\eqref{eqn:CIDP} over the square $SENW$ yields
\begin{equation}\label{eqn:integratedCIDPDE}
    \int\limits_{SENW} \frac{\partial^2\GlCID}{\partial v\partial u} \,\textrm{d}v\,\textrm{d}u+\int\limits_{SENW} Q(r)\GlCID \,\textrm{d}v\,\textrm{d}u=0.
\end{equation}
The first integral in the above equation is exact and trivially performed:
\begin{equation}\label{eqn:exactHalfIntegral}
    \int\limits_{SENW} \frac{\partial^2\GlCID}{\partial v\partial u} \,\textrm{d}u\,\textrm{d}v=
    \GlCID^N-\GlCID^E-\GlCID^W+\GlCID^S,
\end{equation}
where  $\GlCID^K$ denotes the value of  
$\GlCID$ at the point  $K=S$, $E$, $W$ or $N$.

\begin{widetext}

In order to perform the second integral, we make some approximations. Taking into account that the stepsize between nodes is small (i.e., $h\ll M$), the integrand $Q\cdot \GlCID$ of the second term in Eq.~\eqref{eqn:integratedCIDPDE} can be expanded about the point in the middle of the $SENW$ square, $O:= (v_0,u_0)$ (see Fig.~\ref{fig:vuGrid}). Expanding $Q\cdot \GlCID$, as well as
$\GlCID$, which  we shall need later, as Taylor series and then truncating them at a desired order, we obtain
\begin{align}
        Q(r)\GlCID(v,u)\,
        =\sum_{\substack{0\leq m,n\leq 3\\m+n\leq3}}\frac{1}{m!\, n!}\left(\frac{\partial^{m+n}}{\partial v^m\partial u^n}(Q\,\GlCID)\right)_O(v-v_0)^m(u-u_0)^n+\mathcal{O}(h^4),
        \label{eqn:OGTaylorSeries}
        \\
        \GlCID(v,u)\,
        =\sum_{\substack{0\leq m,n\leq 3\\m+n\leq3}}\frac{1}{m!\, n!}\left(\frac{\partial^{m+n}}{\partial v^m\partial u^n}\GlCID\right)_O(v-v_0)^m(u-u_0)^n+\mathcal{O}(h^4).
        \label{eqn:potentialTermTaylorSeries}
\end{align}

Thus the second integral in Eq. (\ref{eqn:integratedCIDPDE}) is given to $\mathcal{O}(h^6)$  by
\begin{equation}\label{eqn:potentialIntegral}
    \begin{split}
        \int\limits_{SENW}Q(r)\GlCID(v,u)\,\textrm{d}v\,\textrm{d}u=\,&4(Q\,\GlCID)_O\,h^2+\frac{2}{3}\left(
        \frac{\partial^2}{\partial v^2}(Q\,\GlCID)
        +
        \frac{\partial^2}{\partial u^2}(Q\,\GlCID)
        \right)_O h^4+\mathcal{O}(h^6).
    \end{split}
\end{equation}
However, with the initial data given in Eq. (\ref{eqn:initialData}), it is not possible to reach up to order $\mathcal{O}(h^6)$. To achieve this order, additional information should be given along $u=u'$ as well as along $v=v'$. 
Specifically, Eq. \eqref{eqn:initialData} readily yields the longitudinal derivatives along these characteristic lines, but we also require the transversal derivatives.
\begin{equation}\label{eqn:derivativesG}
    \begin{split}
        \int\frac{\partial}{\partial v}\left(\frac{\partial \GlCID}{\partial u}\right)_{u'}\,\textrm{d}v\,&=\left(\frac{\partial \GlCID}{\partial u}\right)_{u'}=-\left(\int Q(r)\GlCID(v,u)\,\textrm{d}v\right)_{u'}\\
        &=\frac{1}{2}\left(\int Q(r)\,\textrm{d}v\right)_{u'}+p(u'),\\
        \int\frac{\partial}{\partial u}\left(\frac{\partial \GlCID}{\partial v}\right)_{v'}\,\textrm{d}u\,&=\left(\frac{\partial \GlCID}{\partial v}\right)_{v'}=-\left(\int Q(r)\GlCID(v,u)\,\textrm{d}u\right)_{v'}\\
        &=\frac{1}{2}\left(\int Q(r)\,\textrm{d}u\right)_{v'}+q(v'),
    \end{split}
\end{equation}
for some functions $p(u')$ and $q(v')$, where we have used Eq. (\ref{eqn:initialData}). We use the notation of $u'$ as a subscript in the brackets to indicate evaluation at $u=u'$; similarly for $v'$ to indicate $v=v'$ and for $v',u'$ below to indicate both $v=v'$ and $u=u'$.

From the Hadamard series \cite{PhysRevD.86.024038} for $\GlCID(v,u)$ and using Eq. (\ref{eqn:initialData}), it is trivial to show that, at coincidence,
\begin{equation}
    \left(\frac{\partial \GlCID}{\partial v}\right)_{v',u'}= \left(\frac{\partial \GlCID}{\partial u}\right)_{v',u'}=0,
\end{equation}
and use it to calculate $p(u')$ and $q(v')$. Thus the integrals in Eq. (\ref{eqn:derivativesG}) reduce to
\begin{equation}\label{eqn:derivativesValues}
    \begin{split}
        \left(\frac{\partial \GlCID}{\partial u}\right)_{u'}\,&=\frac{1}{2}\left(\int Q(r)\,\textrm{d}v\right)_{u'}-\left(\frac{1}{2}\int Q(r)\,\textrm{d}v\right)_{v',u'}\\
        &=\frac{1}{4}\left(\frac{\ell(\ell+1)}{r}+\frac{M}{r^2}\right)_{v',u'}-\frac{1}{4}\left(\frac{\ell(\ell+1)}{r}+\frac{M}{r^2}\right)_{u'},\\
        \left(\frac{\partial \GlCID}{\partial v}\right)_{v'}\,&=\frac{1}{2}\left(\int Q(r)\,\textrm{d}u\right)_{v'}-\left(\frac{1}{2}\int Q(r)\,\textrm{d}u\right)_{v',u'}\\
        &=-\frac{1}{4}\left(\frac{\ell(\ell+1)}{r}+\frac{M}{r^2}\right)_{v',u'}+\frac{1}{4}\left(\frac{\ell(\ell+1)}{r}+\frac{M}{r^2}\right)_{u'}.
    \end{split}
\end{equation}
These two equations are used to evaluate the derivatives of $\GlCID$ at the points $E$, $W$ and $S$ (later on we apply a similar reasoning to calculate the derivatives at the point $N$). Now, we construct a system of 12 equations by evaluating the Taylor series for $\GlCID$ and its derivatives at the points $E$, $W$, $N$, $S$. The 12 unknowns of this system are the 10 coefficients of the Taylor series in Eq. (\ref{eqn:potentialTermTaylorSeries}), which are  evaluated at $O$, together with the two first-order derivatives of $\GlCID$ evaluated at the point $N$. Once the system is solved, the coefficients of the Taylor series turn out to be
\begin{align}\label{eqn:GdGofEWNSFirst}
    4\GlCID^O&=2\GlCID^E+2\GlCID^W+h\left(\frac{\partial \GlCID}{\partial u}-\frac{\partial \GlCID}{\partial v}\right)_E-h\left(\frac{\partial \GlCID}{\partial u}-\frac{\partial \GlCID}{\partial v}\right)_W+\mathcal{O}(h^4),\\
    8h\left(\frac{\partial \GlCID}{\partial u}\right)_O&=-5\GlCID^S-\GlCID^E+5\GlCID^W+\GlCID^N
    -2h\left(\frac{\partial \GlCID}{\partial u}
    +\frac{\partial \GlCID}{\partial v}\right)_S
    -2h\left(\frac{\partial \GlCID}{\partial u}
    -\frac{\partial \GlCID}{\partial v}\right)_W+\mathcal{O}(h^4),\\
    8h\left(\frac{\partial \GlCID}{\partial v}\right)_O&=-5\GlCID^S+5\GlCID^E+\GlCID^W-\GlCID^N
    -2h\left(\frac{\partial \GlCID}{\partial u}
    +\frac{\partial \GlCID}{\partial v}\right)_S
    +2h\left(\frac{\partial \GlCID}{\partial u}
    -\frac{\partial \GlCID}{\partial v}\right)_E+\mathcal{O}(h^4),\\
    4h^2\left(\frac{\partial^2\GlCID}{\partial u^2}\right)_O&=\GlCID^S-\GlCID^E-\GlCID^W+\GlCID^N+2h\left(\frac{\partial \GlCID}{\partial u}\right)_E-2h\left(\frac{\partial \GlCID}{\partial u}\right)_W+\mathcal{O}(h^4),\\\label{eqn:GdGofEWNSLast}
    4h^2\left(\frac{\partial^2\GlCID}{\partial v^2}\right)_O&=\GlCID^S-\GlCID^E-\GlCID^W+\GlCID^N-2h\left(\frac{\partial \GlCID}{\partial v}\right)_E+2h\left(\frac{\partial \GlCID}{\partial v}\right)_W+\mathcal{O}(h^4),\\
    4h^2\left(\frac{\partial^2\GlCID}{\partial v\partial u}\right)_O&=\GlCID^N+\GlCID^S-\GlCID^E-\GlCID^W+\mathcal{O}(h^4),\\
    \frac{2}{3}h^3\left(\frac{\partial^3\GlCID}{\partial v^3}\right)_O&=\GlCID^S-\GlCID^E+h\left(\frac{\partial \GlCID}{\partial v}\right)_S+h\left(\frac{\partial \GlCID}{\partial v}\right)_E+\mathcal{O}(h^4),\\ \frac{2}{3}h^3\left(\frac{\partial^3\GlCID}{\partial u^3}\right)_O&=\GlCID^S-\GlCID^W+h\left(\frac{\partial \GlCID}{\partial u}\right)_S+h\left(\frac{\partial \GlCID}{\partial u}\right)_W+\mathcal{O}(h^4),\\
    4h^3\left(\frac{\partial^2\GlCID}{\partial v^2\partial u}\right)_O&=\GlCID^N+\GlCID^S-\GlCID^E-\GlCID^W+2h\left(\frac{\partial \GlCID}{\partial v}\right)_S-2h\left(\frac{\partial \GlCID}{\partial v}\right)_W+\mathcal{O}(h^4),\\\label{eqn:GdGofEWNSNLast}
    4h^3\left(\frac{\partial^2\GlCID}{\partial v\partial u^2}\right)_O&=\GlCID^N+\GlCID^S-\GlCID^E-\GlCID^W+2h\left(\frac{\partial \GlCID}{\partial u}\right)_S-2h\left(\frac{\partial \GlCID}{\partial u}\right)_E+\mathcal{O}(h^4).
\end{align}
Inserting Eqs. (\ref{eqn:exactHalfIntegral}), (\ref{eqn:potentialIntegral}) and (\ref{eqn:GdGofEWNSFirst}) - (\ref{eqn:GdGofEWNSLast}) into Eq. (\ref{eqn:integratedCIDPDE}), we obtain
\begin{equation}\label{eq:GellN}
    \begin{split}
    (6+2Qh^2)_O\,\GlCID^N=\,&\left[6-10Qh^2-2\left(\frac{\partial^2Q}{\partial v^2}+\frac{\partial^2Q}{\partial u^2}\right)h^4\right]_O(\GlCID^E+\GlCID^W)-(6+2Qh^2)_O\,\GlCID^S\\
    &+\left[4Q+\left(\frac{\partial^2Q}{\partial v^2}+\frac{\partial^2Q}{\partial u^2}\right)h^2\right]_O\left[\left(\frac{\partial G}{\partial v}-\frac{\partial \GlCID}{\partial
    u}\right)_E-\left(\frac{\partial \GlCID}{\partial v}-\frac{\partial \GlCID}{\partial u}\right)_W\right]h^3\\
    &+\left[\left(1-\frac{2M}{r}\right)\frac{\textrm{d} Q}{\textrm{d}r}\right]_O\left[\left(\frac{\partial \GlCID}{\partial v}-\frac{\partial \GlCID}{\partial
    u}\right)_E+\left(\frac{\partial \GlCID}{\partial v}-\frac{\partial \GlCID}{\partial u}\right)_W\right]h^4\\
    &-3\left[\left(1-\frac{2M}{r}\right)\left(\frac{\textrm{d} Q}{\textrm{d}r}\right)\right]_O(\GlCID^E-\GlCID^W)h^3+\mathcal{O}(h^6).
    \end{split}
\end{equation}
Since, from Eqs.~\eqref{eqn:initialData} and \eqref{eqn:derivativesValues}, the values at the points $E$, $W$ and $S$ of $\GlCID$ and its derivatives are known, we are able to calculate, via Eq.~\eqref{eq:GellN}, the value of $\GlCID$ at the point $N$. Additionally, we need to calculate the first order derivatives of $\GlCID$ at the point $N$. Those derivatives are easily obtained by integrating Eq. \eqref{eqn:CIDP} once along $u=u_0+h$ and once along $v=v_0+h$:
\begin{equation}
    \begin{split}
        \left(\frac{\partial \GlCID}{\partial v}\right)_N&=\left(\frac{\partial \GlCID}{\partial v}\right)_E+\left(\,\int\limits_{u_0-h}^{u_0+h}\frac{\partial^2\GlCID}{\partial v\partial u}\,\textrm{d}u\right)_{v_0+h}=\left(\frac{\partial \GlCID}{\partial v}\right)_E-\left(\,\int\limits_{u_0-h}^{u_0+h}Q(r)\,\GlCID\,\textrm{d}u\right)_{v_0+h},\\
        \left(\frac{\partial \GlCID}{\partial u}\right)_N&=\left(\frac{\partial \GlCID}{\partial u}\right)_W+\left(\,\int\limits_{v_0-h}^{v_0+h}\frac{\partial^2\GlCID}{\partial v\partial u}\,\textrm{d}v\right)_{u_0+h}=\left(\frac{\partial \GlCID}{\partial u}\right)_W-\left(\,\int\limits_{v_0-h}^{v_0+h}Q(r)\,\GlCID\,\textrm{d}v\right)_{u_0+h}.
    \end{split}
\end{equation}
We then use Eq. \eqref{eqn:OGTaylorSeries} and Eqs. \eqref{eqn:GdGofEWNSFirst} - \eqref{eqn:GdGofEWNSNLast} to calculate the above integrals to $\mathcal{O}(h^4)$. Once these derivatives are known, we can continue to
apply this procedure consecutively throughout the whole grid in order to obtain $\GlCID(v,u)$ at the various nodes.

\section{Derivation of time-mirror symmetry}\label{app:timeinversion}
This appendix gives the derivation of the time-mirror symmetry introduced and discussed in Sec.~\ref{sec:section_timemirrorsymmetry}. 
Given a signaling scenario (with worldlines $\coord{x}_\dd(t)$, switching functions $\eta_\dd(t)$ for $\dd=\da,\db$, and signal terms $C_2$ and $D_2$), the time-mirrored scenario has worldlines $\coord{x}_\dd'(t)=\coord{x}_\dd(-t) $ and switching functions $\eta'_\dd(t)=\eta_\dd(-t)$.
We assume, without loss of generality, that  the proper times of the detectors are given by $\tau_\dd'(t)=-\tau_\dd(-t)$ for both detectors   in the inverted scenario.
The detector frequencies are the same in the original and the mirrored scenario. However, since $\da$ acts as the receiver in the mirrored scenario, their frequency $\Omega_\da$ enters with a positive sign in the imaginary exponent of the coefficient $C_2'$ for the mirrored scenario, whereas it enters with a negative sign in the original coefficient $C_2$. 
Then, if the Green function of the spacetime obeys \eqref{eq:timeconditiononG}, we have that the coefficient for the mirrored scenario %
\begin{align}
 C_2'  &= \frac{-\ii}{4\pi} \integral{t_1}{}{} \integral{t_2}{}{t_1} \difffrac{\lambda_\da'}{t_1}  \eta_\da'(t_1) \difffrac{\tau_\db'}{t_2} \eta_\db'(t_2)  \ee{\ii (\Omega_\da\tau'_A(t_1)- \Omega_\db \tau'_B(t_2))} G_{ret}(t_1,\bm{x}'_A(t_1),t_2,\bm{x}'_B(t_2)) \nn
   &= \frac{-\ii}{4\pi} \integral{s_2}{}{}\integral{s_1}{}{s_2} \difffrac{\tau_\da'(-s_1)}{s_1}  \eta_\da'(-s_1)  \difffrac{\tau_\db'(-s_2)}{s_2} \eta_\db'(-s_2)  \ee{\ii (\Omega_\da\tau'_A(-s_1)- \Omega_\db \tau'_B(-s_2))} G_{ret}(-s_1,\bm{x}'_A(-s_1),-s_2,\bm{x}'_B(-s_2)) \nn
   &= \frac{-\ii}{4\pi} \integral{s_2}{}{}\integral{s_1}{}{s_2} \difffrac{\tau_\da(s_1)}{s_1}  \eta_\da(s_1)  \difffrac{\tau_\db(s_2)}{s_2} \eta_\db(s_2)  \ee{-\ii (\Omega_\da\tau_\da(s_1)- \Omega_\db \tau_\db(s_2))} G_{ret}(-s_1,\bm{x}_A(s_1),-s_2,\bm{x}_B(s_2)) \nn
   &= \frac{-\ii}{4\pi} \integral{s_2}{}{}\integral{s_1}{}{s_2} \difffrac{\tau_\da(s_1)}{s_1}  \eta_\da(s_1)  \difffrac{\tau_\db(s_2)}{s_2} \eta_\db(s_2)  \ee{-\ii (\Omega_\da\tau_\da(s_1)- \Omega_\db \tau_\db(s_2))} G_{ret}(s_2,\bm{x}_B(s_2),s_1,\bm{x}_A(s_1))=C_2 
\end{align}\label{eq:C2,C2'}
\end{widetext}
is identical to the coefficient $C_2$ for the original scenario. (And  we introduced integration variables $s_1=-t_1,\, s_2=-t_2$.) %
For the  signal term $D_2$, we analogously find $D_2'=-D_2^*$. 
This we can also deduce, from the general relation $C_2(\Omega_\da,\Omega_\db)=-D_2(\Omega_\da,-\Omega_\db)$, which implies
\begin{align}
   D_2'( \Omega_\db, \Omega_\da)&=-C_2'(\Omega_\db, -\Omega_\da)=-C_2(-\Omega_\da, \Omega_\db)\nn 
   &=D_2(- \Omega_\da, -\Omega_\db)
\end{align}
and, indeed, in \eqref{eq:D2} we see that  $D_2(- \Omega_\da, -\Omega_\db)=-D_2(\Omega_\da, \Omega_\db)^*$.

\section{Change of integration variable in tail contribution for static detectors}\label{app:varchange}
This appendix discusses how, for static detectors in a static spacetime, a change of integration variables in the double integrals of $C_2$ and $D_2$ makes it possible to separate the expression into a product of one integral containing the Green function and another integral involving the switching functions. The latter  can often be performed analytically thus leaving only the first integral to be performed numerically.

Since the spatial coordinates of  detectors at rest do not change, the value of the Green function in the integrand of $C_2$ only depends on the coordinate time difference  between Alice and Bob.
Following the definitions at the beginning of Section \ref{sec:staticspacetimes}, the coordinate time difference is
\begin{align}
    t(\tau_\db)-t(\tau_\da)%
    =\frac1{v(\spacoord{x}_\da)}\left(\nu \tau_\db-\tau_\da\right)+\Delta t_{\da\to\db}.
\end{align}
In a static spacetime, the retarded Green function only depends on the time coordinate difference between its arguments. We use this and define
\begin{align}\label{eq:define_s}
s=\nu \tau_\db-\tau_\da
\end{align}
such that
\begin{align}
&G_{ret}(\coord{x_B}(\tau_\db),\coord{x_A}(\tau_\da)) = G_{ret}(t(\tau_\db)-t(\tau_\da),\spacoord{x}_\db,\spacoord{x_A})\\
&= G_{ret}(s/v(\spacoord{x}_\da)+ \Delta t_{\da\to\db},\spacoord{x}_\db,\spacoord{x_\da}).
\end{align}

Next, we can change the integration variables in $C_2$ in \eqref{eq:C2Gret} from $(\tau_\da,\tau_\db)$ to $(s,\tau_\db)$.
To this end, denote the support of the switching functions in terms of detector proper times as $\text{supp}\,\eta_\da(\tau_\da)=[A_1,A_2]$ and $\text{supp}\,\eta_\db(\tau_\db)=[B_1,B_2]$. 
The  integrand of the double-integral in $C_2$ then has support only in the region
$B_1\leq\tau_\db\leq B_2, A_1\leq \tau_\da\leq\min[A_2,\nu\tau_\db]$.

\label{page:interpretation_s}
One can visualize the role of $s$  in a 2D-plot of the integration region. We put $\tau_\da$ on the y-axis and $\tau_\db$ on the left axis. Then $s$ is constant along straight lines cutting through the first quadrant of the coordinate system. Their angle depends on $\nu$, i.e., the redshift between Alice and Bob. $s$ increases as one moves to the bottom right in the plot, i.e., for increasing $\tau_\db$.
Points in the integration region  which lie on a line of constant $s$ correspond to point pairs on the worldline of Alice and Bob which are separated by the same amount of coordinate time. This means they are mapped into each other by translations along the Killing field of coordinate time. E.g., there is the line of $s=0$ which has all points connected by a direct null geodesic on it. And there is the line of constant $s$ which has all the points connected by a secondary null geodesic on it, and so on.

Under the change of integration variables from $(\tau_\db,\tau_\da)$ to $(\tau_\db, s)$, the integral thus transforms to
\begin{align}
    &\integral{\tau_\db}{B_1}{B_2}\integral{\tau_\da}{A_1}{\min[A_2,\nu\tau_\db]} \nn 
    &\qquad=  \integral{s}{\max\left[\nu B_1-A_2,0\right]}{\nu B_2-A_1} \integral{\tau_\db}{\max\left[B_1, (s+A_1)/\nu\right]}{\min\left[ B_2, (s+A_2)/\nu\right]},
\end{align}
such that
\begin{widetext}
\begin{align}\label{eqn:C2ForStaticObservers}
    C_2    &=\frac{-\ii}{4\pi}   \integral{s}{\max\left[\nu B_1-A_2,0\right]}{\nu B_2-A_1} \ee{\ii\, \Omega_\da s}     
     G_{ret}(s/v(\spacoord{x}_\da)+ \Delta t_{\da\to\db},\spacoord{x}_\db,\spacoord{x}_\da)
    \integral{\tau_\db}{\max\left[B_1, (s+A_1)/\nu\right]}{\min\left[ B_2, (s+A_2)/\nu\right]} \eta_\db\left(\tau_\db\right)  \eta_\da\left(\nu\tau_\db-s\right) \ee{\ii\, (\Omega_\db-\nu\Omega_\da) \tau_\db}
\end{align}
The inner integral over $\tau_\db$ is typically easy to solve analytically.
In particular, sharp switching functions $\eta_\da(\tau_\da)=\eta_{[A_1,A_2]}(\tau_\da)$ and $\eta_\db(\tau_\db)=\eta_{[B_1,B_2]}(\tau_\db)$, as defined in \eqref{eq:sharpswitchingfn}), yield
\begin{align}
    C_2    &=\frac{-\ii}{4\pi}   \integral{s}{\max\left[\nu B_1-A_2,0\right]}{\nu B_2-A_1} \ee{\ii\, \Omega_\da s}     
     G_{ret}(s/v(\spacoord{x}_\da)+ \Delta t_{\da\to\db},\spacoord{x}_\db,\spacoord{x_A})
    \integral{\tau_\db}{\max\left[B_1, (s+A_1)/\nu\right]}{\min\left[ B_2, (s+A_2)/\nu\right]}  \ee{\ii\, (\Omega_\db-\nu\Omega_\da) \tau_\db}\label{eq:C2static_afterchange_sharpswitching}\\ 
    &=\integral{s}{\max\left[\nu B_1-A_2,0\right]}{\nu B_2-A_1} \frac{\ee{\ii\, \Omega_\da s}     
     G_{ret}(s/v(\spacoord{x}_\da)+ \Delta t_{\da\to\db},\spacoord{x}_\db,\spacoord{x_A})}{4\pi(\Omega_\db-\nu\Omega_\da)} 
     \left( \ee{\ii\, (\Omega_\db-\nu\Omega_\da) {\max\left[B_1, (s+A_1)/\nu\right]}}-\ee{\ii\, (\Omega_\db-\nu\Omega_\da) {\min\left[ B_2, (s+A_2)/\nu\right]}} \right)%
\end{align}

\section{Fourier transformation of switching functions for sharp switching}\label{app:FTofswitching}
To evaluate the signal term using the Fourier technique explained in Section \ref{sec:Fouriertrick} we need
\begin{align}
    \mathcal{F}[\Theta(t_1,t_2)](k_1,k_2)=\integral{t_1}{-\infty}{\infty}\integral{t_2}{-\infty}{\infty} \ee{-\ii(k_1t_1+k_2t_2)} \eta_\db(t_1)\eta_\da(t_2) \theta\left(t_1-t_2-\Delta t_{\da\to\db}\right),
\end{align}
i.e., the two-dimensional Fourier transform of $\Theta(t_1,t_2)= \eta_\db(t_1)\eta_\da(t_2) \theta\left(t_1-t_2-\Delta t_{\da\to\db}\right)$ as defined in \eqref{eq:thetaswitchingfuncprod}. In this appendix we evaluate it for sharp switching functions of the form
\begin{align}
    \eta_\da(t)&=\begin{cases} 1,\quad T_0\leq t \leq T_A \\ 0,\quad \text{otherwise}\end{cases}, \qquad%
    \eta_\db(t)=\begin{cases} 1,\quad T_1\leq t \leq T_2 \\ 0,\quad \text{otherwise}\end{cases}.
\end{align}

To evaluate the Fourier transform it is convenient to distinguish between two cases according to the spacetime separation between Alice's and Bob's switching times. First, 
the case where Bob's coupling is strictly timelike separated from Alice's couplings, i.e.,
\begin{align}
    T_2>T_1>T_A+\Delta t_{\da\to\db}.
\end{align}
Second, 
the case where the switchings are exactly null separated, i.e., 
\begin{align}
    T_0+\Delta t_{\da\to\db}=T_1,\quad T_A+\Delta t_{\da\to\db}=T_2.
\end{align}
More general cases can be split up into sums of integrals with the switchings  either timelike or exactly null separated.

The first case, with exact null separation, is simple to evaluate because the Heaviside in the integrand is equal to 1 everywhere in the support of the switching functions. Hence, the Fourier transform is given by the product of the Fourier transform of Alice and Bob switching functions.
\begin{align}
    \mathcal{F}[\Theta(t_1,t_2)](k_1,k_2)&= \ee{-\ii k_2 (T_A+T_0)/2} \ee{-\ii k_1 (T_2+T_1)/2} (T_A-T_0)(T_2-T_1) \sinc\left(k_2\frac{T_A-T_0}2\right) \sinc\left(k_1\frac{T_2-T_1}2\right).
\end{align}

In the second case, of exact null separation, we need take into account the Heaviside function in the integrand. This can conveniently be done by a change of integration variables to 
$v=t_1-t_2-\Delta t_{\da\to\db}$ and $u=\frac12(t_1+t_2)$, such that $ \Theta(t_2-t_1-\Delta t_{\da\to\db}) =\Theta(v)$. 
Then
\begin{align}
    &\mathcal{F}[\Theta(t_1,t_2)](k_1,k_2)\nn 
    &= \ee{-\ii(k_1-k_2)\Delta t_{\da\to\db}/2} \integral{v}0{T_A-T_0} \ee{-\ii (k_1-k_2) v/2}\integral{u}{T_0+\Delta t_{\da\to\db}/2+v/2}{T_A-v/2+\Delta t_{\da\to\db}/2} \ee{-\ii (k_A+k_B) u} \nn
    &= \frac{\ii (T_A-T_0)}{k_2+k_1}  \ee{-\ii k_1 (\Delta t_{\da\to\db}+T_A)}\left(\ee{-\ii k_2 T_A}  \sinc\left(k_2\frac{T_A-T_0}2\right) -  \ee{-\ii k_2 T_0}  \sinc\left(k_1 \frac{T_A-T_0}2\right)\right).
\end{align}

\section{Signal contribution from principal value distribution}\label{app:secondary_geodesics_signal}
This appendix discusses what qualitative features of the signal strength are expected to arise from secondary light rays.
Due to the singularity structure of the Green function in Schwarzschild spacetime,  discussed in Sec.~\ref{sec:distantpast}, the leading order behaviour of the Green function between points that are connected by secondary null geodesics corresponds to the product of a principal value $\pv\frac1\sigma$ distribution and a regular function.
In order to understand the qualitative behaviour of the signal contribution from secondary null geodesics between stationary detectors, we essentially ignore that pre-factor function and thus replace the Green function in \eqref{eqn:C2ForStaticObservers} by
\begin{align}\label{eq:Gret->PV}
 G_{ret}\left(s/v(\spacoord{x}_\da)+ \Delta t_{\da\to\db},\spacoord{x}_\db,\spacoord{x_A}\right)\to\frac1L\pv\frac1{s-s_2},
\end{align}
where $L$ is some length scale.
The value of $s_2$ corresponds to the time it takes secondary null geodesics to propagate from Alice to Bob, in terms of the integration variable $s$ in \eqref{eqn:C2ForStaticObservers}. 
Concretely, it mimicks the scenario where a secondary null geodesic emanating from Alice at her proper time $\tau_\da'$ arrives at Bob's location at his proper time
\begin{align}
\tau'_\db=\frac{s_2+\tau_\da'}\nu.
\end{align}
(See \eqref{eq:define_s} and discussion on p.~\pageref{page:interpretation_s} for the definition of $s$.)

Assuming sharp switching functions, just as in App.~\ref{app:varchange}, we find that the expression for the contribution from the principal value distribution, which we obtain by inserting \eqref{eq:Gret->PV} into \eqref{eq:C2static_afterchange_sharpswitching}, is
\begin{align}\label{eq:sec_C2}
 C_2    &=\frac{-\ii}{4\pi L}   \integral{s}{\max\left[\nu B_1-A_2,0\right]}{\nu B_2-A_1} \ee{\ii\, \Omega_\da s}     
     \pv\frac1{s-s_2}
   \underbrace{ \integral{\tau_\db}{\max\left[B_1, (s+A_1)/\nu\right]}{\min\left[ B_2, (s+A_2)/\nu\right]}  \ee{\ii\, (\Omega_\db-\nu\Omega_\da) \tau_\db}}_{=:f(s)}.
\end{align}
(Remember \eqref{eq:D2 from C2}, i.e. $D_2(\Omega_\da,\Omega_\db)=-C_2(\Omega_\da,- \Omega_\db)$.)
The inner integral, which we denoted by $f(s)$ is straightforward to evaluate, with the resonant case $\Omega_\db=\nu \Omega_\da$ requiring a separate treatment.

The appearance of the $\pv\frac1{s-s_2}$-distribution in \eqref{eq:sec_C2} raises the question of whether $C_2$ is well-defined and finite when $s_2$ coincides with one of the boundaries of the $s$-integral.
It turns out that the expression is well-defined. The reason being that, in this specific case that $s_2$ coincides with either of the boundaries of the outer integral, the absolute value of the inner integral $|f(s)|= \mathcal{O}(|s-s_2|)$ 
goes to zero linearly, thus rendering the value of $C_2$ finite.
Hence, even for sharp switching functions, the leading order signal contributions from secondary null geodesics  are finite, just as they are for the primary direct null geodesics. (Note, that this also holds true when taking into account the regular pre-factor in the Green function which we are not taking into account in this appendix.)

\begin{figure}[tb]
    
    \subfloat[$\Omega_\da=1/M$\label{fig:PV_Signal_short}]{\includegraphics[width=0.4\textwidth]{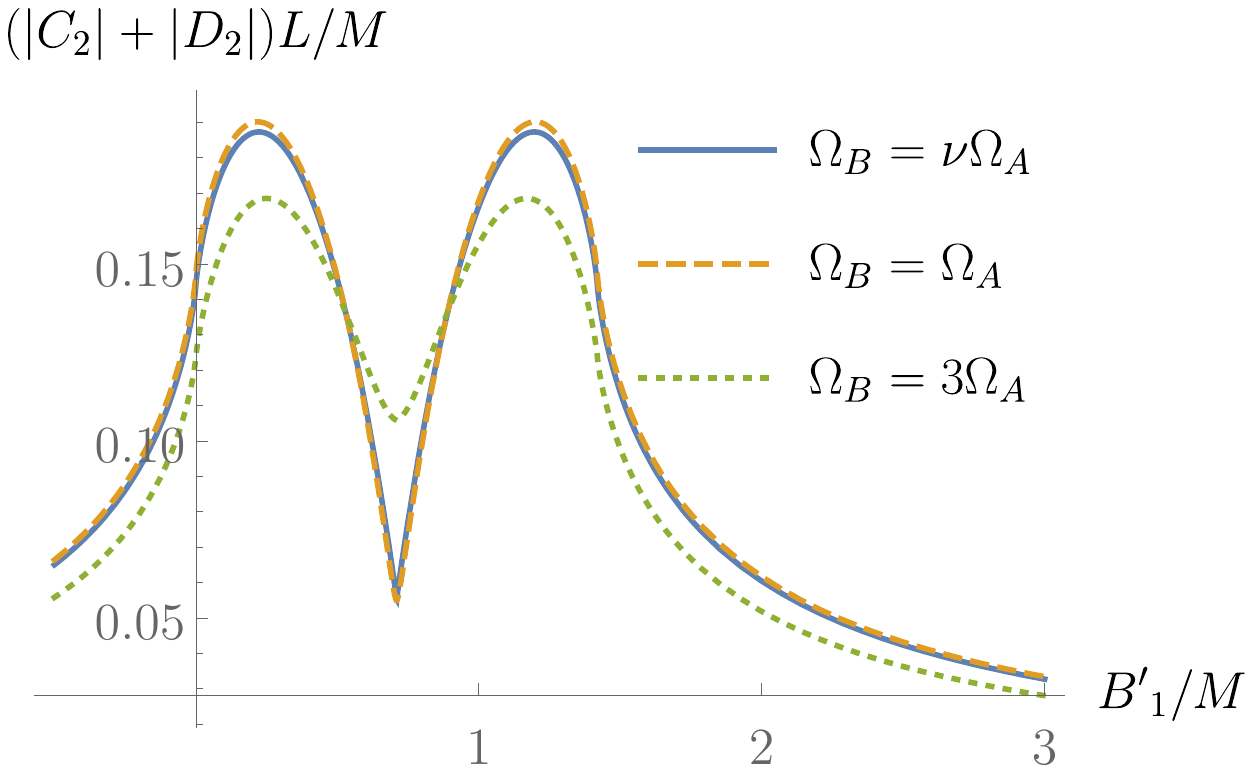}} 
    \subfloat[$\Omega_\da=10/M$\label{fig:PV_Signal_long}]{\includegraphics[width=0.4\textwidth]{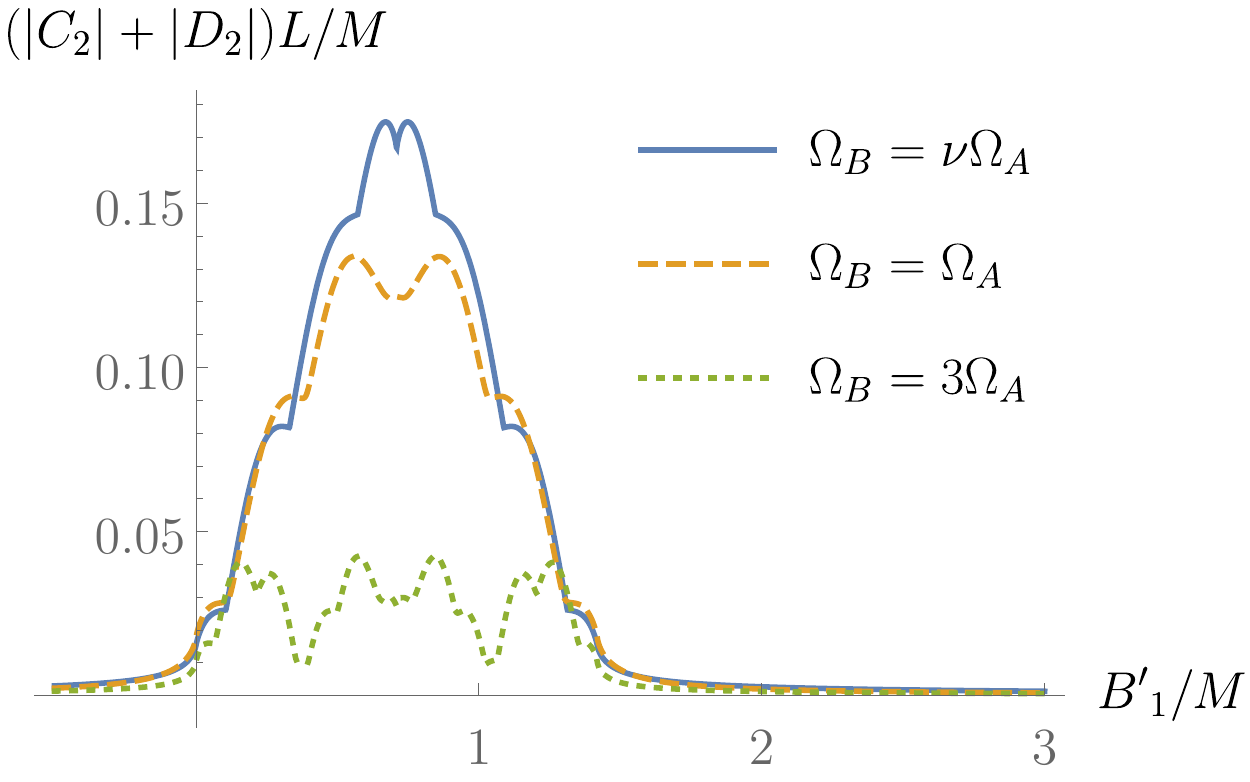}}    
\caption{Contribution to the leading order signal strength from a $\pv\frac1\sigma$-distribution, as resulting from \eqref{eq:C2_pv_shift}, with $\nu=\sqrt{(1-2/6)/(1-2/3.01)}\approx 1.40954$, $A_1=0$ and $A_2=M$.
The horizontal axes show $B_1':=B_1-\frac{s_2-A_2+2A_1}\nu$  which is the switch-on time $B_1$ of Bob shifted so that, for $B'_1<0$, Alice  and Bob  are  not connected by the singularity of the $\pv\frac1\sigma$-distribution while coupling to the field. This corresponds to Bob switching off his detector before  any secondary null geodesics emanating from Alice arrive at his location.
The graphs are symmetric about the point $B_1'=\frac{A-2-A_1}\nu$. This corresponds to the switching-on and switching-off of sender and receiver  being exactly  connected by secondary null geodesics.
The parameters in Fig.~\ref{fig:PV_Signal_short}  match those in Fig.~\ref{fig:C2NonDirectB1ShiftPlot}, thus the curve  with  equal frequencies ($\Omega_\da=\Omega_\db$) here reproduces the features due to secondary null geodesics seen there.
}
\label{fig:PV_Signal} 
\end{figure}

We can reproduce the features of Fig.~\ref{fig:C2NonDirectB1ShiftPlot}  which are due to secondary null geodesics.
To this end, we  let Alice  couple to the field during the fixed proper time interval $A_1\leq\tau_\da\leq A_2$.
Bob begins to couple at varying proper times $\tau_\db=B_1$. However, he always switches off after a time corresponding to the red-shifted duration of Alice's signal, i.e., $B_2-B_1=(A_2-A_1)/\nu$.
In particular, if $B_1=(A_1+s_2)/\nu$, both the switching-on and switching-off of both detectors are connected by secondary null geodesics.
For non-resonant detectors in this scenario, \eqref{eq:sec_C2} has the solution
\begin{align}\label{eq:C2_pv_shift}
&C_2 = 
\frac{-\ii}{4\pi L}   \integral{s}{\max\left[\nu B_1-A_2,0\right]}{\nu B_1+A_2-2A_1} \ee{\ii\, \Omega_\da s}     
     \pv\frac1{s-s_2}
    \integral{\tau_\db}{\max\left[B_1, (s+A_1)/\nu\right]}{\min\left[ B_1+(A_2-A_1)/\nu, (s+A_2)/\nu\right]}  \ee{\ii\, (\Omega_\db-\nu\Omega_\da) \tau_\db} \nn
 &   =
\frac{-1}{4\pi L( \Omega_\db-\nu \Omega_\da)}  \left(R(\Omega_\db/\nu,\nu B_1-A_2,s_m,s_2) \ee{\ii (\Omega_\db-\nu \Omega_\da)A_2/\nu} -R( \Omega_\da,\nu B_1-A_2,s_m,s_2)\ee{\ii(\Omega_\db-\nu \Omega_\da)B_1}\right.  \nn
&\quad+  \left.R(\Omega_\da,s_m,\nu B_2-A_1,s_2)  \ee{\ii ( \Omega_\db-\nu \Omega_\da)B_2}- R(\Omega_\db/\nu,s_m,\nu B_2-A_1,s_2) \ee{\ii (\Omega_\db-\nu \Omega_\da) A_1/\nu} \right),
\end{align}
with $s_m:= \nu B_1-A_1(=\nu B_2-A_2)$ and, for resonant detectors, i.e., $\Omega_\db=\nu \Omega_\da$, it has the solution
\begin{align}
C_2 
&=-\frac\ii{2\pi L}\frac{  (B_2-B_1) (s_2-\nu B_1+A_2) }{ \nu(B_2-B_1)+A_2-A_1 }R(\Omega_\da,\nu B_1-A_2,s_m,s_2) 
\nn*
&\quad    -\frac\ii{4\pi L} \left( (B_2-B_1) -\frac{2(B_2-B_1)(s_2-\nu B_1+A_1)}{\nu(B_2-B_1)+A_2-A_1}\right) R(\Omega_\da,s_m,\nu B_2-A_1,s_2)   \nn*
&\qquad +\frac{(B_2-B_1)}{2L\Omega_\da\left(\nu(B_2-B_1)+A_2-A_1\right)} \left(\ee{\ii \Omega_\da (\nu B_2-A_1)}-2\ee{\ii \Omega_\da s_m}+\ee{\ii \Omega_\da (\nu B_1-A_2)}\right).
\end{align}
Here we defined, using   $\ci(s)=-\integral{u}{s}\infty\cos( u)/u$ and   $\si( s)=\integral{u}{0}{s}\sin( u)/u$,
\begin{align}\label{eq:PV int}
&R(\omega,X,Y,s_2):=\pv\integral{s}{X}{Y}\frac{\ee{\ii\omega s}}{s-s_2}
&=\ee{\ii \omega s_2}\left(\ci\left(|\omega(Y-s_2)|\right)-\ci\left(|\omega(X-s_2)|\right)+\ii\, \si(\omega(Y-s_2))-\ii\, \si(\omega(X-s_2)) \right).
\end{align}
While this function is singular as $s_2\to X$ or $s_2\to Y$, all expressions for $C_2$   are finite and well-defined. This is because they contain a combination of terms with $R$-functions of different arguments, such that the singularities between the different terms exactly cancel out.

Some contributions to the signal strength, resulting from  different coupling durations and detector frequencies, are plotted in Fig.~\ref{fig:PV_Signal}.
As seen there, the signal strength is symmetric about the point where Alice's and Bob's switchings are exactly connected by secondary null geodesics. I.e., it is the point for which Bob switches on his detector when the secondary null geodesic emanating from Alice's switch-on arrives and he switches off when the secondary null geodesic from Alice's switch-off arrives. At this point the signal strength has a local minimum. Overall, if the detectors are resonant or close to resonance, the resulting signal strength rises to its highest levels around this symmetry point. For resonant detectors this maximum scales roughly linearly with the duration of the signal. For non-resonant detectors the signal strength exhibits a mostly periodic behaviour without a distinct maximum in the region where the detectors are connected by some secondary null geodesics.
Outside of this region, i.e., when Bob couples to the field strictly before or after any of Alice's secondary null geodesics arrive, the signal strength exhibits a decaying tail which results from the $\pv\frac1\sigma$ behaviour of the Green function. The tail appears independently of whether the detectors are resonant or not.

\begin{figure}[tb]
    
    \subfloat[$\Omega_\da=1/M,A_2=M$, %
    $(s_2+A_2)/\nu\approx 4.26M$.\label{fig:PV_Cumu_short}]{\includegraphics[width=0.33\textwidth]{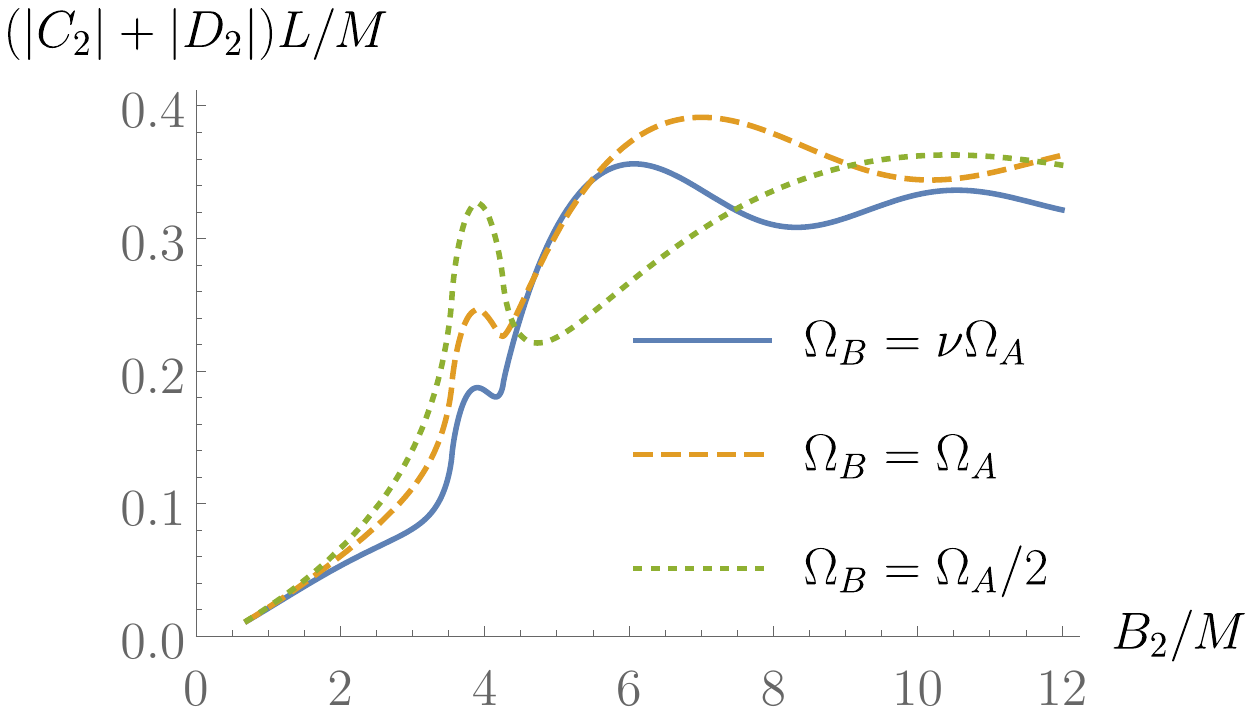}} 
    \subfloat[$\Omega_\da=10/M,A_2=M$, %
    $(s_2+A_2)/\nu\approx 4.26M$.\label{fig:PV_Cumu_medium}]{\includegraphics[width=0.33\textwidth]{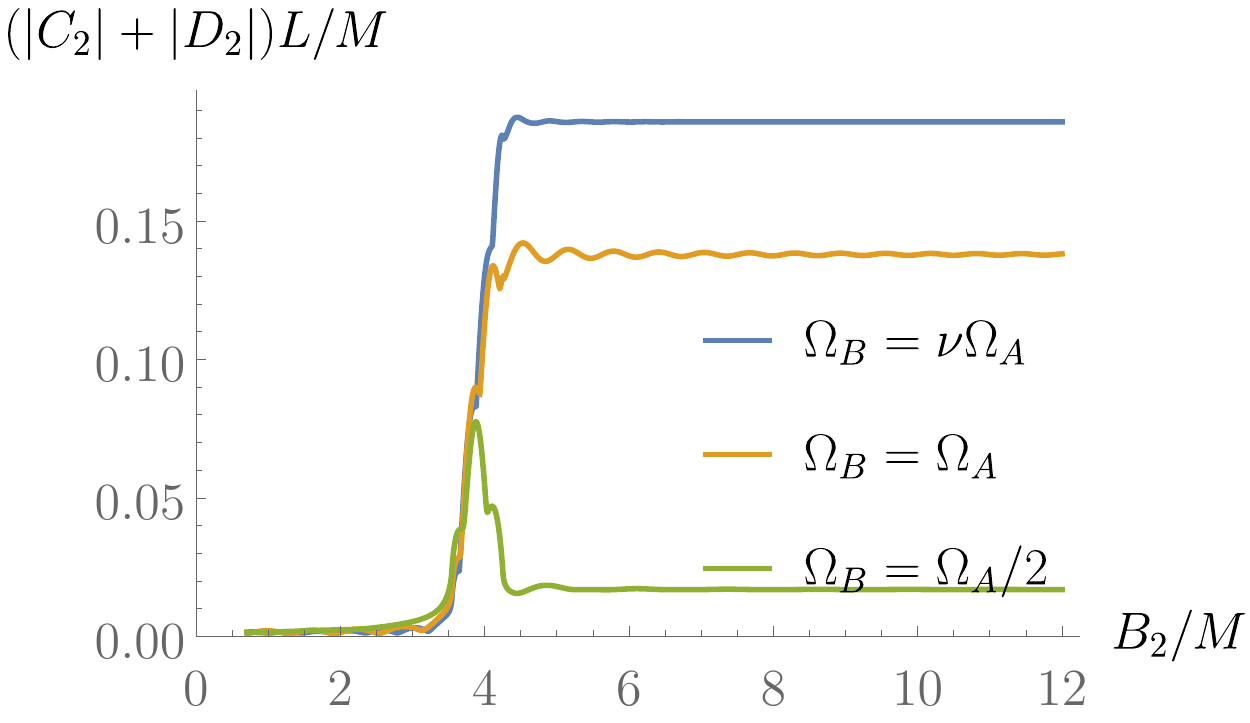}}
    \subfloat[$\Omega_\da=10/M,A_2=3.5M$, %
    $(s_2+A_2)/\nu\approx 6.03M$.\label{fig:PV_Cumu_long}]{\includegraphics[width=0.33\textwidth]{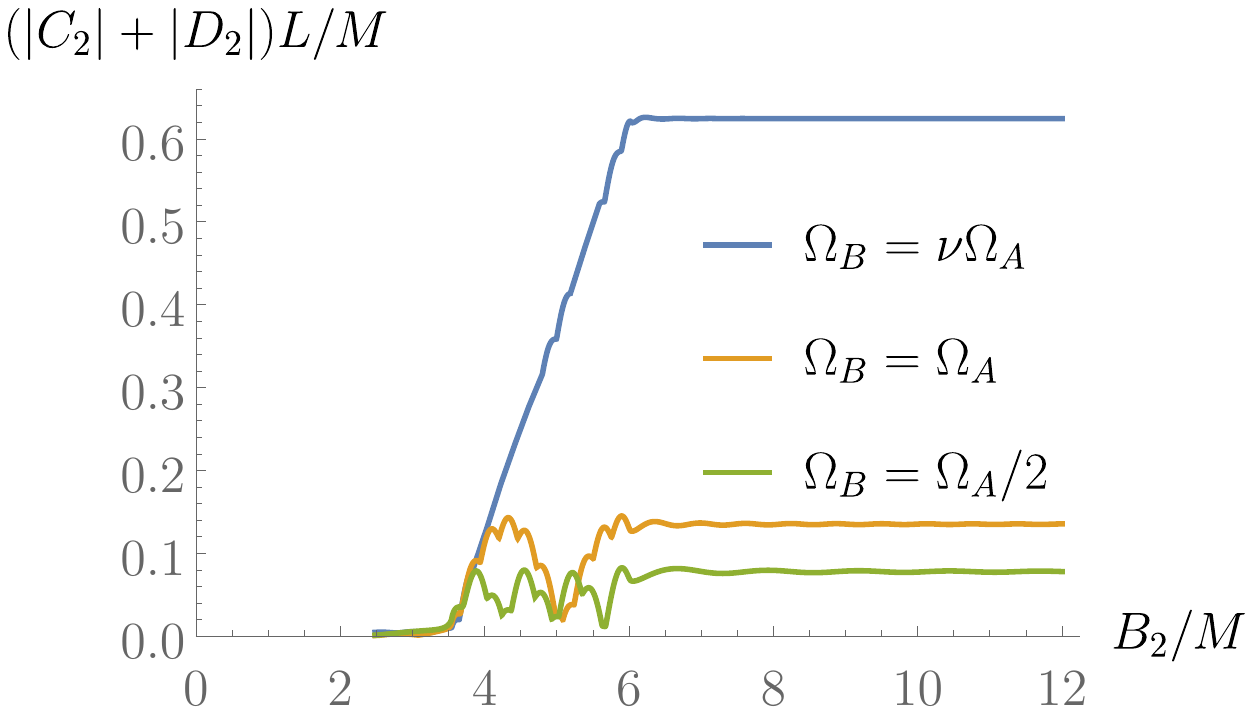}}    
\caption{Contribution to the leading order signal strength from $\pv\frac1\sigma$-distribution, as resulting from \eqref{eq:C2_pv_cumu}, with $\nu=\sqrt{(1-2/6)/(1-2/3.01)}\approx 1.40954$, $A_1=0$ and $s_2=5M$.
(For this cumulative signal strength we always have $B_1=A_1/\nu=0$.)
Alice's first light ray, emanating at $A_1=0$, is connected by the $\pv\frac1\sigma$-singularity to the point on Bob's worldline with proper time $\tau_\db=(s_2+A_1)/\nu\approx 3.55M$. Alice's last light ray, emanating at her proper time $A_2$, is connected to Bob's worldline at his proper time $\tau_\db=(s_2+A_2)/\nu$ which is different for the two figures.
The curve in Fig.~\ref{fig:PV_Cumu_short} with   $\Omega_\db= \Omega_\da/2$  reproduces the features discussed in Figs.~\ref{fig:C2,contour_new} and \ref{fig:C2D2_NonDirect_RainbowPlots}.
In all three figures the oscillations in all curves decay as $B_2\to\infty$ and  asymptote to   constant values.
(Note for non-colour print: The curve in Figs.~\ref{fig:PV_Cumu_medium} and \ref{fig:PV_Cumu_long} asymptote in the same order as they appear in the legend.)}
\label{fig:PV_Cumu} 
\end{figure}

Similarly, we can also isolate the contribution from secondary null geodesics in the plots in Figs. \ref{fig:totalsignal_static} and \ref{fig:C2D2_NonDirect_RainbowPlots}.
In these plots we have $B_1=A_1/\nu$, i.e., Bob always switches on his detector when the first signal from Alice arrives. (Note that this signal is predominantly  carried by the $\delta(\sigma)$-contribution from primary null geodesics, which we discard here.)
Then we ask how the signal strength depends on $B_2$, i.e., the point in time at which Bob switches his detector off again.
For reasons of simplicity, let us assume that Bob couples at least  for a time such that $B_2\geq A_2/\nu$. Then the contribution to the signal strength from the $\pv\frac1\sigma$-distribution is, for general detector frequencies,
\begin{align}\label{eq:C2_pv_cumu}
  C_2&=\frac{-\ii}{4\pi L}   \integral{s}{\max\left[\nu B_1-A_2,0\right]}{\nu B_2-A_1} \ee{\ii\, \Omega_\da s}     
     \pv\frac1{s-s_2}
    \integral{\tau_\db}{(s+A_1)/\nu}{\min\left[ B_2, (s+A_2)/\nu\right]}  \ee{\ii\, (\Omega_\db-\nu\Omega_\da) \tau_\db}\nn
  &=\frac{-1}{4\pi L( \Omega_\db-\nu \Omega_\da)} \left( \ee{\ii (\Omega_\db/\nu- \Omega_\da)A_2}R( \Omega_\db/\nu,0,s_n,s_2)-\ee{\ii (\Omega_\db/\nu- \Omega_\da)A_1} R(\Omega_\db/\nu,0,s_n,s_2) \right.\nn 
	&\qquad\left. +\ee{\ii (\Omega_\db-\nu \Omega_\da)B_2} R(\Omega_\da,s_n,\nu B_2-A_1,s_2) - \ee{\ii (\Omega_\db/\nu- \Omega_\da)A_1} R( \Omega_\db/\nu,s_n,\nu B_2-A_1,s_2)  \right),
\end{align}
where $s_n:=\nu B_2-A_2$, and for resonant detectors, i.e., $\Omega_\db=\nu \Omega_\da$, it is
\begin{align}
 C_2=\frac{-\ii}{4\pi L \nu}\left( (A_2-A_1)  R(\Omega_\da,0,s_n,s_2) +\left(B_2- A_1-s_2\right)  R( \Omega_\da,s_n ,\nu B_2-A_1,s_2) +\ii \left(\ee{\ii \Omega_\da (\nu B_2-A_1)}-\ee{\ii \Omega_\da s_n}\right) \right).
\end{align}
Fig.~\ref{fig:PV_Cumu} plots the resulting signal strength in various scenarios. 
In particular, Fig.~\ref{fig:PV_Cumu_short}  reproduces the features discussed in Figs.~\ref{fig:C2,contour_new} and \ref{fig:C2D2_NonDirect_RainbowPlots}.
The plots show that the resulting signal strength changes significanlty as a function of the switch-off time, if the switch-off happens while secondary null geodesics arrive at Bob's position that emanated from Alice while she was coupled to the field. For non-resonant detectors the signal strength exhibits an oscillatory and periodic behaviour within this time interval. For resonant detectors, however, the signal strength grows roughly linearly in this region.
After this region, where the switch-off time is such that all secondary null geodesics that emanate from Alice arrive at Bob while he is coupled to the field, the signal strength only shows an oscillatory behaviour for later switch-off times. The oscillations decay and asymptote to a final value as $B_2\to\infty$. The limit value appears to be determined by the duration of the original signal emitted by Alice.
\end{widetext}

\bibliography{schwarzschild_refs}

\end{document}